\documentclass[12pt]{article}
\usepackage{amssymb}

\begin{document}

\author{C. Bizdadea\thanks{%
e-mail address: bizdadea@central.ucv.ro}, E. M. Cioroianu\thanks{%
e-mail address: manache@central.ucv.ro}, A. C. Lungu\thanks{%
e-mail address: ac\_lungu@yahoo.com} \\
Faculty of Physics, University of Craiova\\
13 A. I. Cuza Str., Craiova 200585, Romania}
\title{No interactions for a collection of Weyl gravitons intermediated by a scalar
field}
\maketitle

\begin{abstract}
The cross-couplings among several Weyl gravitons (described in the free
limit by a sum of linearized Weyl actions) in the presence of a scalar field
are studied with the help of the deformation theory based on local BRST
cohomology. Under the hypotheses of locality, smoothness of the interactions
in the coupling constant, Poincar\'{e} invariance, (background) Lorentz
invariance, and the preservation of the number of derivatives on each field,
together with the supplementary assumption that the internal metric defined
by the sum of Weyl lagrangians is positively defined, we prove that there
are no consistent cross-interactions among different Weyl gravitons in the
presence of a scalar field. The couplings of a single Weyl graviton to a
scalar field are also discussed.

\end{abstract}

\section{Introduction}

The study of Weyl gravitons is important in view of the remarkable
properties of conformal supergravity~\cite{fradtseyt}, as well as by the
renewed interest in Weyl gravity~\cite{0007211} in connection with the
ADS/CFT correspondence. Recently, the impossibility of direct
cross-interactions among several Weyl gravitons under certain assumptions
has been proved in~\cite{marcann} by means of a cohomological approach based
on the lagrangian BRST symmetry~\cite
{brst1,brst2,batvilk1,batvilk2,batvilk3,henproc,hencarte}.

The purpose of this paper is to analyze the indirect cross-couplings among
several Weyl gravitons (described in the free limit by a sum of linearized
Weyl actions) in the presence of a scalar field. Thus, under the hypotheses
of locality, smoothness of the interactions in the coupling constant,
Poincar\'{e} invariance, (background) Lorentz invariance, and the
preservation of the number of derivatives on each field, together with the
supplementary assumption that the internal metric defined by the sum of Weyl
lagrangians is positively defined, we prove that there are no consistent
cross-interactions among different Weyl gravitons in the presence of a
scalar field. Our result is obtained in the context of the deformation
technique~\cite{def} combined with the local BRST cohomology~\cite{gen1}.

In order to make the presentation as clear as possible, we initially
consider the case of the couplings between a single Weyl field and a scalar
field, and compute the interaction terms to order two in the coupling
constant. In this manner we obtain that the first two orders of the
interacting lagrangian resulting from our setting originate in the
development of the full interacting lagrangian (in four spacetime
dimensions)
\[
\mathcal{L}^{(\mathrm{int})}=\sqrt{-g}\left[ \frac{1}{2}\left( g^{\mu \nu
}\left( \partial _{\mu }\phi \right) \partial _{\nu }\phi +\frac{1}{6}\phi
^{2}R\right) +\lambda \kappa \phi ^{4}\right] ,
\]
where $g_{\mu \nu }$ is the full metric, $g^{\mu \nu }$ denotes its inverse,
$R$ represents the full scalar curvature, $\kappa $ is an arbitrary real
constant, and $\lambda $ is the coupling constant. The term $\lambda \kappa
\phi ^{4}$ is usually omitted in the literature. It appears for instance in
the partial gauge-fixing procedure with $\phi =1$ \cite{vanproeyen} that
leads from $\mathcal{L}^{(\mathrm{int})}$ to the standard Einstein-Hilbert
action with a cosmological term. This term is consistent with the gauge
symmetries of the lagrangian $\mathcal{L}^{\left( \mathrm{W}\right) }+%
\mathcal{L}^{\left( \mathrm{int}\right) }$, where $\mathcal{L}^{\left(
\mathrm{W}\right) }$ is the full Weyl lagrangian. Based on this result, we
begin with a finite sum of linearized Weyl actions and a scalar field, and
prove that there are no consistent cross-interactions between different Weyl
gravitons in the presence of a scalar field under the hypotheses explained
in the above.

This paper is organized in seven sections. In Section 2 we construct the
BRST symmetry of a free model with a single Weyl field and a scalar field.
Section 3 briefly addresses the deformation procedure based on BRST
symmetry. In Section 4 we compute the first two orders of the interactions
between one Weyl graviton and a scalar field. In Section 5 we analyze the
deformed theory obtained in the previous section. Section 6 is devoted to
the proof of the fact that there are no consistent cross-interactions among
different Weyl gravitons in the presence of a scalar field. Section 7
exposes the main conclusions of the paper. The paper also contains one
appendix section, in which a statement mentioned in the body of the paper is
proved.

\section{Free model: lagrangian formulation\newline
and BRST symmetry}

Our starting point is represented by a free lagrangian action, written as
the sum between the linearized Weyl gravity action and the action for a
massive real scalar field
\begin{eqnarray}
S_{0}^{\mathrm{L}}\left[ h_{\mu \nu },\phi \right] &=&\frac{1}{2}\int
d^{4}x\left[ \mathcal{W}_{\mu \nu \alpha \beta }\mathcal{W}^{\mu \nu \alpha
\beta }+\left( \partial _{\mu }\phi \right) \left( \partial ^{\mu }\phi
\right) -m^{2}\phi ^{2}\right] \equiv  \nonumber \\
&\equiv &\int d^{4}x\left( \mathcal{L}_{0}^{\left( \mathrm{W}\right) }+%
\mathcal{L}_{0}^{\left( \mathrm{\phi }\right) }\right) ,  \label{fract}
\end{eqnarray}
where $\mathcal{W}_{\mu \nu \alpha \beta }$ is the linearized Weyl tensor in
four spacetime dimensions, given in terms of the linearized Riemann tensor $%
\mathcal{R}_{\mu \nu \alpha \beta }$ and of its traces by
\begin{equation}
\mathcal{W}_{\mu \nu \alpha \beta }=\mathcal{R}_{\mu \nu \alpha \beta }-%
\frac{1}{2}\left( \sigma _{\mu [\alpha }\mathcal{R}_{\beta ]\nu }-\sigma
_{\nu [\alpha }\mathcal{R}_{\beta ]\mu }\right) +\frac{1}{6}\mathcal{R}%
\sigma _{\mu [\alpha }\sigma _{\beta ]\nu }.  \label{1}
\end{equation}
Throughout the paper we work with the flat metric of `mostly plus' signature
$\sigma _{\mu \nu }=\left( -+++\right) $. The notation $\left[ \mu \cdots
\nu \right] $ signifies antisymmetrization with respect to all indices
between brackets without normalization factors (i.e., the independent terms
appear only once and are not multiplied by overall numerical factors). The
linearized Riemann tensor is expressed by
\begin{equation}
\mathcal{R}_{\mu \nu \alpha \beta }=\frac{1}{2}\left( \partial _{\mu
}\partial _{\beta }h_{\nu \alpha }+\partial _{\nu }\partial _{\alpha }h_{\mu
\beta }-\partial _{\nu }\partial _{\beta }h_{\mu \alpha }-\partial _{\mu
}\partial _{\alpha }h_{\nu \beta }\right) ,  \label{2}
\end{equation}
while its simple and respectively double traces read as
\begin{equation}
\mathcal{R}_{\mu \nu }=\sigma ^{\alpha \beta }\mathcal{R}_{\mu \alpha \nu
\beta },\qquad \mathcal{R}=\sigma ^{\mu \nu }\mathcal{R}_{\mu \nu }.
\label{3}
\end{equation}
The linearized Weyl tensor can be expressed in terms of the symmetric tensor
$\mathcal{K}_{\mu \nu }$ like
\begin{equation}
\mathcal{W}_{\mu \nu \alpha \beta }=\mathcal{R}_{\mu \nu \alpha \beta
}-\left( \sigma _{\mu [\alpha }\mathcal{K}_{\beta ]\nu }-\sigma _{\nu
[\alpha }\mathcal{K}_{\beta ]\mu }\right) ,  \label{3a}
\end{equation}
where
\begin{equation}
\mathcal{K}_{\mu \nu }=\frac{1}{2}\left( \mathcal{R}_{\mu \nu }-\frac{1}{6}%
\sigma _{\mu \nu }\mathcal{R}\right) ,\qquad \mathcal{K=}\frac{1}{6}\mathcal{%
R}.  \label{3b}
\end{equation}
The theory described by the action (\ref{fract}) possesses an irreducible
and abelian generating set of gauge transformations
\begin{equation}
\delta _{\epsilon }h_{\mu \nu }=\partial _{(\mu }\epsilon _{\nu )}+2\sigma
_{\mu \nu }\epsilon ,\qquad \delta _{\epsilon }\phi =0,  \label{5and6}
\end{equation}
where the gauge parameters $\epsilon _{\mu }$ and $\epsilon $ are bosonic
functions. The scalar gauge parameter $\epsilon $ is responsible for the
so-called conformal invariance of Weyl theory, while $\epsilon _{\mu }$
appear also in the Pauli-Fierz theory and will be called `Pauli-Fierz' gauge
parameters. The notation $\left( \mu \nu \right) $ signifies symmetry with
respect to the indices between parentheses without the factor $1/2$.

In order to construct the BRST symmetry for the model under study,
we introduce the fermionic ghosts $\eta _{\mu }$ and $\xi $
respectively associated with the gauge parameters $\epsilon _{\mu }$
and $\epsilon $. The antifield spectrum is organized into the
antifields $\left\{ h^{*\mu \nu },\phi ^{*}\right\} $ (of the
original fields $\left\{ h_{\mu \nu },\phi \right\} $) and $\left\{
\eta ^{*\mu },\xi ^{*}\right\} $ (of the ghosts $\left\{ \eta _{\mu
}, \xi \right\} $), with the statistics of the antifields opposite
to that of the associated fields/ghosts.

Since the gauge generators of the free theory are field independent, it
follows that the BRST differential simply reduces to
\begin{equation}
s=\delta +\gamma ,  \label{7}
\end{equation}
where $\delta $ represents the Koszul-Tate differential, graded by the
antighost number $\mathrm{agh}$ ($\mathrm{agh}\left( \delta \right) =-1$),
and $\gamma $ stands for the exterior derivative along the gauge orbits,
whose degree is named pure ghost number $\mathrm{pgh}$ ($\mathrm{pgh}\left(
\gamma \right) =1$). These two degrees do not interfere ($\mathrm{pgh}\left(
\delta \right) =0$, $\mathrm{agh}\left( \gamma \right) =0$). The overall
degree that grades the BRST complex is known as the ghost number $\mathrm{gh}
$ and is defined like the difference between the pure ghost number and the
antighost number, such that $\mathrm{gh}\left( \delta \right) =\mathrm{gh}%
\left( \gamma \right) =1$. If we denote by
\begin{equation}
\Phi ^{\alpha _{0}}=\left( h_{\mu \nu },\phi \right) ,\qquad \eta ^{\alpha
_{1}}=\left( \eta _{\mu },\xi \right)  \label{8}
\end{equation}
the fields and ghosts of the free theory ($\eta _{\mu }$ will be called the
`Pauli-Fierz' ghosts), and by
\begin{equation}
\Phi _{\alpha _{0}}^{*}=\left( h^{*\mu \nu },\phi ^{*}\right) ,\qquad \eta
_{\alpha _{1}}^{*}=\left( \eta ^{*\mu },\xi ^{*}\right)  \label{9}
\end{equation}
the corresponding antifields, then, according to the standard rules of the
BRST formalism, the corresponding degrees of the generators from the BRST
complex are valued like
\begin{eqnarray}
\mathrm{agh}\left( \Phi ^{\alpha _{0}}\right) &=&0,\qquad \mathrm{agh}\left(
\eta ^{\alpha _{1}}\right) =0,  \label{10} \\
\mathrm{agh}\left( \Phi _{\alpha _{0}}^{*}\right) &=&1,\qquad \mathrm{agh}%
\left( \eta _{\alpha _{1}}^{*}\right) =2,  \label{10aaa} \\
\mathrm{pgh}\left( \Phi ^{\alpha _{0}}\right) &=&0,\qquad \mathrm{pgh}\left(
\eta ^{\alpha _{1}}\right) =1,  \label{11aaa} \\
\mathrm{pgh}\left( \Phi _{\alpha _{0}}^{*}\right) &=&0,\qquad \mathrm{pgh}%
\left( \eta _{\alpha _{1}}^{*}\right) =0.  \label{11}
\end{eqnarray}
The actions of the differentials $\delta $ and $\gamma $ on the generators (%
\ref{8})--(\ref{9}) from the BRST complex are given by
\begin{eqnarray}
\delta h^{*\mu \nu } &=&2\partial _{\alpha }\partial _{\beta }\mathcal{W}%
^{\mu \alpha \nu \beta },\qquad \delta \phi ^{*}=\left( \Box +m^{2}\right)
\phi ,  \label{12} \\
\delta \eta ^{*\mu } &=&-2\partial _{\nu }h^{*\mu \nu },\qquad \delta \xi
^{*}=2h^{*},  \label{13} \\
\delta \Phi ^{\alpha _{0}} &=&0,\qquad \delta \eta ^{\alpha _{1}}=0,
\label{14} \\
\gamma \Phi _{\alpha _{0}}^{*} &=&0,\qquad \gamma \eta _{\alpha _{1}}^{*}=0,
\label{15} \\
\gamma h_{\mu \nu } &=&\partial _{(\mu }\eta _{\nu )}+2\sigma _{\mu \nu }\xi
,\qquad \gamma \phi =0,  \label{16} \\
\gamma \eta _{\mu } &=&0,\qquad \gamma \xi =0,  \label{17}
\end{eqnarray}
being understood that both operators act like right derivations. The
notation $h^{*}$ signifies the trace of $h^{*\mu \nu }$,
$h^{*}=\sigma _{\mu \nu }h^{*\mu \nu }$. The BRST differential is
known to have a canonical action in a structure named antibracket
and denoted by the symbol $\left( ,\right) $ ($s\cdot =\left( \cdot
,\bar{S}\right) $), which is obtained by setting the fields and
ghosts respectively conjugated to the corresponding antifields. The
generator of the BRST symmetry is a bosonic functional, of
ghost number zero, which is solution to the classical master equation $%
\left( \bar{S},\bar{S}\right) =0$. In our case the solution to the master
equation reads as
\begin{equation}
\bar{S}=S_{0}^{\mathrm{L}}\left[ h_{\mu \nu },\phi \right] +\int
d^{4}x\,h^{*\mu \nu }\left( \partial _{(\mu }\eta _{\nu )}+2\sigma _{\mu \nu
}\xi \right) .  \label{18}
\end{equation}

\section{Deformation of the solution to the master equation: a brief review}

We begin with a ``free'' gauge theory, described by a lagrangian action $%
S_{0}^{\mathrm{L}}\left[ \Phi ^{\alpha _{0}}\right] $, invariant under some
gauge transformations
\begin{equation}
\delta _{\epsilon }\Phi ^{\alpha _{0}}=Z_{\;\;\alpha _{1}}^{\alpha
_{0}}\epsilon ^{\alpha _{1}},\qquad \frac{\delta S_{0}^{\mathrm{L}}}{\delta
\Phi ^{\alpha _{0}}}Z_{\;\;\alpha _{1}}^{\alpha _{0}}=0,  \label{2.1}
\end{equation}
and consider the problem of constructing consistent interactions among the
fields $\Phi ^{\alpha _{0}}$ such that the couplings preserve the field
spectrum and the original number of gauge symmetries. This matter is
addressed by means of reformulating the problem of constructing consistent
interactions as a deformation problem of the solution to the master equation
corresponding to the ``free'' theory \cite{def}. Such a reformulation is
possible due to the fact that the solution to the master equation contains
all the information on the gauge structure of the theory. If an interacting
gauge theory can be consistently constructed, then the solution $\bar{S}$ to
the master equation associated with the ``free'' theory, $\left( \bar{S},%
\bar{S}\right) =0$, can be deformed into a solution $S$
\begin{eqnarray}
\bar{S}\rightarrow S &=&\bar{S}+\lambda S_{1}+\lambda ^{2}S_{2}+\cdots =
\nonumber \\
&=&\bar{S}+\lambda \int d^{D}x\,a+\lambda ^{2}\int d^{D}x\,b+\cdots
\label{2.2}
\end{eqnarray}
of the master equation for the deformed theory
\begin{equation}
\left( S,S\right) =0,  \label{2.3}
\end{equation}
such that both the ghost and antifield spectra of the initial theory are
preserved. The equation (\ref{2.3}) splits, according to the various orders
in the coupling constant (deformation parameter) $\lambda $, into
\begin{eqnarray}
\left( \bar{S},\bar{S}\right) &=&0  \label{2.4} \\
2\left( S_{1},\bar{S}\right) &=&0  \label{2.5} \\
2\left( S_{2},\bar{S}\right) +\left( S_{1},S_{1}\right) &=&0  \label{2.6} \\
\left( S_{3},\bar{S}\right) +\left( S_{1},S_{2}\right) &=&0  \label{2.7} \\
&&\vdots  \nonumber
\end{eqnarray}

The equation (\ref{2.4}) is fulfilled by hypothesis. The next one requires
that the first-order deformation of the solution to the master equation, $%
S_{1}$, is a co-cycle of the ``free'' BRST differential $s$, $sS_{1}=0$.
However, only cohomologically nontrivial solutions to (\ref{2.5}) should be
taken into account, as the BRST-exact ones can be eliminated by some (in
general nonlinear) field redefinitions. This means that $S_{1}$ pertains to
the ghost number zero cohomological space of $s$, $H^{0}\left( s\right) $,
which is generically nonempty due to its isomorphism to the space of
physical observables of the ``free'' theory. It has been shown (on behalf of
the triviality of the antibracket map in the cohomology of the BRST
differential) that there are no obstructions in finding solutions to the
remaining equations ((\ref{2.6})--(\ref{2.7}), etc.). However, the resulting
interactions may be nonlocal, and there might even appear obstructions if
one insists on their locality. The analysis of these obstructions can be
done with the help of cohomological techniques.

\section{Consistent interactions between the Weyl graviton and the real
scalar field}

\label{cons}

The aim of this section is to investigate the cross-couplings that can be
introduced between a single Weyl graviton and a scalar field. This matter is
addressed in the context of the antifield-BRST deformation procedure
described in the above and relies on computing the solutions to the
equations (\ref{2.5})--(\ref{2.7}), etc., with the help of the BRST
cohomology of the free theory. The interactions are obtained under the
following (reasonable) assumptions: smoothness in the deformation parameter,
locality, (background) Lorentz invariance, Poincar\'{e} invariance, and
preservation of the number of derivatives on each field. `Smoothness of the
deformations' refers to the fact that the deformed solution to the master
equation (\ref{2.2}) is smooth in the coupling constant $\lambda $ and
reduces to the original solution (\ref{18}) in the free limit $\lambda =0$.
The requirement that the interacting theory is Poincar\'{e} invariant means
that one does not allow an explicit dependence of the deformed solution to
the master equation on the spacetime coordinates. The conservation of the
number of derivatives on each field with respect to the free theory means
here that the following two requirements are simultaneously satisfied: (i)
the derivative order of the equations of motion on each field is the same
for the free and respectively for the interacting theory; (ii) the maximum
number of derivatives in the interaction vertices is equal to four, i.e. the
maximum number of derivatives from the free lagrangian. Conditions of this
type are frequently imposed in the literature at the level of constructing
interacting theories; for instance, see the case of cross-interactions for a
collection of Pauli-Fierz fields~\cite{multi}, the couplings between the
Pauli-Fierz and the massless Rarita-Schwinger fields~\cite{boulcqg}, or the
direct cross-interactions for a collection of Weyl gravitons~\cite{marcann}.

\subsection{Standard material: $H\left( \gamma \right) $ and $H\left( \delta
|d\right) $}

\label{stand}

The equation (\ref{2.5}), which we have seen that controls the first-order
deformation, takes the local form
\begin{equation}
sa=\partial _{\mu }m^{\mu },\qquad \mathrm{gh}\left( a\right) =0,\qquad
\varepsilon \left( a\right) =0,  \label{3.1}
\end{equation}
for some local $m^{\mu }$, and it shows that the nonintegrated density of
the first-order deformation pertains to the local cohomology of the free
BRST\ differential in ghost number zero, $a\in H^{0}\left( s|d\right) $,
where $d$ denotes the exterior spacetime differential. The solution to the
equation (\ref{3.1}) is unique up to $s$-exact pieces plus divergences
\begin{equation}
a\rightarrow a+sb+\partial _{\mu }n^{\mu },  \label{3.1a}
\end{equation}
where
\[
\mathrm{gh}\left( b\right) =-1,\qquad \varepsilon \left( b\right) =1,\qquad
\mathrm{gh}\left( n^{\mu }\right) =0,\qquad \varepsilon \left( n^{\mu
}\right) =0.
\]
At the same time, if the general solution of (\ref{3.1}) is found to be
completely trivial, $a=sb+\partial _{\mu }n^{\mu }$, then it can be made to
vanish $a=0$.

In order to analyze the equation (\ref{3.1}), we develop $a$ according to
the antighost number
\begin{equation}
a=\sum\limits_{i=0}^{I}a_{i},\qquad \mathrm{agh}\left( a_{i}\right)
=i,\qquad \mathrm{gh}\left( a_{i}\right) =0,\qquad \varepsilon
\left( a_{i}\right) =0,  \label{3.2}
\end{equation}
and assume, without loss of generality, that the decomposition (\ref{3.2})
stops at some finite value of $I$. This can be shown for instance like in
Appendix A of~\cite{marcann}. Replacing the decomposition (\ref{3.2}) into
the equation (\ref{3.1}) and projecting it on the various values of the
antighost number by means of the splitting (\ref{7}), we obtain the tower of
equations
\begin{eqnarray}
\gamma a_{I} &=&\partial _{\mu }\stackrel{\left( I\right) }{m}^{\mu },
\label{3.3} \\
\delta a_{I}+\gamma a_{I-1} &=&\partial _{\mu }\stackrel{\left( I-1\right) }{%
m}^{\mu },  \label{3.4} \\
\delta a_{i}+\gamma a_{i-1} &=&\partial _{\mu }\stackrel{\left( i-1\right) }{%
m}^{\mu },\qquad 1\leq i\leq I-1,  \label{3.5}
\end{eqnarray}
where $\left( \stackrel{\left( i\right) }{m}^{\mu }\right) _{i=\overline{0,I}%
}$ are some local currents with $\mathrm{agh}\left( \stackrel{\left(
i\right) }{m}^{\mu }\right) =i$. Moreover, according to the general result
from~\cite{marcann} in the absence of collection indices, the equation (\ref
{3.3}) can be replaced\footnote{%
This is because the presence of matter fields (in our case a real scalar
field) does not modify the general results on $H\left( \gamma \right) $
presented in~\cite{marcann}.} in strictly positive antighost numbers by
\begin{equation}
\gamma a_{I}=0,\qquad I>0.  \label{3.6}
\end{equation}
Due to the second-order nilpotency of $\gamma $ ($\gamma ^{2}=0$), the
solution to the equation (\ref{3.6}) is unique up to $\gamma $-exact
contributions
\begin{equation}
a_{I}\rightarrow a_{I}+\gamma b_{I},\qquad \mathrm{agh}\left( b_{I}\right)
=I,\qquad \mathrm{pgh}\left( b_{I}\right) =I-1,\qquad \varepsilon \left(
b_{I}\right) =1.  \label{r68}
\end{equation}
Meanwhile, if it turns out that $a_{I}$ reduces to $\gamma $-exact terms
only, $a_{I}=\gamma b_{I}$, then it can be made to vanish, $a_{I}=0$. In
other words, the nontriviality of the first-order deformation $a$ is
translated at its highest antighost number component into the requirement
that $a_{I}\in H^{I}\left( \gamma \right) $, where $H^{I}\left( \gamma
\right) $ denotes the cohomology of the exterior longitudinal derivative $%
\gamma $ in pure ghost number equal to $I$. So, in order to solve the
equation (\ref{3.1}) (equivalent with (\ref{3.6}) and (\ref{3.4})--(\ref{3.5}%
)), we need to compute the cohomology of $\gamma $, $H\left( \gamma \right) $%
, and, as it will be made clear below, also the local cohomology of $\delta $%
, $H\left( \delta |d\right) $.

Using the results on the cohomology of $\gamma $ in the Weyl sector~\cite
{marcann} and the definitions of $\gamma $ acting on the scalar field $\phi $
and on its antifield $\phi ^{*}$, we can state that $H\left( \gamma \right) $
is generated on the one hand by $\Phi _{\alpha _{0}}^{*}$, $\eta _{\alpha
_{1}}^{*}$, $\phi $, and $\mathcal{W}_{\mu \nu \alpha \beta }$, as well as
by their spacetime derivatives, and, on the other hand, by the ghosts and
their first-order derivatives $\eta _{\mu }$, $\partial _{[\mu }\eta _{\nu
]} $, $\xi $, and $\partial _{\mu }\xi $. So, the most general (and
nontrivial), local solution to (\ref{3.6}) can be written, up to $\gamma $%
-exact contributions, as
\begin{equation}
a_{I}=\alpha _{I}\left( \left[ \phi \right] ,\left[ \mathcal{W}_{\mu \nu
\alpha \beta }\right] ,\left[ \Phi _{\alpha _{0}}^{*}\right] ,\left[ \eta
_{\alpha _{1}}^{*}\right] \right) \omega ^{I}\left( \eta _{\mu },\partial
_{[\mu }\eta _{\nu ]},\xi ,\partial _{\mu }\xi \right) ,  \label{3.10}
\end{equation}
where the notation $f\left( \left[ q\right] \right) $ means that $f$ depends
on $q$ and its derivatives up to a finite order, while $\omega ^{I}$ denotes
the elements with pure ghost number $I$ of a basis in the space of
polynomials in the corresponding ghosts and their first-order derivatives.
The objects $\alpha _{I}$ (obviously nontrivial in $H^{0}\left( \gamma
\right) $) were taken to have a finite antighost number and a bounded number
of derivatives, and therefore they are polynomials in the antifields $\Phi
_{\alpha _{0}}^{*}$ and $\eta _{\alpha _{1}}^{*}$, in the linearized Weyl
tensor $\mathcal{W}_{\mu \nu \alpha \beta }$, in the first-order derivatives
of the scalar field $\partial _{\mu }\phi $, as well as in their subsequent
derivatives. However, $\alpha _{I}$ may contain infinite, formal series in
the undifferentiated scalar field $\phi $. They are required to fulfill the
property $\mathrm{agh}\left( \alpha _{I}\right) =I$ in order to ensure that
the ghost number of $a_{I}$ is equal to zero. Due to their $\gamma $%
-closeness, $\gamma \alpha _{I}=0$ and to their (partial) polynomial
character, $\alpha _{I}$ will be called ``invariant polynomials''. In zero
antighost number the invariant polynomials are polynomials in the linearized
Weyl tensor, in its derivatives, and in the derivatives of the real scalar
field, with coefficients that may be infinite series in the undifferentiated
real scalar field $\phi $.

Substituting (\ref{3.10}) in (\ref{3.4}), we obtain that a necessary (but
not sufficient) condition for the existence of (nontrivial) solutions $%
a_{I-1}$ is that the invariant polynomials $\alpha _{I}$ are (nontrivial)
objects from the local cohomology of the Koszul-Tate differential $H\left(
\delta |d\right) $ in antighost number $I>0$ and in pure ghost number zero,
\begin{equation}
\delta \alpha _{I}=\partial _{\mu }\stackrel{\left( I-1\right) }{j}^{\mu
},\qquad \mathrm{agh}\left( \stackrel{\left( I-1\right) }{j}^{\mu }\right)
=I-1,\qquad \mathrm{pgh}\left( \stackrel{\left( I-1\right) }{j}^{\mu
}\right) =0.  \label{3.10a}
\end{equation}
We recall that the local cohomology $H\left( \delta |d\right) $ is
completely trivial in both strictly positive antighost \textit{and} pure
ghost number (for instance, see Theorem 5.4 in~\cite{gen1} and also~\cite
{commun1}). Using the fact that the Cauchy order of the free theory under
study is equal to two, together with the general results from~\cite{gen1},
according to which the local cohomology of the Koszul-Tate differential in
pure ghost number zero is trivial in antighost numbers strictly greater than
its Cauchy order, we can state that
\begin{equation}
H_{J}\left( \delta |d\right) =0\qquad \mathrm{for\;all\;}J>2,  \label{3.11}
\end{equation}
where $H_{J}\left( \delta |d\right) $ denotes the local cohomology of the
Koszul-Tate differential in antighost number $J$ and in pure ghost number
equal to zero. It has been shown in detail in~\cite{marcann} (Theorem 5.1)
that any invariant polynomial from the Weyl sector that is trivial in $%
H_{J}\left( \delta |d\right) $ with $J\geq 2$ can be taken to be trivial
also in $H_{J}^{\mathrm{inv}}\left( \delta |d\right) $. [$H_{J}^{\mathrm{inv}%
}\left( \delta |d\right) $ denotes the invariant characteristic cohomology
in antighost number $J$ (the local cohomology of the Koszul-Tate
differential in the space of invariant polynomials).] This property is still
valid here since the real scalar field has no gauge invariance of its own,
and thus the scalar field sector can intervene nontrivially only in the
cohomology of the Koszul-Tate differential in antighost number one. Thus:
\begin{equation}
\left( \alpha _{J}=\delta b_{J+1}+\partial _{\mu }\stackrel{(J)}{c}^{\mu },\;%
\mathrm{agh}\left( \alpha _{J}\right) =J\geq 2\right) \Rightarrow \alpha
_{J}=\delta \beta _{J+1}+\partial _{\mu }\stackrel{(J)}{\gamma }^{\mu },
\label{3.12ax}
\end{equation}
with both $\beta _{J+1}$ and $\stackrel{(J)}{\gamma }^{\mu }$ invariant
polynomials. The results (\ref{3.11}) and (\ref{3.12ax}) yield the
conclusion that
\begin{equation}
H_{J}^{\mathrm{inv}}\left( \delta |d\right) =0\qquad \mathrm{for\;all\;}J>2.
\label{3.12x}
\end{equation}
The antifield of the scalar field brings only trivial contributions to both $%
H_{J}\left( \delta |d\right) $ (in pure ghost number zero) and $H_{J}^{%
\mathrm{inv}}\left( \delta |d\right) $ for $J\geq 2$, so the results from~%
\cite{marcann} regarding $H_{2}\left( \delta |d\right) $ (in pure ghost
number zero) and $H_{2}^{\mathrm{inv}}\left( \delta |d\right) $ remain valid
here. Both cohomologies are still spanned by the undifferentiated antifields
corresponding to the `Pauli-Fierz' ghosts
\begin{equation}
H_{2}\left( \delta |d\right) \qquad \mathrm{and}\qquad H_{2}^{\mathrm{inv}%
}\left( \delta |d\right) :\left( \eta ^{*\mu }\right) .  \label{3.12b}
\end{equation}
In contrast to the groups $\left( H_{J}\left( \delta |d\right) \right)
_{J\geq 2}$ and $\left( H_{J}^{\mathrm{inv}}\left( \delta |d\right) \right)
_{J\geq 2}$, which are finite-dimensional, the cohomology $H_{1}\left(
\delta |d\right) $ in pure ghost number zero, that is related to global
symmetries and ordinary conservation laws, is infinite-dimensional since the
theory is free. Fortunately, it will not be needed in the sequel.

The previous results on $H\left( \delta |d\right) $ and $H^{\mathrm{inv}%
}\left( \delta |d\right) $ in strictly positive antighost numbers are
important because they control the obstructions to removing the antifields
from the first-order deformation. Based on the formulas (\ref{3.11})--(\ref
{3.12x}), one can successively eliminate all the pieces of antighost number
strictly greater that two from the nonintegrated density of the first-order
deformation by adding only trivial terms, so one can take, without loss of
nontrivial objects, the condition $I\leq 2$ in the decomposition (\ref{3.2}%
). The proof of this statement can be realized like in the subsection 6.1
from~\cite{marcann}.

\subsection{First-order deformation}

\label{firstord}

In the case $I=2$ the nonintegrated density of the first-order deformation (%
\ref{3.2}) becomes
\begin{equation}
a=a_{0}+a_{1}+a_{2}.  \label{3.12}
\end{equation}
We can further decompose $a$ in a natural manner as
\begin{equation}
a=a^{\left( \mathrm{W}\right) }+a^{\left( \mathrm{int}\right) }+a^{\left(
\mathrm{\phi }\right) },  \label{3.12a}
\end{equation}
where $a^{\left( \mathrm{W}\right) }$ contains only fields/ghosts/antifields
from the Weyl sector, $a^{\left( \mathrm{int}\right) }$ describes the
cross-interactions between the Weyl graviton and the scalar field (so it
effectively mixes both sectors), and $a^{\left( \mathrm{\phi }\right) }$
involves only the scalar field sector. The component $a^{\left( \mathrm{W}%
\right) }$ is completely known~\cite{marcann} and satisfies individually an
equation of the type (\ref{3.1}). It admits a decomposition similar to (\ref
{3.12})
\begin{equation}
a^{\left( \mathrm{W}\right) }=a_{0}^{\left( \mathrm{W}\right)
}+a_{1}^{\left( \mathrm{W}\right) }+a_{2}^{\left( \mathrm{W}\right) },
\label{3.12w}
\end{equation}
where
\begin{equation}
a_{2}^{(\mathrm{W})}=\eta ^{*\mu }\left( \frac{1}{2}\eta ^{\nu }\partial
_{[\mu }\eta _{\nu ]}+\eta _{\mu }\xi \right) -\xi ^{*}\eta _{\mu }\partial
^{\mu }\xi ,  \label{w1}
\end{equation}
\begin{eqnarray}
a_{1}^{(\mathrm{W})} &=&-\frac{1}{2}h^{*\mu \nu }\eta ^{\rho }\left(
\partial _{(\mu }h_{\nu )\rho }-2\partial _{\rho }h_{\mu \nu }\right) +
\nonumber \\
&&+2h^{*\mu \nu }h_{\mu \nu }\xi +\frac{1}{2}h^{*\mu \nu }h_{\rho (\mu
}\partial _{\nu )}\eta ^{\rho },  \label{w4}
\end{eqnarray}
and $a_{0}^{(\mathrm{W})}$ is the cubic vertex of the Weyl lagrangian%
\footnote{%
The terms $a_{2}^{\left( \mathrm{W}\right) }$ and $a_{1}^{\left( \mathrm{W}%
\right) }$ given in (\ref{w1}) and (\ref{w4}) differ from the corresponding
ones in~\cite{marcann} by a $\gamma $-exact and respectively a $\delta $%
-exact contribution. However, the difference between our $a_{2}^{\left(
\mathrm{W}\right) }+$ $a_{1}^{\left( \mathrm{W}\right) }$ and the
corresponding sum from~\cite{marcann} is a $s$-exact modulo $d$ quantity.
The associated component of antighost number zero, $a_{0}^{\left( \mathrm{W}%
\right) }$, is nevertheless the same in both formulations. As a consequence,
the object $a^{\left( \mathrm{W}\right) }$ and the first-order deformation
in~\cite{marcann} belong to the same cohomological class from $H^{0}\left(
s|d\right) $.}. Since $a^{\left( \mathrm{int}\right) }$ mixes variables from
Weyl and matter sectors, while $a^{\left( \mathrm{\phi }\right) }$ depends
only on the matter field, it follows that these are subject to two separate
equations
\begin{eqnarray}
sa^{\left( \mathrm{int}\right) } &=&\partial _{\mu }m^{\left( \mathrm{int}%
\right) \mu },  \label{uw1} \\
sa^{\left( \mathrm{\phi }\right) } &=&\partial _{\mu }m^{\left( \mathrm{\phi
}\right) \mu },  \label{uw2}
\end{eqnarray}
for some local $m^{\mu }$'s. In the sequel we analyze the general solutions
to these equations.

Like we mentioned before, the scalar field sector may appear nontrivially
only in antighost number one or zero, so one can always take $a_{2}^{\left(
\mathrm{int}\right) }=0$ and hence work with
\begin{equation}
a^{\left( \mathrm{int}\right) }=a_{0}^{\left( \mathrm{int}\right)
}+a_{1}^{\left( \mathrm{int}\right) },  \label{3.25}
\end{equation}
where the components of $a^{\left( \mathrm{int}\right) }$ are subject to the
equations
\begin{eqnarray}
\gamma a_{1}^{\left( \mathrm{int}\right) } &=&0,  \label{3.25q} \\
\delta a_{1}^{\left( \mathrm{int}\right) }+\gamma a_{0}^{\left( \mathrm{int}%
\right) } &=&\partial _{\mu }\stackrel{\left( 0\right) }{m}^{\left( \mathrm{%
int}\right) \mu }.  \label{3.25w}
\end{eqnarray}
According to (\ref{3.10}) in pure ghost number equal to one, it results that
the most general form of $a_{1}^{\left( \mathrm{int}\right) }$ as solution
to (\ref{3.25q}) that might provide effective cross-interactions is written
like
\begin{eqnarray}
a_{1}^{\left( \mathrm{int}\right) } &=&\phi ^{*}\left( M^{\mu }\partial
_{\mu }\xi +\bar{M}^{\mu \nu }\partial _{[\mu }\eta _{\nu ]}+\hat{M}\xi +%
\hat{M}^{\mu }\eta _{\mu }\right) +  \nonumber \\
&&+h^{*\alpha \beta }\left( M_{\alpha \beta }^{\mu }\partial _{\mu }\xi +%
\bar{M}_{\alpha \beta }^{\mu \nu }\partial _{[\mu }\eta _{\nu ]}+\hat{M}%
_{\alpha \beta }\xi +\hat{M}_{\alpha \beta }^{\mu }\eta _{\mu }\right) ,
\label{3.26}
\end{eqnarray}
where the $M$-like functions may depend on the scalar field, on the
linearized Weyl tensor, as well as on their spacetime derivatives, and
satisfy obvious `symmetry' properties. Using the definitions of $\delta $
and $\gamma $, after some computations we obtain that
\begin{equation}
\delta a_{1}^{(\mathrm{int})}=\gamma b_{0}+\partial _{\mu }j_{0}^{\mu
}+c_{0},  \label{new2}
\end{equation}
where we used the notations
\begin{equation}
b_{0}=\sum\limits_{i=1}^{8}b_{0}^{(i)},  \label{new3}
\end{equation}
\begin{eqnarray}
j^{\mu } &=&\left[ \phi \partial ^{\mu }M^{\nu }-M^{\nu }\partial ^{\mu
}\phi +4\phi \bar{M}^{\mu \nu }+\sigma ^{\mu \nu }\phi \hat{M}-2M_{\alpha
\beta }^{\nu }\partial _{\rho }\mathcal{W}^{\mu \alpha \rho \beta }+\right.
\nonumber \\
&&\left. +2\mathcal{W}^{\mu \alpha \rho \beta }\left( \partial _{\rho
}M_{\alpha \beta }^{\nu }+4\sigma _{\rho \lambda }\bar{M}_{\alpha \beta
}^{\lambda \nu }+\delta _{\rho }^{\nu }\hat{M}_{\alpha \beta }\right)
\right] \partial _{\nu }\xi +  \nonumber \\
&&+\left[ \phi \left( \partial ^{\mu }M-\hat{M}^{\mu }\right) -\hat{M}%
\partial ^{\mu }\phi -2\hat{M}_{\alpha \beta }\partial _{\nu }\mathcal{W}%
^{\mu \alpha \nu \beta }+\right.  \nonumber \\
&&\left. +2\mathcal{W}^{\mu \alpha \nu \beta }\left( \partial _{\nu }\hat{M}%
_{\alpha \beta }-\sigma _{\nu \rho }\hat{M}_{\alpha \beta }^{\rho }\right)
\right] \xi +  \nonumber \\
&&+\left[ \phi \partial ^{\mu }\bar{M}^{\rho \lambda }-\bar{M}^{\rho \lambda
}\partial ^{\mu }\phi +\frac{1}{4}\sigma ^{\mu [\rho }\hat{M}^{\lambda
]}\phi -2\bar{M}_{\alpha \beta }^{\rho \lambda }\partial _{\nu }\mathcal{W}%
^{\mu \alpha \nu \beta }+\right.  \nonumber \\
&&\left. +2\mathcal{W}^{\mu \alpha \nu \beta }\left( \partial _{\nu }\bar{M}%
_{\alpha \beta }^{\rho \lambda }+\frac{1}{4}\delta _{\nu }^{[\rho }\hat{M}%
_{\alpha \beta }^{\lambda ]}\right) \right] \partial _{[\rho }\eta _{\lambda
]}+  \nonumber \\
&&+\left( \phi \partial ^{\mu }\hat{M}^{\lambda }-\hat{M}^{\lambda }\partial
^{\mu }\phi -2\hat{M}_{\alpha \beta }^{\lambda }\partial _{\nu }\mathcal{W}%
^{\mu \alpha \nu \beta }+2\mathcal{W}^{\mu \alpha \nu \beta }\partial _{\nu }%
\hat{M}_{\alpha \beta }^{\lambda }\right) \eta _{\lambda },  \label{new4}
\end{eqnarray}
\begin{eqnarray}
c_{0} &=&\left[ -\phi \left( \left( \square +m^{2}\right) M^{\lambda
}+8\partial _{\rho }\bar{M}^{\rho \lambda }+2\hat{M}^{\lambda }+2\partial
^{\lambda }\hat{M}\right) -\right.  \nonumber \\
&&-2\mathcal{W}^{\mu \alpha \nu \beta }\left( \partial _{\mu }\partial _{\nu
}M_{\alpha \beta }^{\lambda }+8\sigma _{\mu \rho }\partial _{\nu }\bar{M}%
_{\alpha \beta }^{\rho \lambda }+2\delta _{\nu }^{\lambda }\partial _{\mu }%
\hat{M}_{\alpha \beta }-\right.  \nonumber \\
&&\left. \left. -2\sigma _{\mu \rho }\delta _{\nu }^{\lambda }\hat{M}%
_{\alpha \beta }^{\rho }\right) \right] \partial _{\lambda }\xi +  \nonumber
\\
&&+\left[ -\phi \left( \left( \square +m^{2}\right) \hat{M}+2\partial
_{\lambda }\hat{M}^{\lambda }\right) -\right.  \nonumber \\
&&\left. -2\mathcal{W}^{\mu \alpha \nu \beta }\left( \partial _{\mu
}\partial _{\nu }\hat{M}_{\alpha \beta }-2\sigma _{\mu \rho }\partial _{\nu }%
\hat{M}_{\alpha \beta }^{\rho }\right) \right] \xi -  \nonumber \\
&&-\left[ \phi \left( \left( \square +m^{2}\right) \bar{M}^{\rho \lambda }+%
\frac{1}{2}\partial ^{[\rho }\hat{M}^{\lambda ]}\right) +\right.  \nonumber
\\
&&\left. +2\mathcal{W}^{\mu \alpha \nu \beta }\left( \partial _{\mu
}\partial _{\nu }\bar{M}_{\alpha \beta }^{\rho \lambda }+\delta _{\nu
}^{\rho }\partial _{\mu }\hat{M}_{\alpha \beta }^{\lambda }\right) \right]
\partial _{[\rho }\eta _{\lambda ]}-  \nonumber \\
&&-\left( \phi \left( \square +m^{2}\right) \hat{M}^{\lambda }+2\mathcal{W}%
^{\mu \alpha \nu \beta }\partial _{\mu }\partial _{\nu }\hat{M}_{\alpha
\beta }^{\lambda }\right) \eta _{\lambda },  \label{new5}
\end{eqnarray}
as well as
\begin{equation}
b_{0}^{(1)}=\left( \phi \partial ^{\mu }M^{\nu }-M^{\mu }\partial ^{\nu
}\phi \right) \mathcal{K}_{\mu \nu },  \label{new6a}
\end{equation}
\begin{equation}
b_{0}^{(2)}=\left( \bar{M}^{\mu \nu }\partial ^{\rho }\phi -\phi \partial
^{\rho }\bar{M}^{\mu \nu }\right) \partial _{[\mu }h_{\nu ]\rho },
\label{new6b}
\end{equation}
\begin{equation}
b_{0}^{(3)}=\phi \hat{M}\mathcal{K},  \label{new6c}
\end{equation}
\begin{equation}
b_{0}^{(4)}=\frac{1}{2}\left( \hat{M}^{\mu }\partial ^{\nu }\phi -\phi
\partial ^{\mu }\hat{M}^{\nu }\right) h_{\mu \nu }-\frac{1}{2}\phi \hat{M}%
^{\mu }\left( \partial ^{\nu }h_{\mu \nu }-\partial _{\mu }h\right) ,
\label{new6d}
\end{equation}
\begin{equation}
b_{0}^{(5)}=-2\left( M_{\alpha \beta }^{\rho }\partial _{\nu }\mathcal{W}%
^{\mu \alpha \nu \beta }-\mathcal{W}^{\mu \alpha \nu \beta }\partial _{\nu
}M_{\alpha \beta }^{\rho }\right) \mathcal{K}_{\mu \rho },  \label{new6e}
\end{equation}
\begin{eqnarray}
b_{0}^{(6)} &=&2\left( \bar{M}_{\alpha \beta }^{\rho \lambda }\partial _{\nu
}\mathcal{W}^{\mu \alpha \nu \beta }-\mathcal{W}^{\mu \alpha \nu \beta
}\partial _{\nu }\bar{M}_{\alpha \beta }^{\rho \lambda }\right) \partial
_{[\rho }h_{\lambda ]\mu }+  \nonumber \\
&&+8\mathcal{W}^{\mu \alpha \nu \beta }\sigma _{\mu \rho }\bar{M}_{\alpha
\beta }^{\rho \lambda }\mathcal{K}_{\lambda \nu },  \label{new6f}
\end{eqnarray}
\begin{equation}
b_{0}^{(7)}=2\mathcal{W}^{\mu \alpha \nu \beta }\hat{M}_{\alpha \beta }%
\mathcal{K}_{\mu \nu },  \label{new6g}
\end{equation}
\begin{eqnarray}
b_{0}^{(8)} &=&\left( \hat{M}_{\alpha \beta }^{\rho }\partial _{\nu }%
\mathcal{W}^{\mu \alpha \nu \beta }-\mathcal{W}^{\mu \alpha \nu \beta
}\partial _{\nu }\hat{M}_{\alpha \beta }^{\rho }\right) h_{\rho \mu }-
\nonumber \\
&&-\mathcal{W}^{\mu \alpha \nu \beta }\hat{M}_{\alpha \beta }^{\rho
}\partial _{[\mu }h_{\rho ]\nu }.  \label{new6h}
\end{eqnarray}
According to (\ref{3.25w}), the terms (\ref{new6a})--(\ref{new6h}) give, up
to a global factor, some of the pieces from the interacting Lagrangian at
order one in the coupling constant. The hypothesis on the conservation of
the number of derivatives on each field [conditions (i)--(ii) expressed in
the beginning of the section \ref{cons}] induces further restrictions on the
type-$M$ functions, as it will be seen below. The second term in (\ref{new6a}%
) outputs a field equation for the scalar field with three derivatives,
which disagrees with the condition (i). Therefore, we must set
\begin{equation}
M^{\mu }=0.  \label{cond1}
\end{equation}
The pieces (\ref{new6b}) satisfy the requirement (i) in relation to the
scalar field if the functions $\bar{M}^{\mu \nu }$ depend only on the
undifferentiated fields $h_{\alpha \beta }$ and $\phi $. Because $\bar{M}%
^{\mu \nu }$ are gauge invariant, it results that these functions actually
depend on the undifferentiated scalar field. Their antisymmetry property $%
\bar{M}^{\mu \nu }=-\bar{M}^{\nu \mu }$ prevents the appearance of such
functions, and thus we take
\begin{equation}
\bar{M}^{\mu \nu }=0.  \label{cond2}
\end{equation}
The most general form of (\ref{new6c}) satisfying (i) is $\hat{M}=M\left(
\phi \right) u\left( \mathcal{W}^{\mu \nu \alpha \beta }\mathcal{W}_{\mu \nu
\alpha \beta }\right) $, with $u$ and $M$ some arbitrary, smooth functions.
On the other hand, the condition (ii) prescribes that $\hat{M}$ comprises at
most two derivatives (as $\mathcal{K}$ already has two). This finally yields
$u=1$ and
\begin{equation}
\hat{M}=M\left( \phi \right) .  \label{cond3}
\end{equation}
The terms (\ref{new6d}) fulfill (i) in relation to the scalar field if the
functions $\hat{M}^{\mu }$ depend at most on the first-order derivatives of $%
\phi $ and $h_{\alpha \beta }$. Since the only gauge-invariant objects
constructed out of the graviton field contain at least two derivatives, it
follows that $\hat{M}^{\mu }$ can only be written as
\begin{equation}
\hat{M}^{\mu }=\partial ^{\mu }N\left( \phi \right) ,  \label{cond4}
\end{equation}
for some smooth function $N$. Due to the fact that (\ref{new6e})--(\ref
{new6f}) already produce a field equation for the scalar field with three
derivatives, the condition (i) imposes that we must take
\begin{equation}
M_{\alpha \beta }^{\rho }=0,\qquad \bar{M}_{\alpha \beta }^{\rho \lambda }=0.
\label{cond5}
\end{equation}
Regarding the element (\ref{new6g}), it outputs at most a second-order
derivative scalar field equation if the functions $\hat{M}_{\alpha \beta }$
are of the type
\begin{equation}
\hat{M}_{\alpha \beta }=Q\left( \phi \right) \mathcal{W}_{\alpha \mu \nu
\rho }\mathcal{W}_{\beta }^{\;\;\mu \nu \rho }w\left( \mathcal{W}_{\mu \nu
\rho \lambda }\mathcal{W}^{\mu \nu \rho \lambda }\right) ,  \label{aleg}
\end{equation}
where $Q$ and $w$ are smooth functions. If we introduce (\ref{aleg}) in (\ref
{new6g}), then the latter provides an interaction vertex of at least order
eight in the derivatives, in disagreement with the condition (ii).
Consequently, we have
\begin{equation}
\hat{M}_{\alpha \beta }=0.  \label{cond7}
\end{equation}
Finally, it is easy to see that (\ref{new6h}) provide a three-order
derivative equation for the scalar field, and therefore we must set
\begin{equation}
\hat{M}_{\alpha \beta }^{\rho }=0.  \label{cond8}
\end{equation}

Substituting the results (\ref{cond1})--(\ref{cond8}) in (\ref{3.26}) and (%
\ref{new3})--(\ref{new5}), we arrive at
\begin{equation}
a_{1}^{(\mathrm{int})}=\phi ^{*}\left[ M\xi +\left( \partial ^{\mu }N\right)
\eta _{\mu }\right] ,  \label{new7}
\end{equation}
\begin{eqnarray}
b_{0} &=&\phi M\left( \phi \right) \mathcal{K+}\frac{1}{2}\left[ \left(
\partial ^{\mu }N\right) \partial ^{\nu }\phi -\phi \partial ^{\mu }\partial
^{\nu }N\right] h_{\mu \nu }-  \nonumber \\
&&-\frac{1}{2}\phi \left( \partial ^{\mu }N\right) \left( \partial ^{\nu
}h_{\mu \nu }-\partial _{\mu }h\right) ,  \label{new9}
\end{eqnarray}
\begin{eqnarray}
j^{\mu } &=&\phi M\partial ^{\mu }\xi +\left[ \phi \partial ^{\mu }\partial
^{\nu }N-\left( \partial ^{\mu }\phi \right) \partial ^{\nu }N\right] \eta
_{\nu }+  \nonumber \\
&&+\left( \phi \partial ^{\mu }M-M\partial ^{\mu }\phi -\phi \partial ^{\mu
}N\right) \xi ,  \label{new8}
\end{eqnarray}
\begin{eqnarray}
c_{0} &=&-2\phi \partial ^{\mu }\left( M+N\right) \partial _{\mu }\xi -\phi
\left[ \left( \square +m^{2}\right) M-2\square N\right] \xi -  \nonumber \\
&&-\phi \left[ \left( \square +m^{2}\right) \partial ^{\mu }N\right] \eta
_{\mu }.  \label{new10}
\end{eqnarray}
After some manipulations, the terms (\ref{new10}) can be brought to
\begin{equation}
c_{0}=\gamma b_{0}^{\prime }+\partial _{\mu }j_{0}^{\prime \mu
}+c_{0}^{\prime },  \label{new11}
\end{equation}
where we employed the notations
\begin{eqnarray}
j_{0}^{\prime \mu } &=&\left( \partial ^{\mu }\phi \right) \left( \partial
^{\nu }\phi \right) \frac{dN\left( \phi \right) }{d\phi }\eta _{\nu }-\left[
\frac{1}{2}\left( \partial ^{\rho }\phi \right) \left( \partial _{\rho }\phi
\right) \frac{dN\left( \phi \right) }{d\phi }+\right.  \nonumber \\
&&\left. +\phi \square N+m^{2}\sum\limits_{n=2}^{\infty }\frac{\left(
-\right) ^{n}}{n!}\phi ^{n}\frac{d^{n-1}N\left( \phi \right) }{d\phi ^{n-1}}%
\right] \eta ^{\mu }-\bar{M}\left( \phi \right) \partial ^{\mu }\xi -
\nonumber \\
&&-\phi \xi \partial ^{\mu }\left( M+2N\right) ,  \label{new12a}
\end{eqnarray}
\begin{eqnarray}
b_{0}^{\prime } &=&\frac{1}{2}\left[ -\left( \partial ^{\mu }\phi \right)
\partial ^{\nu }N+\sigma ^{\mu \nu }\left( \frac{1}{2}\left( \partial ^{\rho
}\phi \right) \partial _{\rho }N+\phi \square N\right) \right] h_{\mu \nu }-
\nonumber \\
&&-\bar{M}\left( \phi \right) \mathcal{K}+\frac{m^{2}}{2}\left(
\sum\limits_{n=2}^{\infty }\frac{\left( -\right) ^{n}}{n!}\phi ^{n}\frac{%
d^{n-1}N\left( \phi \right) }{d\phi ^{n-1}}\right) h,  \label{new12b}
\end{eqnarray}
and
\begin{eqnarray}
c_{0}^{\prime } &=&\frac{1}{2}\left( \partial ^{\mu }\phi \right) \left(
\partial _{\mu }\phi \right) \left( \partial _{\nu }\phi \right) \frac{%
d^{2}N\left( \phi \right) }{d\phi ^{2}}\eta ^{\nu }+  \nonumber \\
&&+\left( \partial ^{\mu }\phi \right) \left( \partial _{\mu }\phi \right)
\frac{d\left( M\left( \phi \right) +N\left( \phi \right) \right) }{d\phi }%
\xi -  \nonumber \\
&&-m^{2}\left( \phi M\left( \phi \right) +4\sum\limits_{n=2}^{\infty }\frac{%
\left( -\right) ^{n}}{n!}\phi ^{n}\frac{d^{n-1}N\left( \phi \right) }{d\phi
^{n-1}}\right) \xi .  \label{new12c}
\end{eqnarray}
In the above $\bar{M}\left( \phi \right) $ is defined via the relation $d%
\bar{M}\left( \phi \right) /d\phi =\phi \left( dM\left( \phi \right) /d\phi
\right) $. Inserting (\ref{new11}) in (\ref{new2}), we deduce
\[
\delta a_{1}^{(\mathrm{int})}=\gamma \left( b_{0}+b_{0}^{\prime }\right)
+\partial _{\mu }\left( j_{0}^{\mu }+j_{0}^{\prime \mu }\right)
+c_{0}^{\prime }.
\]
Because the piece $c_{0}^{\prime }$ is nontrivial in $H\left( \gamma \right)
$ in pure ghost number one, the existence of $a_{0}^{(\mathrm{int})}$ as
solution to the equation (\ref{3.25w}) demands that $c_{0}^{\prime }$ must
vanish, which further implies the following relations that must be satisfied
by the functions $N\left( \phi \right) $ and $M\left( \phi \right) $:
\begin{eqnarray}
\frac{d^{2}N\left( \phi \right) }{d\phi ^{2}} &=&0,  \label{c1} \\
\frac{d\left( M\left( \phi \right) +N\left( \phi \right) \right) }{d\phi }
&=&0,  \label{c2} \\
m^{2}\left( \phi M\left( \phi \right) +4\sum\limits_{n=2}^{\infty }\frac{%
\left( -\right) ^{n}}{n!}\phi ^{n}\frac{d^{n-1}N\left( \phi \right) }{d\phi
^{n-1}}\right) &=&0.  \label{c3}
\end{eqnarray}
From (\ref{c1}) we find that
\begin{equation}
N\left( \phi \right) =c_{1}-k\phi ,  \label{n}
\end{equation}
with $c_{1}$ and $k$ two arbitrary, real constants. Inserting (\ref{n}) in (%
\ref{c2}), we have that
\begin{equation}
M\left( \phi \right) =\bar{k}+k\phi ,  \label{m}
\end{equation}
with $\bar{k}$ another arbitrary, real constant. Introducing the solutions (%
\ref{n})--(\ref{m}) in (\ref{c3}), the last one becomes
\begin{equation}
-m^{2}k\phi ^{2}+m^{2}\bar{k}\phi =0,  \label{c4}
\end{equation}
and leads to
\begin{equation}
m^{2}k=0,\qquad m^{2}\bar{k}=0.  \label{c5}
\end{equation}
There are two types of solutions to (\ref{c5}). The former has the form
\begin{equation}
k=\bar{k}=0,\qquad m^{2}\neq 0,  \label{c9}
\end{equation}
and obviously gives no cross-couplings between the Weyl graviton and the
scalar field. For this reason this solution is not interesting, and
therefore it will be omitted in the sequel. The latter solution is written
like
\begin{equation}
k\neq 0,\qquad \bar{k}\neq 0,\qquad m^{2}=0,  \label{c6}
\end{equation}
and provides the first-order deformation
\begin{equation}
a_{1}^{(\mathrm{int})}=\bar{k}\phi ^{*}\xi +k\phi ^{*}\left[ \phi \xi
-\left( \partial ^{\mu }\phi \right) \eta _{\mu }\right] ,  \label{c7}
\end{equation}
\begin{equation}
a_{0}^{(\mathrm{int})}=-\left( \bar{k}\phi +\frac{k}{2}\phi ^{2}\right)
\mathcal{K}+\frac{k}{2}\left[ \left( \partial ^{\mu }\phi \right) \partial
^{\nu }\phi -\frac{1}{2}\sigma ^{\mu \nu }\left( \partial ^{\rho }\phi
\right) \partial _{\rho }\phi \right] h_{\mu \nu }+\bar{a}_{0}^{\left(
\mathrm{int}\right) },  \label{c8}
\end{equation}
where $\bar{a}_{0}^{\left( \mathrm{int}\right) }$ is solution to the
`homogeneous' equation
\begin{equation}
\gamma \bar{a}_{0}^{\left( \mathrm{int}\right) }=\partial _{\mu }\bar{m}%
^{\left( \mathrm{int}\right) \mu },  \label{3.39}
\end{equation}
for some local $\bar{m}^{\left( \mathrm{int}\right) \mu }$. The solutions to
the `homogeneous' equation correspond to $\bar{a}_{1}^{\left( \mathrm{int}%
\right) }=0$, and thus they do not deform the gauge transformations, but
only the interacting lagrangian at order one in the coupling constant. The
constant $c_{1}$ appearing in (\ref{n}) brings no contribution to the
first-order deformation and will be set zero in the sequel.

Next, we analyze the equation (\ref{3.39}). There are two main types of
solutions to this equation. The first type, to be denoted by $\bar{a}%
_{0}^{\prime \left( \mathrm{int}\right) }$, corresponds to $\bar{m}^{\left(
\mathrm{int}\right) \mu }=0$ and is given by gauge-invariant, nonintegrated
densities constructed out of the original fields and their spacetime
derivatives, which, according to (\ref{3.10}), are of the form
\begin{equation}
\bar{a}_{0}^{\prime \left( \mathrm{int}\right) }=\bar{a}_{0}^{\prime \left(
\mathrm{int}\right) }\left( \left[ \phi \right] ,\left[ \mathcal{W}_{\mu \nu
\alpha \beta }\right] \right) ,  \label{3.41}
\end{equation}
up to the condition that they effectively describe cross-couplings between
the two types of fields and cannot be written in a divergence-like form. The
sole possibility that complies with all the hypotheses mentioned in the
beginning of this section is
\begin{equation}
\bar{a}_{0}^{\prime \left( \mathrm{int}\right) }=\frac{1}{2}\mathcal{W}_{\mu
\nu \rho \lambda }\mathcal{W}^{\mu \nu \rho \lambda }v\left( \phi \right) ,
\label{3.41a}
\end{equation}
where $v\left( \phi \right) $ is a smooth function of the undifferentiated
scalar field. The second kind of solutions, to be denoted by $\bar{a}%
_{0}^{\prime \prime \left( \mathrm{int}\right) }$, is associated with $\bar{m%
}^{\left( \mathrm{int}\right) \mu }\neq 0$ in (\ref{3.39}), being understood
that we discard the divergence-like quantities and maintain the condition on
the conservation of the number of derivatives on each field. At this point
it is useful to decompose the exterior derivative along the gauge orbits $%
\gamma $ like in~\cite{marcann}
\begin{equation}
\gamma =\gamma _{0}+\gamma _{1},  \label{gama1}
\end{equation}
where $\gamma _{0}$ and $\gamma _{1}$ act nontrivially only on the Weyl
field through
\begin{equation}
\gamma _{0}h_{\mu \nu }=\partial _{(\mu }\eta _{\nu )},\qquad \gamma
_{1}h_{\mu \nu }=2\sigma _{\mu \nu }\xi .  \label{gama2}
\end{equation}
The grading associated with this splitting is the number of the ghosts $\xi $
and of their derivatives ($\gamma _{1}$ increases this number by one unit,
while $\gamma _{0}$ does not affect it). If one plugs this decomposition
into the equation (\ref{3.39}) corresponding to $\bar{m}^{\left( \mathrm{int}%
\right) \mu }\neq 0$, it becomes equivalent to two equations, namely
\begin{equation}
\gamma _{0}\bar{a}_{0}^{\prime \prime \left( \mathrm{int}\right) }=\partial
_{\mu }\bar{m}_{0}^{\left( \mathrm{int}\right) \mu },\qquad \gamma _{1}\bar{a%
}_{0}^{\prime \prime \left( \mathrm{int}\right) }=\partial _{\mu }\bar{m}%
_{1}^{\left( \mathrm{int}\right) \mu }.  \label{gama3}
\end{equation}
As it is shown in Appendix \ref{apphomog}, the solution to the former
equation in (\ref{gama3}) reads as
\begin{equation}
\bar{a}_{0}^{\prime \prime \left( \mathrm{int}\right) }=\left( c_{1}\mathcal{R}%
_{\mu \alpha \nu \beta }\mathcal{R}^{\mu \alpha \nu \beta }+c_{2}\mathcal{R}%
_{\mu \nu }\mathcal{R}^{\mu \nu }+c_{3}\mathcal{R}^{2}\right) \bar{f}\left(
\phi \right) +\mathcal{R}f\left( \phi \right) ,  \label{gama4}
\end{equation}
with $f\left( \phi \right) $ and $\bar{f}\left( \phi \right) $ arbitrary,
smooth functions of the undifferentiated scalar field, while $c_{1}$, $c_{2}$%
, and $c_{3}$ are some arbitrary, real constants. Inserting (\ref{gama4})
into the latter equation from (\ref{gama3}), we obtain that
\begin{eqnarray}
&&-\bar{f}\left( \phi \right) \left[ 4\left( 2c_{1}+c_{2}\right) \mathcal{R}%
^{\mu \nu }\partial _{\mu }\partial _{\nu }\xi +2\left( c_{2}+6c_{3}\right)
\mathcal{R}\Box \xi \right] -  \nonumber \\
&&-6f\left( \phi \right) \Box \xi =\partial _{\mu }\bar{m}_{1}^{\left(
\mathrm{int}\right) \mu }.  \label{zzz}
\end{eqnarray}
The left-hand side of (\ref{zzz}) reduces to a divergence if
\begin{equation}
c_{1}=c_{2}=c_{3}=0,\qquad f\left( \phi \right) =\mathrm{constant}.
\label{gama5}
\end{equation}
The relations (\ref{gama4}) and (\ref{gama5}) lead to an $\bar{a}%
_{0}^{\prime \prime \left( \mathrm{int}\right) }$ that cannot describe
cross-couplings, and consequently we must take\footnote{%
Apparently, a more general solution to the equation (\ref{zzz}) would be $%
c_{1}=-\frac{1}{2}c_{2}$, $c_{3}=-\frac{1}{6}c_{2}$, $f\left( \phi \right) =$
$\mathrm{constant}$. However, this solution provides a vertex of the type (%
\ref{3.41a}), which has already been considered.}
\begin{equation}
\bar{a}_{0}^{\prime \prime \left( \mathrm{int}\right) }=0.  \label{xwz}
\end{equation}

Finally, we focus on the solutions to the equation (\ref{uw2}). As the
scalar field action from (\ref{fract}) has no nontrivial gauge invariance,
it follows that $a^{\left( \mathrm{\phi }\right) }$ can only reduce to its
component of antighost number zero
\begin{equation}
a^{\left( \mathrm{\phi }\right) }\equiv \bar{a}^{\left( \mathrm{\phi
}\right) }=\bar{a}_{0}^{\left( \mathrm{\phi }\right) }\left( \left[
\phi \right] \right) ,  \label{vw1}
\end{equation}
which is automatically solution to the equation $s\bar{a}^{\left( \mathrm{%
\phi }\right) }\equiv \gamma \bar{a}_{0}^{\left( \mathrm{\phi }\right) }=0$.
It comes from $\bar{a}_{1}^{\left( \mathrm{\phi }\right) }=0$ and does not
deform the gauge transformations (\ref{5and6}), but merely modifies the
scalar field action. The hypothesis on the conservation of the number of
derivatives on each field is translated at the level of $\bar{a}^{\left(
\mathrm{\phi }\right) }$ into
\begin{equation}
\bar{a}_{0}^{\left( \mathrm{\phi }\right) }=J\left( \phi \right) \left(
\partial _{\mu }\phi \right) \left( \partial ^{\mu }\phi \right) \left(
\partial _{\rho }\phi \right) \partial ^{\rho }\phi +F\left( \phi \right)
\left( \partial _{\rho }\phi \right) \partial ^{\rho }\phi +G\left( \phi
\right) ,  \label{vw2}
\end{equation}
where $J\left( \phi \right) $, $F\left( \phi \right) $ and $G\left( \phi
\right) $ are some arbitrary functions of the undifferentiated scalar field.
We can summarize the above results by concluding that the `interacting' part
of the first-order deformation of the solution to the master equation can be
written as
\begin{eqnarray}
S_{1}^{(\mathrm{int})} &=&\int d^{4}x\left\{ \phi ^{*}\left[ \bar{k}\xi
+k\left( \phi \xi -\left( \partial ^{\mu }\phi \right) \eta _{\mu }\right)
\right] -\bar{k}\phi \sigma ^{\mu \nu }\mathcal{K}_{\mu \nu }-\right.
\nonumber \\
&&-\frac{k}{2}\left[ \phi ^{2}\sigma ^{\mu \nu }\mathcal{K}_{\mu \nu
}-h_{\mu \nu }\left( \left( \partial ^{\mu }\phi \right) \partial ^{\nu
}\phi -\frac{1}{2}\sigma ^{\mu \nu }\left( \partial ^{\rho }\phi \right)
\partial _{\rho }\phi \right) \right] +  \nonumber \\
&&+\frac{1}{2}\mathcal{W}_{\mu \nu \rho \lambda }\mathcal{W}^{\mu \nu \rho
\lambda }v\left( \phi \right) +J\left( \phi \right) \left( \partial _{\mu
}\phi \right) \left( \partial ^{\mu }\phi \right) \left( \partial _{\rho
}\phi \right) \partial ^{\rho }\phi +  \nonumber \\
&&\left. +F\left( \phi \right) \left( \partial _{\rho }\phi \right) \partial
^{\rho }\phi +G\left( \phi \right) \right\} .  \label{def1}
\end{eqnarray}

\subsection{Second-order deformation}

\label{secondord}

So far we have seen that the first-order deformation can be written like the
sum between the first-order deformation for the Weyl theory $S_{1}^{\left(
\mathrm{W}\right) }$ (exposed in~\cite{marcann}) and the interacting part $%
S_{1}^{\left( \mathrm{int}\right) }$, given in (\ref{def1}).

In this section we investigate the consistency of the first-order
deformation, described by the equation (\ref{2.6}). Along the same line as
before, we can write the second-order deformation like the sum between the
Weyl contribution and the interacting part
\begin{equation}
S_{2}=S_{2}^{\left( \mathrm{W}\right) }+S_{2}^{\left( \mathrm{int}\right) }.
\label{4.2}
\end{equation}
The piece $S_{2}^{\left( \mathrm{W}\right) }$ can be deduced from~\cite
{marcann}, while $S_{2}^{\left( \mathrm{int}\right) }$ must satisfy the
equation
\begin{equation}
\frac{1}{2}\left( S_{1},S_{1}\right) ^{\left( \mathrm{int}\right)
}+sS_{2}^{\left( \mathrm{int}\right) }=0,  \label{4.3}
\end{equation}
where
\begin{equation}
\left( S_{1},S_{1}\right) ^{\left( \mathrm{int}\right) }=\left(
S_{1}^{\left( \mathrm{int}\right) },S_{1}^{\left( \mathrm{int}\right)
}\right) +2\left( S_{1}^{\left( \mathrm{W}\right) },S_{1}^{\left( \mathrm{int%
}\right) }\right) .  \label{4.4}
\end{equation}
If we denote by $\Lambda ^{\left( \mathrm{int}\right) }$ and $b^{\left(
\mathrm{int}\right) }$ the nonintegrated densities of the functionals $%
\left( S_{1},S_{1}\right) ^{\left( \mathrm{int}\right) }$ and respectively $%
S_{2}^{\left( \mathrm{int}\right) }$, the local form of (\ref{4.3}) becomes
\begin{equation}
\Lambda ^{\left( \mathrm{int}\right) }=-2sb^{\left( \mathrm{int}\right)
}+\partial _{\mu }n^{\mu },  \label{4.5}
\end{equation}
with
\begin{equation}
\mathrm{gh}\left( \Lambda ^{\left( \mathrm{int}\right) }\right) =1,\qquad
\mathrm{gh}\left( b^{\left( \mathrm{int}\right) }\right) =0,\qquad \mathrm{gh%
}\left( n^{\mu }\right) =1,  \label{45a}
\end{equation}
for some local currents $n^{\mu }$. Direct computation shows that $\Lambda
^{\left( \mathrm{int}\right) }$ decomposes like
\begin{equation}
\Lambda ^{\left( \mathrm{int}\right) }=\Lambda _{0}^{\left( \mathrm{int}%
\right) }+\Lambda _{1}^{\left( \mathrm{int}\right) },\qquad \mathrm{agh}%
\left( \Lambda _{I}^{\left( \mathrm{int}\right) }\right) =I,\qquad I=0,1,
\label{4.6}
\end{equation}
with
\begin{eqnarray}
\Lambda _{1}^{\left( \mathrm{int}\right) } &=&-2\left( 1+k\right) \left[
\left( \bar{k}\phi ^{*}+k\phi ^{*}\phi \right) \eta ^{\mu }\partial _{\mu
}\xi +k\phi ^{*}\left( \partial ^{\mu }\phi \right) \eta _{\mu }\xi +\right.
\nonumber \\
&&\left. +\frac{k}{2}\phi ^{*}\left( \partial ^{\mu }\phi \right) \eta ^{\nu
}\partial _{[\mu }\eta _{\nu ]}\right] +\gamma \left( k^{2}\phi ^{*}\left(
\partial ^{\mu }\phi \right) \eta ^{\nu }h_{\mu \nu }\right) ,  \label{4.11}
\end{eqnarray}
and
\begin{eqnarray}
\Lambda _{0}^{\left( \mathrm{int}\right) } &=&-2\left[ \bar{k}\left( \bar{k}%
+2k\phi \right) +k^{2}\phi ^{2}\right] \mathcal{K}\xi +2k\left( h_{\mu \nu }-%
\frac{1}{2}\sigma _{\mu \nu }h\right) \times  \nonumber \\
&&\times \left\{ \bar{k}\left( \partial ^{\mu }\phi \right) \partial ^{\nu
}\xi +k\left[ \left( \partial ^{\mu }\phi \right) \partial ^{\nu }\left(
\phi \xi \right) -\left( \partial ^{\mu }\phi \right) \partial ^{\nu }\left(
\left( \partial ^{\rho }\phi \right) \eta _{\rho }\right) \right] \right\} +
\nonumber \\
&&+2k\left( \bar{k}+k\phi \right) \mathcal{K}\left( \partial ^{\rho }\phi
\right) \eta _{\rho }+k\left[ \left( \partial ^{\mu }\phi \right) \partial
^{\nu }\phi -\frac{1}{2}\sigma ^{\mu \nu }\left( \partial ^{\rho }\phi
\right) \partial _{\rho }\phi \right] \times  \nonumber \\
&&\times \left( -\eta ^{\lambda }\partial _{\mu }h_{\nu \lambda }+\eta
^{\lambda }\partial _{\lambda }h_{\mu \nu }+2\xi h_{\mu \nu }+h_{\lambda \mu
}\partial _{\nu }\eta ^{\lambda }\right) +  \nonumber \\
&&+\frac{1}{6}\left( 2\bar{k}+k\phi \right) \phi \left( \partial ^{\mu
}\partial ^{\nu }-\sigma ^{\mu \nu }\square \right) \times  \nonumber \\
&&\times \left( \eta ^{\lambda }\partial _{\mu }h_{\nu \lambda }-\eta
^{\lambda }\partial _{\lambda }h_{\mu \nu }-2\xi h_{\mu \nu }-h_{\lambda \mu
}\partial _{\nu }\eta ^{\lambda }\right) +  \nonumber \\
&&+2\bar{k}\left[ \left( \frac{dF}{d\phi }\left( \partial _{\rho }\phi
\right) \partial ^{\rho }\phi +\frac{dG}{d\phi }\right) \xi +2F\left(
\partial _{\rho }\phi \right) \partial ^{\rho }\xi \right] +  \nonumber \\
&&+2k\left[ \left( \frac{dF}{d\phi }\left( \partial _{\rho }\phi \right)
\partial ^{\rho }\phi +\frac{dG}{d\phi }\right) \left( \phi \xi -\left(
\partial ^{\mu }\phi \right) \eta _{\mu }\right) +\right.  \nonumber \\
&&\left. +2F\left( \partial _{\rho }\phi \right) \partial ^{\rho }\left(
\phi \xi -\left( \partial ^{\mu }\phi \right) \eta _{\mu }\right) \right] +
\nonumber \\
&&+2\left[ \left( \partial _{\mu }\phi \right) \left( \partial ^{\mu }\phi
\right) \right] ^{2}\xi \left[ k\left( \frac{dJ\left( \phi \right) }{d\phi }%
\phi +4J\left( \phi \right) \right) +\bar{k}\frac{dJ\left( \phi \right) }{%
d\phi }\right] +  \nonumber \\
&&+8\left( \partial _{\mu }\phi \right) \left( \partial ^{\mu }\phi \right)
\left( \partial _{\nu }\phi \right) \left( \partial ^{\nu }\xi \right)
J\left( \phi \right) \left( \bar{k}+k\phi \right) +  \nonumber \\
&&+\gamma \left[ \left( -\frac{1}{2}\mathcal{W}^{\mu \nu \rho \lambda }%
\mathcal{W}_{\mu \nu \rho \lambda }h+4\mathcal{W}^{\mu \alpha \nu \beta
}\sigma ^{\rho \lambda }\stackrel{(1)}{\Gamma }_{\rho \mu \nu }\stackrel{(1)%
}{\Gamma }_{\lambda \alpha \beta }\right) v\left( \phi \right) +\right.
\nonumber \\
&&+8\left( \mathcal{W}^{\mu \alpha \nu \beta }h_{\mu \nu }\mathcal{K}%
_{\alpha \beta }+\frac{1}{2}\mathcal{W}^{\mu \alpha \nu \beta }\mathcal{W}%
_{\mu \rho \nu \beta }h_{\;\;\alpha }^{\rho }\right) v\left( \phi \right) +
\nonumber \\
&&\left. +kJ\left( \phi \right) \left( \partial _{\rho }\phi \right) \left(
\partial ^{\rho }\phi \right) \left( \partial _{\mu }\phi \right) \left(
\partial _{\nu }\phi \right) \left( \sigma ^{\mu \nu }h-4h^{\mu \nu }\right)
\right] +  \nonumber \\
&&+\mathcal{W}^{\mu \alpha \nu \beta }\mathcal{W}_{\mu \alpha \nu \beta
}\left( \bar{k}+k\phi \right) \frac{dv\left( \phi \right) }{d\phi }\xi -
\nonumber \\
&&-\left( k+1\right) \left( \mathcal{W}^{\mu \alpha \nu \beta }\mathcal{W}%
_{\mu \alpha \nu \beta }\frac{dv\left( \phi \right) }{d\phi }\partial _{\rho
}\phi \right) \eta ^{\rho },  \label{delta0}
\end{eqnarray}
where
\begin{equation}
\stackrel{(1)}{\Gamma }_{\mu \nu \rho }=\frac{1}{2}\left( \partial _{\nu
}h_{\rho \mu }+\partial _{\rho }h_{\nu \mu }-\partial _{\mu }h_{\nu \rho
}\right) .  \label{wgt}
\end{equation}
Since the first-order deformation in the interacting sector starts in
antighost number one, we can take, without loss of generality, the
corresponding second-order deformation to start in antighost number two
\begin{eqnarray}
b^{\left( \mathrm{int}\right) } &=&b_{0}^{\left( \mathrm{int}\right)
}+b_{1}^{\left( \mathrm{int}\right) }+b_{2}^{\left( \mathrm{int}\right)
},\qquad \mathrm{agh}\left( b_{I}^{\left( \mathrm{int}\right) }\right)
=I,\qquad I=0,1,2,  \label{d1} \\
n^{\mu } &=&n_{0}^{\mu }+n_{1}^{\mu }+n_{2}^{\mu },\qquad \mathrm{agh}\left(
n_{I}^{\mu }\right) =I,\qquad I=0,1,2.  \label{d2}
\end{eqnarray}
By projecting the equation (\ref{4.5}) on various antighost numbers, we
infer the following tower:
\begin{eqnarray}
\gamma b_{2}^{\left( \mathrm{int}\right) } &=&\partial _{\mu }\left( \frac{1%
}{2}n_{2}^{\mu }\right) ,  \label{d3} \\
\Lambda _{1}^{\left( \mathrm{int}\right) } &=&-2\left( \delta b_{2}^{\left(
\mathrm{int}\right) }+\gamma b_{1}^{\left( \mathrm{int}\right) }\right)
+\partial _{\mu }n_{1}^{\mu },  \label{d4} \\
\Lambda _{0}^{\left( \mathrm{int}\right) } &=&-2\left( \delta b_{1}^{\left(
\mathrm{int}\right) }+\gamma b_{0}^{\left( \mathrm{int}\right) }\right)
+\partial _{\mu }n_{0}^{\mu }.  \label{d5}
\end{eqnarray}
As argued in the subsection~\ref{stand}, the equation (\ref{d3}) can be
always be replaced with
\begin{equation}
\gamma b_{2}^{\left( \mathrm{int}\right) }=0.  \label{d6}
\end{equation}
If we make the notation
\begin{eqnarray}
\Gamma &=&-2\left( 1+k\right) \left[ \left( \bar{k}\phi ^{*}+k\phi ^{*}\phi
\right) \eta ^{\mu }\partial _{\mu }\xi +k\phi ^{*}\left( \partial ^{\mu
}\phi \right) \eta _{\mu }\xi +\right.  \nonumber \\
&&\left. +\frac{k}{2}\phi ^{*}\left( \partial ^{\mu }\phi \right) \eta ^{\nu
}\partial _{[\mu }\eta _{\nu ]}\right] ,  \label{d7}
\end{eqnarray}
and inspect (\ref{4.11}), it results that $\Lambda _{1}^{\left( \mathrm{int}%
\right) }$ can be written like in (\ref{d4}) if
\begin{equation}
\Gamma =\delta \psi +\gamma \Pi +\partial _{\mu }\rho ^{\mu },  \label{d8}
\end{equation}
for some local nonintegrated quantities $\psi $, $\Pi $, and $\rho ^{\mu }$.
By applying $\delta $ on (\ref{d8}), we infer that
\begin{equation}
\delta \Gamma =\gamma \left( -\delta \Pi \right) +\partial _{\mu }\left(
\delta \rho ^{\mu }\right) .  \label{d9}
\end{equation}
Let us suppose that (\ref{d9}) holds. Acting now with $\delta $ on (\ref{d7}%
), we arrive at
\begin{eqnarray}
\delta \Gamma &=&\partial _{\mu }\left\{ -2\left( 1+k\right) \left[ \bar{k}%
\left( \left( \partial ^{\mu }\phi \right) \eta ^{\nu }-\frac{1}{2}\phi
\partial ^{[\mu }\eta ^{\nu ]}\right) \partial _{\nu }\xi +\right. \right.
\nonumber \\
&&+k\phi \left( \left( \partial ^{\mu }\phi \right) \eta ^{\nu }-\frac{1}{4}%
\phi \partial ^{[\mu }\eta ^{\nu ]}+\frac{1}{2}\phi \xi \sigma ^{\mu \nu
}\right) \partial _{\nu }\xi -  \nonumber \\
&&\left. \left. -kT^{\mu \nu }\eta _{\nu }\xi -\frac{k}{2}T^{\mu \nu }\eta
^{\rho }\partial _{[\nu }\eta _{\rho ]}\right] \right\} +  \nonumber \\
&&+\gamma \left\{ -2\left( 1+k\right) \left[ \bar{k}\left( \partial ^{\mu
}\phi \right) \left( \eta ^{\nu }\mathcal{K}_{\mu \nu }+\frac{1}{2}\left(
\partial ^{\nu }\xi \right) h_{\mu \nu }\right) -\right. \right.  \nonumber
\\
&&-\frac{\bar{k}}{2}\phi \left( \partial ^{\mu }h_{\mu \nu }-\partial _{\nu
}h\right) \partial ^{\nu }\xi +k\phi \left( \partial ^{\mu }\phi \right)
\left( \eta ^{\nu }\mathcal{K}_{\mu \nu }+\frac{1}{2}\left( \partial ^{\nu
}\xi \right) h_{\mu \nu }\right) +  \nonumber \\
&&+\frac{k}{2}\phi ^{2}\left( \xi \mathcal{K}+\frac{1}{2}\left( \partial
^{\mu }h_{\mu \nu }-\partial _{\nu }h\right) \partial ^{\nu }\xi \right) -
\nonumber \\
&&-\frac{k}{2}T^{\mu \nu }\left( \xi h_{\mu \nu }+\eta ^{\rho }\partial
_{\rho }h_{\mu \nu }-\eta ^{\rho }\partial _{\mu }h_{\nu \rho }\right) +
\nonumber \\
&&\left. \left. +\frac{k}{4}\left( \partial ^{\mu }\phi \right) \left(
\partial _{\rho }\phi \right) h^{\rho \nu }\partial _{[\mu }\eta _{\nu
]}\right] \right\} ,  \label{d10}
\end{eqnarray}
where
\begin{equation}
T^{\mu \nu }=\frac{1}{2}\sigma ^{\mu \nu }\left( \partial _{\rho }\phi
\right) \left( \partial ^{\rho }\phi \right) -\left( \partial ^{\mu }\phi
\right) \left( \partial ^{\nu }\phi \right)  \label{nw1}
\end{equation}
is the stress-energy tensor of the scalar field. Due to the fact that
neither of $\eta ^{\mu }$, $\partial ^{[\mu }\eta ^{\nu ]}$, $\xi $, and $%
\partial _{\mu }\xi $ are $\delta $-exact, the right-hand side of (\ref{d10}%
) can be written like in the right-hand side of (\ref{d9}) if
\begin{equation}
\phi =\delta \Omega ,  \label{1l}
\end{equation}
for some local $\Omega $. Indeed, if $\Omega $ were nonlocal, then (\ref{d9}%
)--(\ref{d10}) would imply that $\delta \Pi $ is also nonlocal. It is clear
that the non-locality of $\delta \Pi $ yields the same with respect to $\Pi $%
. In this setting, from (\ref{4.11}), (\ref{d4}), and (\ref{d7})--(\ref{d8})
we get that
\begin{equation}
b_{1}^{\left( \mathrm{int}\right) }=-\frac{1}{2}\left( k^{2}\phi ^{*}\left(
\partial ^{\mu }\phi \right) \eta ^{\nu }h_{\mu \nu }+\Pi \right) ,
\label{nw2}
\end{equation}
which indicates that the second-order deformation would contain a nonlocal
term, and thus contradicts the hypothesis on the locality of the
deformations. We observe that the requirement on $\Omega $ to be a local
function is a direct consequence of the required locality of the
interactions. On behalf of the second relation in (\ref{12}) and the last
equation from (\ref{c6}) we find that $\phi $ is $\delta $-exact
\begin{equation}
\phi =\delta \left( \frac{1}{\Box }\phi ^{*}\right) ,  \label{vm3}
\end{equation}
but only in the space of nonlocal nonintegrated densities. As a consequence,
the equation (\ref{1l}) cannot take place in the space of local
nonintegrated densities, where the admitted deformations belong. This
further yields that neither (\ref{d8}) can hold in this space. So, $\Gamma $
of the form (\ref{d7}) must be made to vanish, which happens if
\begin{equation}
\bar{k}\left( 1+k\right) =0,\qquad k\left( 1+k\right) =0.  \label{4.12}
\end{equation}
The nontrivial solution to (\ref{4.12}) reads as
\begin{equation}
k=-1,  \label{4.13}
\end{equation}
and $\bar{k}$ remains an arbitrary real constant. Replacing (\ref{4.13}) in (%
\ref{4.11}), and then in (\ref{d4}), we infer that
\begin{eqnarray}
b_{2}^{\left( \mathrm{int}\right) } &=&0,  \label{wm1} \\
b_{1}^{\left( \mathrm{int}\right) } &=&-\frac{1}{2}\phi ^{*}\left( \partial
^{\mu }\phi \right) \eta ^{\nu }h_{\mu \nu }.  \label{4.14}
\end{eqnarray}
Inserting now (\ref{4.13}) in (\ref{delta0}), we obtain
\begin{eqnarray}
\Lambda _{0}^{\left( \mathrm{int}\right) }+2\delta b_{1}^{\left( \mathrm{int}%
\right) } &=&\gamma \left\{ -\frac{1}{8}\left( h^{2}-2h_{\mu \nu }h^{\mu \nu
}\right) \left( \partial _{\rho }\phi \right) \partial ^{\rho }\phi -\right.
\nonumber \\
&&-\left( h_{\mu \rho }h_{\;\;\nu }^{\rho }-\frac{1}{2}hh_{\mu \nu }\right)
\left( \partial ^{\nu }\phi \right) \partial ^{\mu }\phi +  \nonumber \\
&&+\phi ^{2}\left[ \frac{1}{3}\mathcal{R}_{\mu \nu }h^{\mu \nu }-\frac{1}{2}%
\mathcal{K}h+\frac{1}{6}\sigma ^{\mu \mu _{1}}\sigma ^{\nu \nu _{1}}\sigma
^{\rho \rho _{1}}\times \right.  \nonumber \\
&&\left. \times \left( \stackrel{(1)}{\Gamma }_{\mu \nu \nu _{1}}\stackrel{%
(1)}{\Gamma }_{\mu _{1}\rho \rho _{1}}-\stackrel{(1)}{\Gamma }_{\mu \nu \rho
}\stackrel{(1)}{\Gamma }_{\mu _{1}\nu _{1}\rho _{1}}\right) \right] -
\nonumber \\
&&-\bar{k}\phi \left[ \frac{2}{3}\mathcal{R}_{\mu \nu }h^{\mu \nu }-\mathcal{%
K}h-\frac{1}{6}\sigma ^{\mu \mu _{1}}\sigma ^{\nu \nu _{1}}\sigma ^{\rho
\rho _{1}}\times \right.  \nonumber \\
&&\times \left( \stackrel{(1)}{\Gamma }_{\mu \nu \rho }\stackrel{(1)}{\Gamma
}_{\mu _{1}\nu _{1}\rho _{1}}-2\stackrel{(1)}{\Gamma }_{\mu \nu \nu _{1}}%
\stackrel{(1)}{\Gamma }_{\mu _{1}\rho \rho _{1}}\right) +  \nonumber \\
&&+\left( -\frac{1}{2}\mathcal{W}^{\mu \nu \rho \lambda }\mathcal{W}_{\mu
\nu \rho \lambda }h+4\mathcal{W}^{\mu \alpha \nu \beta }\sigma ^{\rho
\lambda }\stackrel{(1)}{\Gamma }_{\rho \mu \nu }\stackrel{(1)}{\Gamma }%
_{\lambda \alpha \beta }+\right.  \nonumber \\
&&\left. \left. +8\left( \mathcal{W}^{\mu \alpha \nu \beta }h_{\mu \nu }%
\mathcal{K}_{\alpha \beta }+\frac{1}{2}\mathcal{W}^{\mu \alpha \nu \beta }%
\mathcal{W}_{\mu \rho \nu \beta }h_{\alpha }^{\rho }\right) \right) v\left(
\phi \right) \right] -  \nonumber \\
&&-Gh-F\left( \phi ^{2}\mathcal{K}+2T^{\mu \nu }h_{\mu \nu }\right) -
\nonumber \\
&&\left. -J\left( \phi \right) \left( \partial _{\rho }\phi \right) \left(
\partial ^{\rho }\phi \right) \left( \partial _{\mu }\phi \right) \left(
\partial _{\nu }\phi \right) \left( \sigma ^{\mu \nu }h-4h^{\mu \nu }\right)
\right\} +  \nonumber \\
&&+2\bar{k}\left[ \left( -\bar{k}\mathcal{K}+\frac{dF}{d\phi }\left(
\partial _{\rho }\phi \right) \partial ^{\rho }\phi +\frac{dG}{d\phi }%
\right) \xi +\right.  \nonumber \\
&&\left. +2F\left( \partial _{\rho }\phi \right) \partial ^{\rho }\xi
\right] -2\left( \frac{dG}{d\phi }\phi -4G\right) \xi +  \nonumber \\
&&+\left( \partial _{\mu }F\right) \left[ 2\phi ^{2}\partial ^{\mu }\xi -\xi
\partial ^{\mu }\left( \phi ^{2}\right) \right] +  \nonumber \\
&&+\mathcal{W}^{\mu \alpha \nu \beta }\mathcal{W}_{\mu \alpha \nu \beta
}\left( \bar{k}-\phi \right) \frac{dv\left( \phi \right) }{d\phi }\xi +
\nonumber \\
&&+2\left[ \left( \partial _{\mu }\phi \right) \left( \partial ^{\mu }\phi
\right) \right] ^{2}\xi \left[ -\left( \frac{dJ\left( \phi \right) }{d\phi }%
\phi +4J\left( \phi \right) \right) +\bar{k}\frac{dJ\left( \phi \right) }{%
d\phi }\right] +  \nonumber \\
&&+8\left( \partial _{\mu }\phi \right) \left( \partial ^{\mu }\phi \right)
\left( \partial _{\nu }\phi \right) \left( \partial ^{\nu }\xi \right)
J\left( \phi \right) \left( \bar{k}-\phi \right) +\partial _{\mu }n_{0}^{\mu
}.  \label{def2}
\end{eqnarray}
By comparing (\ref{def2}) with (\ref{d5}), we remark that the consistency of
the first-order deformation requires that
\begin{eqnarray}
\Theta &=&2\bar{k}\left\{ \left[ -\bar{k}\mathcal{K}+\frac{dF}{d\phi }\left(
\partial _{\rho }\phi \right) \partial ^{\rho }\phi +\frac{dG}{d\phi }%
\right] \xi +2F\left( \partial _{\rho }\phi \right) \partial ^{\rho }\xi
\right\} -  \nonumber \\
&&-2\left( \frac{dG}{d\phi }\phi -4G\right) \xi +\left( \partial _{\mu
}F\right) \left[ 2\phi ^{2}\partial ^{\mu }\xi -\xi \partial ^{\mu }\left(
\phi ^{2}\right) \right] +  \nonumber \\
&&+\mathcal{W}^{\mu \alpha \nu \beta }\mathcal{W}_{\mu \alpha \nu \beta
}\left( \bar{k}-\phi \right) \frac{dv\left( \phi \right) }{d\phi }\xi +
\nonumber \\
&&+2\left[ \left( \partial _{\mu }\phi \right) \left( \partial ^{\mu }\phi
\right) \right] ^{2}\xi \left[ -\left( \frac{dJ\left( \phi \right) }{d\phi }%
\phi +4J\left( \phi \right) \right) +\bar{k}\frac{dJ\left( \phi \right) }{%
d\phi }\right] +  \nonumber \\
&&+8\left( \partial _{\mu }\phi \right) \left( \partial ^{\mu }\phi \right)
\left( \partial _{\nu }\phi \right) \left( \partial ^{\nu }\xi \right)
J\left( \phi \right) \left( \bar{k}-\phi \right) ,  \label{wq2}
\end{eqnarray}
must be of the form
\begin{equation}
\Theta =\gamma \theta +\partial _{\mu }\chi ^{\mu },  \label{wq1}
\end{equation}
for some local $\theta $ and $\chi ^{\mu }$. Assume that (\ref{wq1}) is
satisfied. Acting with $\gamma $ on it, it follows that
\begin{equation}
\gamma \Theta =\partial _{\mu }\left( \gamma \chi ^{\mu }\right) .
\label{wq3}
\end{equation}
Using (\ref{wq2}), by direct computation we get that
\begin{equation}
\gamma \Theta =\partial _{\mu }\left( 2\bar{k}^{2}\xi \partial ^{\mu }\xi
\right) .  \label{wq4}
\end{equation}
By means of (\ref{wq3})--(\ref{wq4}), we arrive at
\begin{equation}
2\bar{k}^{2}\xi \partial ^{\mu }\xi =\gamma \chi ^{\mu }+\partial _{\nu
}\lambda ^{\nu \mu },  \label{wq5}
\end{equation}
with $\lambda ^{\nu \mu }$ some antisymmetric, but otherwise arbitrary
functions
\begin{equation}
\lambda ^{\nu \mu }=-\lambda ^{\mu \nu }.  \label{wxq5}
\end{equation}
The relation (\ref{wq5}) expresses the compatibility between the equations (%
\ref{wq3}) and (\ref{wq4}). On behalf of
\begin{equation}
\xi =\gamma \left( \frac{1}{8}h\right) -\frac{1}{4}\partial ^{\nu }\eta
_{\nu },  \label{wyq5}
\end{equation}
we determine
\begin{eqnarray}
2\xi \partial ^{\mu }\xi &=&\gamma \left[ -\frac{1}{2}\left( \frac{1}{2}%
h\partial ^{\mu }\xi +\eta _{\nu }\mathcal{K}^{\mu \nu }\right) \right] +
\nonumber \\
&&+\partial _{\nu }\left( -\frac{1}{4}\eta ^{\left[ \nu \right. }\partial
^{\left. \mu \right] }\xi \right) +\partial _{\nu }\left( -\frac{1}{4}\eta
^{\left( \nu \right. }\partial ^{\left. \mu \right) }\xi \right) .
\label{wzq5}
\end{eqnarray}
The presence of the term $\partial _{\nu }\left( -\frac{1}{4}\eta ^{\left(
\nu \right. }\partial ^{\left. \mu \right) }\xi \right) $ in the right-hand
side of (\ref{wzq5}) indicates that the equations (\ref{wq3}) and (\ref{wq4}%
) are compatible if
\begin{equation}
\bar{k}=0.  \label{pq5}
\end{equation}
Taking into account the formula (\ref{pq5}), we conclude that the
first-order deformation is consistent if
\begin{eqnarray}
\Theta ^{\prime } &=&-2\left( \frac{dG}{d\phi }\phi -4G\right) \xi +\left(
\partial _{\mu }F\right) \left[ 2\phi ^{2}\partial ^{\mu }\xi -\xi \partial
^{\mu }\left( \phi ^{2}\right) \right] -  \nonumber \\
&&-2\left[ \left( \partial _{\mu }\phi \right) \left( \partial ^{\mu }\phi
\right) \right] ^{2}\left( \frac{dJ\left( \phi \right) }{d\phi }\phi
+4J\left( \phi \right) \right) \xi -  \nonumber \\
&&-8\left( \partial _{\mu }\phi \right) \left( \partial ^{\mu }\phi \right)
\left( \partial _{\nu }\phi \right) \left( \partial ^{\nu }\xi \right) \phi
J\left( \phi \right) -  \nonumber \\
&&-\phi \mathcal{W}^{\mu \alpha \nu \beta }\mathcal{W}_{\mu \alpha \nu \beta
}\frac{dv\left( \phi \right) }{d\phi }\xi ,  \label{pqx}
\end{eqnarray}
can be written as
\begin{equation}
\Theta ^{\prime }=\gamma \theta ^{\prime }+\partial _{\mu }\chi ^{\prime \mu
},  \label{pqz5}
\end{equation}
for some local $\theta ^{\prime }$ and $\chi ^{\prime \mu }$. Using again (%
\ref{wyq5}), after some computation we find that $\Theta ^{\prime }$ may be
expressed in the form (\ref{pqz5}) if
\begin{eqnarray}
\partial _{\mu }F &=&0,\qquad \frac{dv\left( \phi \right) }{d\phi }=0,\qquad
\frac{dG}{d\phi }\phi -4G=\bar{C},  \label{pqy5} \\
J\left( \phi \right) &=&0,  \label{new1}
\end{eqnarray}
with $\bar{C}$ a real constant. The equations (\ref{pqy5}) provide the
solutions
\begin{equation}
F\left( \phi \right) =C_{1},\qquad v\left( \phi \right) =C_{2},\qquad
G\left( \phi \right) =\kappa \phi ^{4}-\frac{1}{4}\bar{C},  \label{neint}
\end{equation}
with $\kappa $, $C_{1}$, and $C_{2}$ some real constants. The first solution
(\ref{neint}) is not interesting here since, in agreement with (\ref{vw2}),
it gives a term proportional with the free scalar field lagrangian at the
level of the first-order deformation, which is trivial
\begin{equation}
C_{1}\left( \partial _{\rho }\phi \right) \partial ^{\rho }\phi =\partial
_{\rho }\left( C_{1}\phi \partial ^{\rho }\phi \right) +s\left( -C_{1}\phi
\phi ^{*}\right) ,  \label{wa1}
\end{equation}
so we can take
\begin{equation}
F\left( \phi \right) =0.  \label{qw1}
\end{equation}
The second solution from (\ref{neint}) leads, via (\ref{3.41a}), to
\begin{equation}
\bar{a}_{0}^{\prime \left( \mathrm{int}\right) }=\frac{1}{2}C_{2}\mathcal{W}%
_{\mu \nu \rho \lambda }\mathcal{W}^{\mu \nu \rho \lambda }.  \label{abc}
\end{equation}
We observe that $\bar{a}_{0}^{\prime \left( \mathrm{int}\right) }$like in (%
\ref{abc}) does not describe the interaction between the scalar field and
the Weyl graviton, so we can choose $C_{2}=0$ and obtain
\begin{equation}
v\left( \phi \right) =0,  \label{aab}
\end{equation}
and
\begin{equation}
\bar{a}_{0}^{\prime \left( \mathrm{int}\right) }=0.  \label{q1}
\end{equation}
The third solution from (\ref{neint}) can be written as $G\left( \phi
\right) =\kappa \phi ^{4}-\partial _{\mu }\left( \frac{1}{16}\bar{C}x^{\mu
}\right) $. Since $\bar{a}_{0}^{\left( \mathrm{\phi }\right) }$ given in (%
\ref{vw2}) is defined up to a total divergence, we can set, without loss of
generality, $\bar{C}=0$, so we get
\begin{equation}
G\left( \phi \right) =\kappa \phi ^{4}.  \label{egal}
\end{equation}
Plugging (\ref{pq5}), (\ref{new1}), (\ref{qw1}), (\ref{aab}), and (\ref{egal}%
) into (\ref{def2}), we infer that
\begin{eqnarray}
b_{0}^{\left( \mathrm{int}\right) } &=&\frac{1}{16}\left( h^{2}-2h_{\mu \nu
}h^{\mu \nu }\right) \left( \partial _{\rho }\phi \right) \left( \partial
^{\rho }\phi \right) +  \nonumber \\
&&+\frac{1}{2}\left( h_{\mu \rho }h_{\nu }^{\rho }-\frac{1}{2}hh_{\mu \nu
}\right) \left( \partial ^{\mu }\phi \right) \left( \partial ^{\nu }\phi
\right) -  \nonumber \\
&&-\frac{1}{2}\phi ^{2}\left( \frac{1}{3}\mathcal{R}_{\mu \nu }h^{\mu \nu }-%
\frac{1}{2}\mathcal{K}h\right) -\frac{1}{12}\phi ^{2}\sigma ^{\mu \mu
_{1}}\sigma ^{\nu \nu _{1}}\sigma ^{\rho \rho _{1}}\times  \nonumber \\
&&\times \left( \stackrel{(1)}{\Gamma }_{\mu \nu \nu _{1}}\stackrel{(1)}{%
\Gamma }_{\mu _{1}\rho \rho _{1}}-\stackrel{(1)}{\Gamma }_{\mu \nu \rho }%
\stackrel{(1)}{\Gamma }_{\mu _{1}\nu _{1}\rho _{1}}\right) +\frac{1}{2}%
\kappa h\phi ^{4}.  \label{awx}
\end{eqnarray}
The formulas (\ref{wm1}),\ (\ref{4.14}),\ and (\ref{awx}) offer us the
complete form of the interacting part from the second-order deformation of
the solution to the master equation, $S_{2}^{(\mathrm{int})}$. Meanwhile,
with the help of the relations (\ref{4.13}), (\ref{pq5}), (\ref{new1}), (\ref
{qw1}), (\ref{aab}), and (\ref{egal}) replaced in (\ref{def1}), we also gain
the final form of the first-order deformation
\begin{eqnarray}
S_{1}^{(\mathrm{int})} &=&\int d^{4}x\left\{ -\phi ^{*}\left[ \phi \xi
-\left( \partial ^{\mu }\phi \right) \eta _{\mu }\right] +\frac{\phi ^{2}}{2}%
\sigma ^{\mu \nu }\mathcal{K}_{\mu \nu }-\right.  \nonumber \\
&&\left. -\frac{1}{2}h_{\mu \nu }\left[ \left( \partial ^{\mu }\phi \right)
\partial ^{\nu }\phi -\frac{1}{2}\sigma ^{\mu \nu }\left( \partial ^{\rho
}\phi \right) \partial _{\rho }\phi \right] +\kappa \phi ^{4}\right\} .
\label{defor1}
\end{eqnarray}
In this manner, so far we have completely determined both the first- and
second-order deformations.

\section{Interacting theory}

In order to identify the interacting theory, we start from the decomposition
of $g_{\mu \nu }$ like
\begin{equation}
g_{\mu \nu }=\sigma _{\mu \nu }+\lambda h_{\mu \nu }.  \label{nx0}
\end{equation}
Then, the inverse of $g_{\mu \nu }$, to be denoted by $g^{\mu \nu }$ ($%
g_{\mu \nu }g^{\nu \rho }=\delta _{\mu }^{\;\;\rho }$), starts like
\begin{equation}
g^{\mu \nu }=\stackrel{(0)}{g}^{\mu \nu }+\lambda \stackrel{(1)}{g}^{\mu \nu
}+\lambda ^{2}\stackrel{(2)}{g}^{\mu \nu }+\cdots =\sigma ^{\mu \nu
}-\lambda h^{\mu \nu }+\lambda ^{2}h_{\;\;\rho }^{\mu }h^{\rho \nu }+\cdots .
\label{nx1}
\end{equation}
Based on the relations (\ref{nx0})--(\ref{nx1}), we obtain that the
expansions of the scalar curvature and $\sqrt{-g}$ begin like
\begin{eqnarray}
R &=&\lambda \stackrel{(1)}{R}+\lambda ^{2}\stackrel{(2)}{R}+\cdots =\lambda
\mathcal{R}-\lambda ^{2}\left[ 2h^{\mu \nu }\mathcal{R}_{\mu \nu }+\right.
\nonumber \\
&&\left. +\sigma ^{\rho \lambda }\sigma ^{\mu \alpha }\sigma ^{\nu \beta
}\left( \stackrel{(1)}{\Gamma }_{\rho \mu \alpha }\stackrel{(1)}{\Gamma }%
_{\lambda \beta \nu }-\stackrel{(1)}{\Gamma }_{\rho \mu \beta }\stackrel{(1)%
}{\Gamma }_{\lambda \alpha \nu }\right) \right] +\cdots ,  \label{nx2}
\end{eqnarray}
\begin{eqnarray}
\sqrt{-g} &=&\sqrt{-\det \left( g_{\mu \nu }\right) }=\stackrel{(0)}{\sqrt{-g%
}}+\stackrel{(1)}{\lambda \sqrt{-g}}+\lambda ^{2}\stackrel{(2)}{\sqrt{-g}}%
+\cdots =  \nonumber \\
&=&1+\frac{\lambda }{2}h+\frac{\lambda ^{2}}{8}\left( h^{2}-2h_{\mu \nu
}h^{\mu \nu }\right) +\cdots .  \label{nx3}
\end{eqnarray}

The piece of antighost number zero in $S_{1}^{(\mathrm{int})}$ is nothing
but the interacting lagrangian at order one in the coupling constant
\begin{equation}
\mathcal{L}_{1}^{(\mathrm{int})}=a_{0}^{\left( \mathrm{int}\right)
}+a_{0}^{\left( \mathrm{\phi }\right) }=\frac{1}{12}\phi ^{2}\mathcal{R}-%
\frac{1}{2}\left( h^{\mu \nu }-\frac{1}{4}h\sigma ^{\mu \nu }\right) \left(
\partial _{\mu }\phi \right) \partial _{\nu }\phi +\kappa \phi ^{4},
\label{nx4}
\end{equation}
which, according to (\ref{nx1})--(\ref{nx3}), can be put in the form
\begin{equation}
\mathcal{L}_{1}^{(\mathrm{int})}=\frac{1}{12}\phi ^{2}\stackrel{(0)}{\sqrt{-g%
}}\stackrel{(1)}{R}+\frac{1}{2}\left( \stackrel{(0)}{\sqrt{-g}}\stackrel{(1)%
}{g}^{\mu \nu }+\stackrel{(1)}{\sqrt{-g}}\stackrel{(0)}{g}^{\mu \nu }\right)
\left( \partial _{\mu }\phi \right) \partial _{\nu }\phi +\stackrel{(0)}{%
\sqrt{-g}}\kappa \phi ^{4}.  \label{nnx4}
\end{equation}
Along the same line, the term of antighost number zero in $S_{2}^{(\mathrm{%
int})}$ produces the cross-coupling lagrangian at order two as
\begin{eqnarray}
\mathcal{L}_{2}^{(\mathrm{int})} &=&b_{0}^{\left( \mathrm{int}\right) }=%
\frac{1}{2}\left[ h^{\mu \rho }h_{\rho }^{\;\;\nu }-\frac{1}{2}hh^{\mu \nu }+%
\frac{1}{8}\left( h^{2}-2h_{\alpha \beta }h^{\alpha \beta }\right) \sigma
^{\mu \nu }\right] \times  \nonumber \\
&&\times \left( \partial _{\mu }\phi \right) \partial _{\nu }\phi +\frac{1}{%
12}\phi ^{2}\left[ \frac{1}{2}h\mathcal{R}-2\mathcal{R}_{\mu \nu }h^{\mu \nu
}-\sigma ^{\mu \mu _{1}}\sigma ^{\nu \nu _{1}}\sigma ^{\rho \rho _{1}}\times
\right.  \nonumber \\
&&\left. \times \left( \stackrel{(1)}{\Gamma }_{\mu \nu \nu _{1}}\stackrel{%
(1)}{\Gamma _{\mu _{1}\rho \rho _{1}}}-\stackrel{(1)}{\Gamma }_{\mu \nu \rho
}\stackrel{(1)}{\Gamma }_{\mu _{1}\nu _{1}\rho _{1}}\right) \right] +\frac{1%
}{2}h\kappa \phi ^{4}.  \label{nx5}
\end{eqnarray}
Using again (\ref{nx1})--(\ref{nx3}), we have that
\begin{eqnarray}
\mathcal{L}_{2}^{(\mathrm{int})} &=&\frac{1}{2}\left( \stackrel{(0)}{\sqrt{-g%
}}\stackrel{(2)}{g}^{\mu \nu }+\stackrel{(1)}{\sqrt{-g}}\stackrel{(1)}{g}%
^{\mu \nu }+\stackrel{(2)}{\sqrt{-g}}\stackrel{(0)}{g}^{\mu \nu }\right)
\left( \partial _{\mu }\phi \right) \partial _{\nu }\phi +  \nonumber \\
&&+\frac{1}{12}\phi ^{2}\left( \stackrel{(1)}{\sqrt{-g}}\stackrel{(1)}{R}+%
\stackrel{(0)}{\sqrt{-g}}\stackrel{(2)}{R}\right) \stackrel{(1)}{+\sqrt{-g}}%
\kappa \phi ^{4}.  \label{nnx5}
\end{eqnarray}
Taking into account (\ref{nnx4}) and (\ref{nnx5}), as well as the expansions
\begin{eqnarray}
\sqrt{-g}g^{\mu \nu } &=&\left( \sqrt{-g}g^{\mu \nu }\right) ^{\left(
0\right) }+\lambda \left( \sqrt{-g}g^{\mu \nu }\right) ^{\left( 1\right)
}+\lambda ^{2}\left( \sqrt{-g}g^{\mu \nu }\right) ^{\left( 2\right) }+\cdots
=  \nonumber \\
&&\stackrel{(0)}{=\sqrt{-g}}\stackrel{(0)}{g}^{\mu \nu }+\lambda \left(
\stackrel{(0)}{\sqrt{-g}}\stackrel{(1)}{g}^{\mu \nu }+\stackrel{(1)}{\sqrt{-g%
}}\stackrel{(0)}{g}^{\mu \nu }\right) +  \nonumber \\
&&+\lambda ^{2}\left( \stackrel{(0)}{\sqrt{-g}}\stackrel{(2)}{g}^{\mu \nu }+%
\stackrel{(1)}{\sqrt{-g}}\stackrel{(1)}{g}^{\mu \nu }+\stackrel{(2)}{\sqrt{-g%
}}\stackrel{(0)}{g}^{\mu \nu }\right) +\cdots ,  \label{nx6}
\end{eqnarray}
\begin{eqnarray}
\sqrt{-g}R &=&\lambda \left( \sqrt{-g}R\right) ^{\left( 1\right) }+\lambda
^{2}\left( \sqrt{-g}R\right) ^{\left( 2\right) }+\cdots =  \nonumber \\
&=&\lambda \stackrel{(0)}{\sqrt{-g}}\stackrel{(1)}{R}+\lambda ^{2}\left(
\stackrel{(0)}{\sqrt{-g}}\stackrel{(2)}{R}+\stackrel{(1)}{\sqrt{-g}}%
\stackrel{(1)}{R}\right) +\cdots ,  \label{nx7}
\end{eqnarray}
we finally find that
\begin{eqnarray}
&&\mathcal{L}_{0}^{\left( \mathrm{\phi }\right) }+\lambda \mathcal{L}_{1}^{(%
\mathrm{int})}+\lambda ^{2}\mathcal{L}_{2}^{(\mathrm{int})}+\cdots =
\nonumber \\
&=&\left( \left( \sqrt{-g}g^{\mu \nu }\right) ^{\left( 0\right) }+\lambda
\left( \sqrt{-g}g^{\mu \nu }\right) ^{\left( 1\right) }+\lambda ^{2}\left(
\sqrt{-g}g^{\mu \nu }\right) ^{\left( 2\right) }+\cdots \right) \times
\nonumber \\
&&\times \left( \partial _{\mu }\phi \right) \partial _{\nu }\phi +\frac{1}{%
12}\phi ^{2}\left( \lambda \left( \sqrt{-g}R\right) ^{\left( 1\right)
}+\lambda ^{2}\left( \sqrt{-g}R\right) ^{\left( 2\right) }+\cdots \right) +
\nonumber \\
&&+\lambda \left( \stackrel{(0)}{\sqrt{-g}}+\stackrel{(1)}{\lambda \sqrt{-g}}%
+\cdots \right) \kappa \phi ^{4}.  \label{nx8}
\end{eqnarray}
The formula (\ref{nx8}) shows us that $\mathcal{L}_{0}^{\left( \mathrm{\phi }%
\right) }+\lambda \mathcal{L}_{1}^{(\mathrm{int})}+\lambda ^{2}\mathcal{L}%
_{2}^{(\mathrm{int})}+\cdots $ comes from the expansion of the fully
deformed lagrangian
\begin{equation}
\mathcal{L}^{(\mathrm{int})}=\sqrt{-g}\left\{ \frac{1}{2}\left[ g^{\mu \nu
}\left( \partial _{\mu }\phi \right) \partial _{\nu }\phi +\frac{1}{6}\phi
^{2}R\right] +\lambda \kappa \phi ^{4}\right\} .  \label{Wint}
\end{equation}

The elements from the deformed solution to the master equation that are
linear in the antifield of the scalar field furnish the deformed gauge
transformation of this field like
\begin{eqnarray}
\delta _{\epsilon }\phi &=&\lambda \left[ \left( \partial _{\mu }\phi
\right) \epsilon ^{\mu }-\phi \epsilon \right] -\frac{1}{2}\lambda
^{2}\left( \partial _{\mu }\phi \right) \epsilon ^{\nu }h_{\nu }^{\;\;\mu
}+\cdots =  \nonumber \\
&=&\lambda \stackrel{(1)}{\delta }_{\epsilon }\phi +\lambda ^{2}\stackrel{(2)%
}{\delta }_{\epsilon }\phi +\cdots .  \label{nx9}
\end{eqnarray}
The first two orders from the above gauge transformation can be put in the
form
\begin{eqnarray}
\stackrel{(1)}{\delta }_{\epsilon }\phi &=&\left( \partial _{\mu }\phi
\right) \stackrel{(0)}{\bar{\epsilon}}^{\mu }-\phi \epsilon ,  \label{nx10}
\\
\stackrel{(2)}{\delta }_{\epsilon }\phi &=&\left( \partial _{\mu }\phi
\right) \stackrel{(1)}{\bar{\epsilon}}^{\mu },  \label{nx11}
\end{eqnarray}
where
\begin{eqnarray}
\stackrel{(0)}{\bar{\epsilon}}^{\mu } &=&\epsilon ^{\mu }=\epsilon
^{a}\delta _{a}^{\;\;\mu },  \label{nx12} \\
\stackrel{(1)}{\bar{\epsilon}}^{\mu } &=&-\frac{1}{2}\epsilon
^{a}h_{a}^{\;\;\mu }.  \label{nx13}
\end{eqnarray}
Thus, the gauge transformation of the scalar field is expressed as
\begin{equation}
\delta _{\epsilon }\phi =\lambda \left[ \left( \partial _{\mu }\phi \right)
\left( \stackrel{(0)}{\bar{\epsilon}}^{\mu }+\lambda \stackrel{(1)}{\bar{%
\epsilon}}^{\mu }+\cdots \right) -\phi \epsilon \right] .  \label{nx14}
\end{equation}
The last formula emphasizes that the gauge transformation (\ref{nx14}) comes
from the perturbative expansion of the fully deformed one
\begin{equation}
\delta _{\epsilon }\phi =\lambda \left[ \left( \partial _{\mu }\phi \right)
\bar{\epsilon}^{\mu }-\phi \epsilon \right] ,  \label{nx15}
\end{equation}
where the gauge parameters $\bar{\epsilon}^{\mu }$ are given by
\begin{equation}
\bar{\epsilon}^{\mu }=\left( \delta _{a}^{\;\;\mu }-\frac{\lambda }{2}%
h_{a}^{\;\;\mu }+\cdots \right) \epsilon ^{a}=e_{a}^{\;\;\mu }\epsilon ^{a},
\label{nx16}
\end{equation}
with $e_{a}^{\;\;\mu }$ the vierbein field. The conclusion of this section
is that the interactions between a single Weyl graviton and a real scalar
field are described by the coupled lagrangian (\ref{Wint}), while the scalar
field is endowed with the gauge symmetry (\ref{nx15}).

\section{Impossibility of cross-couplings between different Weyl gravitons
via a scalar field}

Now, we start from the action
\begin{equation}
S_{0}^{\mathrm{L}}\left[ h_{\mu \nu }^{A},\phi \right] =\frac{1}{2}\int
d^{4}x\left[ \mu _{AB}\mathcal{W}_{\mu \nu \alpha \beta }^{A}\mathcal{W}%
^{B\mu \nu \alpha \beta }+\left( \partial _{\mu }\phi \right) \left(
\partial ^{\mu }\phi \right) -m^{2}\phi ^{2}\right] ,  \label{ww1}
\end{equation}
where $\mu _{AB}$ are some constants and $\mathcal{W}_{\mu \nu \alpha \beta
}^{A}$ is the linearized Weyl tensor in four spacetime dimensions
corresponding to the field $h_{\mu \nu }^{A}$ from the collection, with $%
A=1,2,\cdots ,n$. The gauge transformations of the action (\ref{ww1}) are
given by
\begin{equation}
\delta _{\epsilon }h_{\mu \nu }^{A}=\partial _{(\mu }\epsilon _{\nu
)}^{A}+2\sigma _{\mu \nu }\epsilon ^{A},\qquad \delta _{\epsilon }\phi =0.
\label{ww2}
\end{equation}
In this section, under the hypotheses of locality, smoothness of the
interactions in the coupling constant, Poincar\'{e} invariance, (background)
Lorentz invariance, and the preservation of the number of derivatives on
each field, together with the supplementary assumption that the internal
metric $\mu _{AB}$ defined by the sum of Weyl lagrangians is positively
defined, we prove that there are no indirect cross-couplings between
different Weyl gravitons in the presence of a scalar field. The assumption
that $\mu _{AB}$ is positively defined was employed in \cite{marcann} in
order to prove that there are no direct cross-couplings among different Weyl
gravitons. In the case of Weyl gravity, there is no strong reason for taking
the internal metric to be positively defined. However, we work under this
presumption, for a proper comparison of our results to those from \cite
{marcann}. This assumption allows us to normalize the metric such that $\mu
_{AB}=\delta _{AB}$. From now on we will use everywhere in the paper the
latter form of the internal metric.

The fields/ghosts and antifields from the BRST complex are denoted by
\begin{eqnarray}
\Phi ^{a_{0}} &=&\left( h_{\mu \nu }^{A},\phi \right) ,\qquad \eta
^{a_{1}}=\left( \eta _{\mu }^{A},\xi ^{A}\right) ,  \label{ww3} \\
\Phi _{a_{0}}^{*} &=&\left( h_{A}^{*\mu \nu },\phi ^{*}\right) ,\qquad \eta
_{a_{1}}^{*}=\left( \eta _{A}^{*\mu },\xi _{A}^{*}\right) .  \label{ww4}
\end{eqnarray}
The BRST differential splits like in (\ref{7}), while the actions of $\delta
$ and $\gamma $ on the generators from the BRST complex are given by
\begin{eqnarray}
\delta h_{A}^{*\mu \nu } &=&2\partial _{\alpha }\partial _{\beta }\mathcal{W}%
_{A}^{\mu \alpha \nu \beta },\qquad \delta \phi ^{*\mu }=\left( \Box
+m^{2}\right) \phi ,  \label{ww5} \\
\delta \eta _{A}^{*\mu } &=&-2\partial _{\nu }h_{A}^{*\mu \nu },\qquad
\delta \xi _{A}^{*}=2h_{A}^{*},  \label{ww6} \\
\delta \Phi ^{a_{0}} &=&0,\qquad \delta \eta ^{a_{1}}=0,  \label{ww7} \\
\gamma \Phi _{a_{0}}^{*} &=&0,\qquad \gamma \eta _{a_{1}}^{*}=0,  \label{ww8}
\\
\gamma h_{\mu \nu }^{A} &=&\partial _{(\mu }\eta _{\nu )}^{A}+2\sigma _{\mu
\nu }\xi ^{A},\qquad \gamma \phi =0,  \label{ww9} \\
\gamma \eta _{\mu }^{A} &=&0,\qquad \gamma \xi ^{A}=0.  \label{ww10}
\end{eqnarray}
The notation $h_{A}^{*}$ signifies the trace of $h_{A}^{*\mu \nu }$, $%
h_{A}^{*}=\sigma _{\mu \nu }h_{A}^{*\mu \nu }$. The solution to the master
equation reads as
\begin{equation}
\bar{S}=S_{0}^{\mathrm{L}}\left[ h_{\mu \nu }^{A},\phi \right] +\int
d^{4}x\,\,h_{A}^{*\mu \nu }\left( \partial _{(\mu }\eta _{\nu )}^{A}+2\sigma
_{\mu \nu }\xi ^{A}\right) .  \label{ww11}
\end{equation}

The first-order deformation of the solution to the master equation can be
decomposed, like in the case of a single Weyl graviton, into
\begin{equation}
\alpha =\alpha ^{\left( \mathrm{W}\right) }+\alpha ^{\left( \mathrm{int}%
\right) }+\alpha ^{\left( \mathrm{\phi }\right) },  \label{ww12}
\end{equation}
where $\alpha ^{\left( \mathrm{W}\right) }$ splits like~\cite{marcann}
\begin{equation}
\alpha ^{\left( \mathrm{W}\right) }=\alpha _{2}^{\left( \mathrm{W}\right)
}+\alpha _{1}^{\left( \mathrm{W}\right) }+\alpha _{0}^{\left( \mathrm{W}%
\right) },  \label{ww13}
\end{equation}
with
\begin{equation}
\alpha _{2}^{(\mathrm{W})}=C_{BC}^{A}\left( \eta _{A}^{*\mu }\left( \frac{1}{%
2}\eta ^{B\nu }\partial _{[\mu }\eta _{\nu ]}^{C}+\eta _{\mu }^{B}\xi
^{C}\right) -\xi _{A}^{*}\eta _{\mu }^{B}\partial ^{\mu }\xi ^{C}\right) .
\label{ww14}
\end{equation}
In formula (\ref{ww14}) the coefficients $C_{BC}^{A}$ are some constants.
The consistency of $\alpha _{2}^{(\mathrm{W})}$ in antighost number one
requires the symmetry of the above coefficients in their lower indices%
\footnote{%
The piece (\ref{ww14}) differs from that used in~\cite{marcann} through a $%
\gamma $-exact term, which does not change the condition (\ref{ww15}).}
\begin{equation}
C_{BC}^{A}=C_{CB}^{A}.  \label{ww15}
\end{equation}
Taking into consideration the relation (\ref{ww15}), we infer $\alpha _{1}^{(%
\mathrm{W})}$ in the form
\begin{equation}
\alpha _{1}^{(\mathrm{W})}=\frac{1}{2}h_{A}^{*\mu \nu }C_{BC}^{A}\left[
-\eta ^{B\rho }\left( \partial _{(\mu }h_{\nu )\rho }^{C}-2\partial _{\rho
}h_{\mu \nu }^{C}\right) +4h_{\mu \nu }^{B}\xi ^{C}+h_{\rho (\mu
}^{B}\partial _{\nu )}\eta ^{C\rho }\right] .  \label{ww16}
\end{equation}
In order to obtain a consistent $\alpha _{0}^{(\mathrm{W})}$ from this $%
\alpha _{1}^{(\mathrm{W})}$, it follows that the constants $C_{BC}^{A}$ are
further subject to the relations~\cite{marcann}\footnote{%
The piece (\ref{ww16}) differs from that given in~\cite{marcann} through a $%
\delta $-exact term, which does not affect (\ref{ww17}).}
\begin{equation}
C_{ABC}=\frac{1}{3}C_{(ABC)},  \label{ww17}
\end{equation}
where, by definition, $C_{ABC}=\delta _{AD}C_{BC}^{D}$. With (\ref{ww17}) at
hand, we find that the resulting $\alpha _{0}^{(\mathrm{W})}$ reads as in~%
\cite{marcann} (where this component is denoted by $a_{0}$ and $C_{ABC}$ by $%
a_{abc}$).

Acting in the same manner like in subsection~\ref{firstord}, we find that $%
\alpha ^{\left( \mathrm{int}\right) }=\alpha _{1}^{\left( \mathrm{int}%
\right) }+\alpha _{0}^{\left( \mathrm{int}\right) }$, with
\begin{equation}
\alpha _{1}^{\left( \mathrm{int}\right) }=\bar{k}_{A}\phi ^{*}\xi
^{A}+k_{A}\phi ^{*}\left[ \phi \xi ^{A}-\left( \partial ^{\mu }\phi \right)
\eta _{\mu }^{A}\right] ,  \label{ww18}
\end{equation}
\begin{eqnarray}
\alpha _{0}^{\left( \mathrm{int}\right) } &=&-\bar{k}_{A}\phi \sigma ^{\mu
\nu }\mathcal{K}_{\mu \nu }^{A}-\frac{1}{2}k_{A}\phi ^{2}\sigma ^{\mu \nu }%
\mathcal{K}_{\mu \nu }^{A}+  \nonumber \\
&&+\frac{1}{2}k_{A}h_{\mu \nu }^{A}\left[ \left( \partial ^{\mu }\phi
\right) \partial ^{\nu }\phi -\frac{1}{2}\sigma ^{\mu \nu }\left( \partial
^{\rho }\phi \right) \partial _{\rho }\phi \right] +  \nonumber \\
&&+\frac{1}{2}\mathcal{W}_{\mu \nu \rho \lambda }^{A}\mathcal{W}_{A}^{\mu
\nu \rho \lambda }v\left( \phi \right) ,  \label{ww19}
\end{eqnarray}
where $k_{A}$ and $\bar{k}_{A}$ are arbitrary real constants. The
deformations (\ref{ww18})--(\ref{ww19}) correspond to the situation where $%
m^{2}=0$. Meanwhile, we get in a direct way that
\begin{equation}
\alpha ^{\left( \mathrm{\phi }\right) }=a_{0}^{\left( \mathrm{\phi }\right)
},  \label{ww20}
\end{equation}
where $a_{0}^{\left( \mathrm{\phi }\right) }$ is given in (\ref{vw2}).

Next, we investigate the consistency of the first-order deformation. In view
of this, it is useful to make the following notations:
\begin{eqnarray}
\hat{S}_{1}^{\left( \mathrm{W}\right) } &=&\int d^{4}x\,\alpha ^{\left(
\mathrm{W}\right) },  \label{ww21} \\
\hat{S}_{1}^{\left( \mathrm{int}\right) } &=&\int d^{4}x\left( \alpha
^{\left( \mathrm{int}\right) }+\alpha ^{\left( \mathrm{\phi }\right)
}\right) ,  \label{ww22} \\
\hat{S}_{1} &=&\hat{S}_{1}^{\left( \mathrm{W}\right) }+\hat{S}_{1}^{\left(
\mathrm{int}\right) }.  \label{ww23}
\end{eqnarray}
By expressing the second-order deformation under the form
\begin{equation}
\hat{S}_{2}=\hat{S}_{2}^{\left( \mathrm{W}\right) }+\hat{S}_{2}^{\left(
\mathrm{int}\right) },  \label{ww24}
\end{equation}
we observe that the equation
\begin{equation}
\left( \hat{S}_{1},\hat{S}_{1}\right) +2s\hat{S}_{2}=0,  \label{ww25}
\end{equation}
controlling the second-order deformation, equivalently splits into
\begin{eqnarray}
\left( \hat{S}_{1}^{\left( \mathrm{W}\right) },\hat{S}_{1}^{\left( \mathrm{W}%
\right) }\right) +2s\hat{S}_{2}^{\left( \mathrm{W}\right) } &=&0,
\label{ww26} \\
2\left( \hat{S}_{1}^{\left( \mathrm{W}\right) },\hat{S}_{1}^{\left( \mathrm{%
int}\right) }\right) +\left( \hat{S}_{1}^{\left( \mathrm{int}\right) },\hat{S%
}_{1}^{\left( \mathrm{int}\right) }\right) +2s\hat{S}_{2}^{\left( \mathrm{int%
}\right) } &=&0.  \label{ww27}
\end{eqnarray}
The equation (\ref{ww26}) emphasizes that the constants $C_{BC}^{A}$ must
satisfy the conditions~\cite{marcann}
\begin{equation}
C_{A\left[ B\right. }^{D}C_{\left. C\right] D}^{E}=0,  \label{ww28}
\end{equation}
so they are the structure constants of a finite-dimensional, commutative,
symmetric, and associative real algebra $\mathcal{A}$. As it has been shown
in~\cite{multi}, such an algebra displays a trivial structure, in the sense
that it can be written like a direct sum of some one-dimensional ideals.
Consequently, we deduce that
\begin{equation}
C_{BC}^{A}=0\qquad \mathrm{if}\;B\neq C,\;\mathrm{or}\;A\neq C,\;\mathrm{or}%
\;A\neq B.  \label{ww29}
\end{equation}
The last formula actually implies that only the constants $C_{AA}^{A}$ may
be non-vanishing.

In the sequel we analyze the equation (\ref{ww27}). If we denote by $\hat{%
\Lambda}^{\left( \mathrm{int}\right) }$ and $\beta ^{\left( \mathrm{int}%
\right) }$ the nonintegrated densities of the functionals $2\left( \hat{S}%
_{1}^{\left( \mathrm{W}\right) },\hat{S}_{1}^{\left( \mathrm{int}\right)
}\right) +\left( \hat{S}_{1}^{\left( \mathrm{int}\right) },\hat{S}%
_{1}^{\left( \mathrm{int}\right) }\right) $ and respectively of $\hat{S}%
_{2}^{\left( \mathrm{int}\right) }$, the local expression of the equation (%
\ref{ww27}) becomes
\begin{equation}
\hat{\Lambda}^{\left( \mathrm{int}\right) }=-2s\beta ^{\left( \mathrm{int}%
\right) }+\partial _{\mu }q^{\mu },  \label{ww30}
\end{equation}
with
\begin{equation}
\mathrm{gh}\left( \hat{\Lambda}^{\left( \mathrm{int}\right) }\right)
=1,\qquad \mathrm{gh}\left( \beta ^{\left( \mathrm{int}\right) }\right)
=0,\qquad \mathrm{gh}\left( q^{\mu }\right) =1.  \label{ww31}
\end{equation}
In our case we obtain the following decomposition of $\hat{\Lambda}^{\left(
\mathrm{int}\right) }$ with respect to the antighost number:
\begin{equation}
\hat{\Lambda}^{\left( \mathrm{int}\right) }=\hat{\Lambda}_{0}^{\left(
\mathrm{int}\right) }+\hat{\Lambda}_{1}^{\left( \mathrm{int}\right) },\qquad
\mathrm{agh}\left( \hat{\Lambda}_{I}^{\left( \mathrm{int}\right) }\right)
=I,\qquad I=0,1,  \label{ww32}
\end{equation}
with
\begin{eqnarray}
\hat{\Lambda}_{1}^{\left( \mathrm{int}\right) } &=&-2\left(
k_{A}C_{BC}^{A}+k_{B}k_{C}\right) \left[ \phi ^{*}\phi \eta ^{B\mu }\partial
_{\mu }\xi ^{C}+\phi ^{*}\left( \partial ^{\mu }\phi \right) \eta _{\mu
}^{B}\xi ^{C}+\right.  \nonumber \\
&&\left. +\frac{1}{2}\phi ^{*}\left( \partial ^{\mu }\phi \right) \eta
^{B\nu }\partial _{[\mu }\eta _{\nu ]}^{C}\right] -\left( \bar{k}_{A}k_{B}-%
\bar{k}_{B}k_{A}\right) \phi ^{*}\xi ^{A}\xi ^{B}-  \nonumber \\
&&-2\left( \bar{k}_{A}C_{BC}^{A}+k_{B}\bar{k}_{C}\right) \phi ^{*}\eta _{\mu
}^{B}\partial ^{\mu }\xi ^{C}+  \nonumber \\
&&+\gamma \left[ k_{A}k_{B}\phi ^{*}\left( \partial ^{\mu }\phi \right) \eta
^{A\nu }h_{\mu \nu }^{B}\right] ,  \label{ww33}
\end{eqnarray}
\begin{eqnarray}
\hat{\Lambda}_{0}^{\left( \mathrm{int}\right) } &=&-2\left( \bar{k}_{A}\bar{k%
}_{B}+k_{A}k_{B}\phi ^{2}\right) \mathcal{K}^{A}\xi ^{B}+2k_{A}\left( h_{\mu
\nu }^{A}-\frac{1}{2}\sigma _{\mu \nu }h^{A}\right) \times  \nonumber \\
&&\times \left[ \bar{k}_{B}\left( \partial ^{\mu }\phi \right) \partial
^{\nu }\xi ^{B}+k_{B}\left( \left( \partial ^{\mu }\phi \right) \partial
^{\nu }\left( \phi \xi ^{B}\right) -\left( \partial ^{\mu }\phi \right)
\partial ^{\nu }\left( \partial ^{\rho }\phi \eta _{\rho }^{B}\right)
\right) \right] -  \nonumber \\
&&-2\left( \bar{k}_{A}k_{B}+k_{A}\bar{k}_{B}\right) \phi \mathcal{K}^{A}\xi
^{B}+2k_{B}\left( \bar{k}_{A}+k_{A}\phi \right) \mathcal{K}^{A}\left(
\partial ^{\rho }\phi \right) \eta _{\rho }^{B}+  \nonumber \\
&&+k_{A}C_{BC}^{A}\left[ \left( \partial ^{\mu }\phi \right) \partial ^{\nu
}\phi -\frac{1}{2}\sigma ^{\mu \nu }\left( \partial ^{\rho }\phi \right)
\partial _{\rho }\phi \right] \left( -\eta ^{B\lambda }\partial _{\mu
}h_{\nu \lambda }^{C}+\eta ^{B\lambda }\partial _{\lambda }h_{\mu \nu
}^{C}+\right.  \nonumber \\
&&\left. +2\xi ^{B}h_{\mu \nu }^{C}+h_{\lambda \mu }^{B}\partial _{\nu }\eta
^{C\lambda }\right) +\frac{1}{6}C_{BC}^{A}\left( 2\bar{k}_{A}+k_{A}\phi
\right) \phi \left( \partial ^{\mu }\partial ^{\nu }-\sigma ^{\mu \nu
}\square \right) \times  \nonumber \\
&&\times \left( \eta ^{B\lambda }\partial _{\mu }h_{\nu \lambda }^{C}-\eta
^{B\lambda }\partial _{\lambda }h_{\mu \nu }^{C}-2\xi ^{B}h_{\mu \nu
}^{C}-h_{\lambda \mu }^{B}\partial _{\nu }\eta ^{C\lambda }\right) +
\nonumber \\
&&+2\bar{k}_{A}\left[ \left( \frac{dF}{d\phi }\left( \partial _{\rho }\phi
\right) \partial ^{\rho }\phi +\frac{dG}{d\phi }\right) \xi ^{A}+2F\left(
\phi \right) \left( \partial _{\rho }\phi \right) \partial ^{\rho }\xi
^{A}\right] +  \nonumber \\
&&+2k_{A}\left[ \left( \frac{dF}{d\phi }\left( \partial _{\rho }\phi \right)
\partial ^{\rho }\phi +\frac{dG}{d\phi }\right) \left( \phi \xi ^{A}-\left(
\partial ^{\mu }\phi \right) \eta _{\mu }^{A}\right) +\right.  \nonumber \\
&&\left. +2F\left( \phi \right) \left( \partial _{\rho }\phi \right)
\partial ^{\rho }\left( \phi \xi ^{A}-\left( \partial ^{\mu }\phi \right)
\eta _{\mu }^{A}\right) \right] +  \nonumber \\
&&+2\left[ \left( \partial _{\mu }\phi \right) \left( \partial ^{\mu }\phi
\right) \right] ^{2}\xi ^{A}\left[ k_{A}\left( \frac{dJ\left( \phi \right) }{%
d\phi }\phi +4J\left( \phi \right) \right) +\bar{k}_{A}\frac{dJ\left( \phi
\right) }{d\phi }\right] +  \nonumber \\
&&+8\left( \partial _{\mu }\phi \right) \left( \partial ^{\mu }\phi \right)
\left( \partial _{\nu }\phi \right) \left( \partial ^{\nu }\xi ^{A}\right)
J\left( \phi \right) \left( \bar{k}_{A}+k_{A}\phi \right) +  \nonumber \\
&&+\gamma \left[ \mathcal{W}_{A}^{\mu \alpha \nu \beta }v\left( \phi \right)
C_{BC}^{A}\left( 8\mathcal{K}_{\mu \nu }^{B}h_{\alpha \beta }^{C}+4\stackrel{%
(1)}{\Gamma }_{\rho \mu \nu }^{B}\stackrel{(1)}{\Gamma }_{\lambda \alpha
\beta }^{C}\sigma ^{\rho \lambda }+\right. \right.  \nonumber \\
&&\left. +4\mathcal{W}_{\mu \rho \nu \beta }^{B}h_{\alpha }^{C\rho }-\frac{1%
}{2}\mathcal{W}_{\mu \alpha \nu \beta }^{B}h^{C}\right) +  \nonumber \\
&&\left. +k_{A}J\left( \phi \right) \left( \partial _{\rho }\phi \right)
\left( \partial ^{\rho }\phi \right) \left( \partial _{\mu }\phi \right)
\left( \partial _{\nu }\phi \right) \left( \sigma ^{\mu \nu }h^{A}-4h^{A\mu
\nu }\right) \right] +  \nonumber \\
&&+\mathcal{W}_{A}^{\mu \alpha \nu \beta }\mathcal{W}_{\mu \alpha \nu \beta
}^{A}\frac{dv\left( \phi \right) }{d\phi }\left( \bar{k}_{B}+k_{B}\phi
\right) \xi ^{B}-  \nonumber \\
&&-\mathcal{W}_{A}^{\mu \alpha \nu \beta }\mathcal{W}_{\mu \alpha \nu \beta
}^{B}\frac{dv\left( \phi \right) }{d\phi }\left( \partial ^{\rho }\phi
\right) \left( C_{BC}^{A}+\delta _{B}^{A}k_{C}\right) \eta _{\rho }^{C}.
\label{ww33x}
\end{eqnarray}
The decomposition (\ref{ww32}) implies that $\beta ^{\left( \mathrm{int}%
\right) }$ and $q^{\mu }$ can be represented like
\begin{eqnarray}
\beta ^{\left( \mathrm{int}\right) } &=&\beta _{0}^{\left( \mathrm{int}%
\right) }+\beta _{1}^{\left( \mathrm{int}\right) }+\beta _{2}^{\left(
\mathrm{int}\right) },\qquad \mathrm{agh}\left( \beta _{I}^{\left( \mathrm{%
int}\right) }\right) =I,\;I=0,1,2,  \label{ww34} \\
q^{\mu } &=&q_{0}^{\mu }+q_{1}^{\mu }+q_{2}^{\mu },\qquad \mathrm{agh}\left(
q_{I}^{\mu }\right) =I,\qquad I=0,1,2.  \label{ww35}
\end{eqnarray}
So, by projecting the equation (\ref{ww30}) on various antighost numbers, we
find the tower of equations
\begin{eqnarray}
\gamma \beta _{2}^{\left( \mathrm{int}\right) } &=&\partial _{\mu }\left(
\frac{1}{2}q_{2}^{\mu }\right) ,  \label{ww36} \\
\hat{\Lambda}_{1}^{\left( \mathrm{int}\right) } &=&-2\left( \delta \beta
_{2}^{\left( \mathrm{int}\right) }+\gamma \beta _{1}^{\left( \mathrm{int}%
\right) }\right) +\partial _{\mu }q_{1}^{\mu },  \label{ww37} \\
\hat{\Lambda}_{0}^{\left( \mathrm{int}\right) } &=&-2\left( \delta \beta
_{1}^{\left( \mathrm{int}\right) }+\gamma \beta _{0}^{\left( \mathrm{int}%
\right) }\right) +\partial _{\mu }q_{0}^{\mu }.  \label{ww38}
\end{eqnarray}
The equation (\ref{ww36}) can always be replaced, via some trivial
redefinitions, with
\begin{equation}
\gamma \beta _{2}^{\left( \mathrm{int}\right) }=0.  \label{ww39}
\end{equation}
We notice that $\hat{\Lambda}_{1}^{\left( \mathrm{int}\right) }$ expressed
by (\ref{ww33}) can be written like in (\ref{ww37}) if
\begin{eqnarray}
\bar{\Gamma} &=&-2\left( k_{A}C_{BC}^{A}+k_{B}k_{C}\right) \left[ \phi
^{*}\phi \eta ^{B\mu }\partial _{\mu }\xi ^{C}+\phi ^{*}\left( \partial
^{\mu }\phi \right) \eta _{\mu }^{B}\xi ^{C}+\right.  \nonumber \\
&&\left. +\frac{1}{2}\phi ^{*}\left( \partial ^{\mu }\phi \right) \eta
^{B\nu }\partial _{[\mu }\eta _{\nu ]}^{C}\right] -\left( \bar{k}_{A}k_{B}-%
\bar{k}_{B}k_{A}\right) \phi ^{*}\xi ^{A}\xi ^{B}-  \nonumber \\
&&-2\left( \bar{k}_{A}C_{BC}^{A}+k_{B}\bar{k}_{C}\right) \phi ^{*}\eta _{\mu
}^{B}\partial ^{\mu }\xi ^{C}  \label{ww40}
\end{eqnarray}
takes the form
\begin{equation}
\bar{\Gamma}=\delta \bar{\psi}+\gamma \bar{\Pi}+\partial _{\mu }\bar{\rho}%
^{\mu },  \label{ww41}
\end{equation}
for some local $\bar{\psi}$, $\bar{\Pi}$, and $\bar{\rho}^{\mu }$. Assuming
that $\bar{\Gamma}$ reads like in (\ref{ww41}) and applying $\delta $ on the
last equation, we get
\begin{equation}
\delta \bar{\Gamma}=\gamma \left( -\delta \bar{\Pi}\right) +\partial _{\mu
}\left( \delta \bar{\rho}^{\mu }\right) .  \label{ww42}
\end{equation}
From the expression of $\bar{\Gamma}$, by direct computation, we arrive at
\begin{eqnarray}
\delta \bar{\Gamma} &=&\partial _{\mu }\left\{ -\left( \bar{k}_{A}k_{B}-\bar{%
k}_{B}k_{A}\right) \xi ^{A}\left( \xi ^{B}\partial ^{\mu }\phi -2\phi
\partial ^{\mu }\xi ^{B}\right) -\right.  \nonumber \\
&&-2\left( \bar{k}_{A}C_{BC}^{A}+k_{B}\bar{k}_{C}\right) \left[ \left(
\partial ^{\mu }\phi \right) \eta _{\nu }^{B}\partial ^{\nu }\xi ^{C}+\phi
\xi ^{B}\partial ^{\mu }\xi ^{C}-\right.  \nonumber \\
&&\left. -\frac{1}{2}\sigma ^{\mu \rho }\phi \left( \partial _{[\rho }\eta
_{\nu ]}^{B}\right) \partial ^{\nu }\xi ^{C}\right] +\left(
k_{A}C_{BC}^{A}+k_{B}k_{C}\right) \left[ -2\phi \left( \partial ^{\mu }\phi
\right) \eta ^{B\nu }\partial _{\nu }\xi ^{C}+\right.  \nonumber \\
&&+\phi ^{2}\left( \frac{1}{2}\sigma ^{\mu \rho }\left( \partial _{[\rho
}\eta _{\lambda ]}^{B}\right) \partial ^{\lambda }\xi ^{C}-\xi ^{B}\partial
^{\mu }\xi ^{C}\right) +  \nonumber \\
&&\left. \left. +2T^{\mu \nu }\eta _{\nu }^{B}\xi ^{C}+T^{\mu \nu }\eta
^{B\rho }\partial _{[\nu }\eta _{\rho ]}^{C}\right] \right\} +  \nonumber \\
&&+\gamma \left\{ 2\left( \bar{k}_{A}k_{B}-\bar{k}_{B}k_{A}\right) \phi \xi
^{A}\mathcal{K}^{B}+2\left( \bar{k}_{A}C_{BC}^{A}+k_{B}\bar{k}_{C}\right)
\times \right.  \nonumber \\
&&\times \left[ \phi \xi ^{B}\mathcal{K}^{C}-\frac{1}{2}\left( \partial
^{\mu }\phi \right) h_{\mu \nu }^{B}\partial ^{\nu }\xi ^{C}-\left( \partial
^{\mu }\phi \right) \eta ^{B\nu }\mathcal{K}_{\mu \nu }^{C}+\right.
\nonumber \\
&&\left. +\frac{1}{2}\phi \left( \partial ^{\mu }h_{\mu \nu }^{B}-\partial
_{\nu }h^{B}\right) \partial ^{\nu }\xi ^{C}\right] +\left(
k_{A}C_{BC}^{A}+k_{B}k_{C}\right) \times  \nonumber \\
&&\times \left[ \phi ^{2}\left( -\xi ^{B}\mathcal{K}^{C}+\frac{1}{2}\left(
\partial ^{\mu }h_{\mu \nu }^{B}-\partial _{\nu }h^{B}\right) \partial ^{\nu
}\xi ^{C}\right) -\phi \partial ^{\mu }\phi \left( h_{\mu \nu }^{B}\partial
^{\nu }\xi ^{C}+\right. \right.  \nonumber \\
&&\left. +\eta ^{B\nu }\mathcal{K}_{\mu \nu }^{C}\right) +T^{\mu \nu }\left(
h_{\mu \nu }^{B}\xi ^{C}-\eta ^{B\rho }\left( \partial _{\mu }h_{\nu \rho
}^{C}-\partial _{\rho }h_{\mu \nu }^{C}\right) \right) -  \nonumber \\
&&\left. \left. -\frac{1}{2}\left( \partial ^{\mu }\phi \right) \left(
\partial ^{\nu }\phi \right) \sigma ^{\rho \lambda }h_{\mu \rho
}^{B}\partial _{[\nu }\eta _{\lambda ]}^{C}\right] \right\} .  \label{ww43}
\end{eqnarray}
Since neither of $\eta _{\mu }^{A}$, $\partial _{\left[ \mu \right. }\eta
_{\left. \nu \right] }^{A}$, $\xi ^{A}$, or $\partial _{\mu }\xi ^{A}$ are $%
\delta $-exact objects, it results that the right-hand side of (\ref{ww43})
can be put in the form of the right-hand side of (\ref{ww42}) if the
relation (\ref{1l}) is fulfilled in the space of local nonintegrated
densities. As it has been shown in the subsection~\ref{secondord}, the
relation (\ref{1l}) cannot hold in this space, so $\bar{\Gamma}$ must vanish
\begin{equation}
\bar{\Gamma}=0.  \label{ww43q}
\end{equation}
This takes place if the constants $k_{A}$ and $\bar{k}_{A}$ are subject to
the equations
\begin{eqnarray}
\bar{k}_{A}k_{B}-\bar{k}_{B}k_{A} &=&0,  \label{ww44} \\
k_{A}C_{BC}^{A}+k_{B}k_{C} &=&0,  \label{ww45} \\
\bar{k}_{A}C_{BC}^{A}+k_{B}\bar{k}_{C} &=&0.  \label{ww46}
\end{eqnarray}
Let us analyze the conditions (\ref{ww44})--(\ref{ww46}). The first one, (%
\ref{ww44}), exhibits two types of solutions, namely
\begin{equation}
\bar{k}_{A}=qk_{A},  \label{ww47}
\end{equation}
and respectively
\begin{equation}
k_{A}=p\bar{k}_{A},  \label{ww48}
\end{equation}
where $p$ and $q$ are some real numbers. Initially, we consider the solution
(\ref{ww47}). If $q\neq 0$, then the equations (\ref{ww46})--(\ref{ww47})
yield
\begin{equation}
\bar{k}_{A}C_{BC}^{A}+\frac{1}{q}\bar{k}_{B}\bar{k}_{C}=0.  \label{ww48x}
\end{equation}
The relation (\ref{ww29}), combined with (\ref{ww45})--(\ref{ww46}) and (\ref
{ww48x}), ensures that
\begin{equation}
k_{B}k_{C}=0,\qquad k_{B}\bar{k}_{C}=0,\qquad \bar{k}_{B}\bar{k}%
_{C}=0,\qquad \mathrm{if}\;B\neq C.  \label{ww48y}
\end{equation}
Whenever the relations (\ref{ww48y}) hold, from (\ref{ww33}), (\ref{ww37}),
and (\ref{ww43q}) it is easy to see that
\begin{eqnarray}
\beta _{2}^{\left( \mathrm{int}\right) } &=&0,  \label{xw48} \\
\beta _{1}^{\left( \mathrm{int}\right) } &=&-\frac{1}{2}k_{A}k_{A}\phi
^{*}\left( \partial ^{\mu }\phi \right) \eta ^{A\nu }h_{\mu \nu }^{A}.
\label{zw48}
\end{eqnarray}
Plugging (\ref{ww29}) and (\ref{ww48y}) into (\ref{ww33x}), we find by
direct calculation that
\begin{eqnarray}
&&\hat{\Lambda}_{0}^{(\mathrm{int})}+2\delta \beta _{1}^{(\mathrm{int}%
)}=\gamma \left\{ k_{A}k_{A}\left[ \frac{1}{4}\left( \partial _{\rho }\phi
\right) \left( \partial ^{\rho }\phi \right) \left( h_{\mu \nu }^{A}h^{A\mu
\nu }-\frac{1}{2}h^{A}h^{A}\right) -\right. \right.  \nonumber \\
&&-\left( \partial ^{\mu }\phi \right) \left( \partial ^{\nu }\phi \right)
\left( \sigma ^{\rho \lambda }h_{\mu \rho }^{A}h_{\nu \lambda }^{A}-\frac{1}{%
2}h^{A}h_{\mu \nu }^{A}\right) +  \nonumber \\
&&+\frac{1}{3}\phi ^{2}\left( \mathcal{R}_{\mu \nu }^{A}h^{A\mu \nu }-\frac{1%
}{4}\mathcal{R}^{A}h^{A}\right) -  \nonumber \\
&&\left. -\frac{1}{6}\sigma ^{\mu \nu }\sigma ^{\rho \lambda }\sigma
^{\alpha \beta }\phi ^{2}\left( \stackrel{(1)}{\Gamma }_{\mu \rho \alpha
}^{A}\stackrel{(1)}{\Gamma }_{\nu \lambda \beta }^{A}-\stackrel{(1)}{\Gamma }%
_{\mu \rho \lambda }^{A}\stackrel{(1)}{\Gamma }_{\nu \alpha \beta
}^{A}\right) \right] +  \nonumber \\
&&+k_{A}\bar{k}_{A}\phi \left[ \frac{2}{3}\left( \mathcal{R}_{\mu \nu
}^{A}h^{A\mu \nu }-\frac{1}{4}\mathcal{R}^{A}h^{A}\right) -\right.  \nonumber
\\
&&\left. -\frac{1}{3}\sigma ^{\mu \nu }\sigma ^{\rho \lambda }\sigma
^{\alpha \beta }\left( \frac{1}{2}\stackrel{(1)}{\Gamma }_{\mu \rho \alpha
}^{A}\stackrel{(1)}{\Gamma }_{\nu \lambda \beta }^{A}-\stackrel{(1)}{\Gamma }%
_{\mu \rho \lambda }^{A}\stackrel{(1)}{\Gamma }_{\nu \alpha \beta
}^{A}\right) \right] +  \nonumber \\
&&+C_{AA}^{A}v\left( \phi \right) \mathcal{W}_{A}^{\mu \alpha \nu \beta
}\left( 8\mathcal{K}_{\mu \nu }^{A}h_{\alpha \beta }^{A}+4\sigma ^{\rho
\lambda }\stackrel{(1)}{\Gamma }_{\rho \mu \nu }^{A}\stackrel{(1)}{\Gamma }%
_{\lambda \alpha \beta }^{A}+\right.  \nonumber \\
&&\left. +4\mathcal{W}_{\mu \rho \nu \beta }^{A}h_{\alpha }^{A\rho }-\frac{1%
}{2}\mathcal{W}_{\mu \alpha \nu \beta }^{A}h^{A}\right) +k_{A}\left[ G\left(
\phi \right) h^{A}-\right.  \nonumber \\
&&-2F\left( \phi \right) \left( \left( \partial ^{\mu }\phi \right) \partial
^{\nu }\phi -\frac{1}{2}\sigma ^{\mu \nu }\left( \partial _{\rho }\phi
\right) \partial ^{\rho }\phi \right) h_{\mu \nu }^{A}+  \nonumber \\
&&\left. \left. +2F\left( \phi \right) \phi ^{2}\mathcal{K}^{A}\right]
+k_{A}J\left( \phi \right) \left( \partial _{\rho }\phi \right) \left(
\partial ^{\rho }\phi \right) \left( \partial _{\mu }\phi \right) \left(
\partial _{\nu }\phi \right) \left( \sigma ^{\mu \nu }h^{A}-4h^{A\mu \nu
}\right) \right\} +  \nonumber \\
&&+2\left\{ \left[ -\bar{k}_{A}\bar{k}_{A}\mathcal{K}^{A}+k_{A}\left( \frac{%
dG\left( \phi \right) }{d\phi }\phi -4G\left( \phi \right) \right) +\right.
\right.  \nonumber \\
&&+\left( \frac{dF\left( \phi \right) }{d\phi }\left( \partial ^{\rho }\phi
\right) \partial _{\rho }\phi +\frac{1}{2}\mathcal{W}_{B}^{\mu \alpha \nu
\beta }\mathcal{W}_{\mu \alpha \nu \beta }^{B}\frac{dv\left( \phi \right) }{%
d\phi }\right) \left( \bar{k}_{A}+k_{A}\phi \right) +  \nonumber \\
&&\left. +\bar{k}_{A}\frac{dG\left( \phi \right) }{d\phi }\right] \xi
^{A}+\left( \partial ^{\rho }\phi \right) \left( 2\bar{k}_{A}F\left( \phi
\right) -k_{A}\frac{dF\left( \phi \right) }{d\phi }\phi ^{2}\right) \partial
_{\rho }\xi ^{A}-  \nonumber \\
&&\left. -\frac{1}{2}\mathcal{W}_{A}^{\mu \alpha \nu \beta }\mathcal{W}_{\mu
\alpha \nu \beta }^{A}\frac{dv\left( \phi \right) }{d\phi }\left( \partial
^{\rho }\phi \right) \left( C_{AA}^{A}\eta _{\rho }^{A}+k_{C}\eta _{\rho
}^{C}\right) \right\} +  \nonumber \\
&&+2\left[ \left( \partial _{\mu }\phi \right) \left( \partial ^{\mu }\phi
\right) \right] ^{2}\xi ^{A}\left[ k_{A}\left( \frac{dJ\left( \phi \right) }{%
d\phi }\phi +4J\left( \phi \right) \right) +\bar{k}_{A}\frac{dJ\left( \phi
\right) }{d\phi }\right] +  \nonumber \\
&&+8\left( \partial _{\mu }\phi \right) \left( \partial ^{\mu }\phi \right)
\left( \partial _{\nu }\phi \right) \left( \partial ^{\nu }\xi ^{A}\right)
J\left( \phi \right) \left( \bar{k}_{A}+k_{A}\phi \right) +\partial _{\mu
}q_{0}^{\mu }.  \label{yw48}
\end{eqnarray}
Comparing now (\ref{yw48}) with (\ref{ww38}), we observe that the
consistency of the first-order deformation requires that
\begin{eqnarray}
\hat{\Theta} &=&2\left\{ \left[ -\bar{k}_{A}\bar{k}_{A}\mathcal{K}%
^{A}+k_{A}\left( \frac{dG\left( \phi \right) }{d\phi }\phi -4G\left( \phi
\right) \right) +\right. \right.  \nonumber \\
&&+\left( \frac{dF\left( \phi \right) }{d\phi }\left( \partial ^{\rho }\phi
\right) \partial _{\rho }\phi +\frac{1}{2}\mathcal{W}_{B}^{\mu \alpha \nu
\beta }\mathcal{W}_{\mu \alpha \nu \beta }^{B}\frac{dv\left( \phi \right) }{%
d\phi }\right) \left( \bar{k}_{A}+k_{A}\phi \right) +  \nonumber \\
&&\left. +\bar{k}_{A}\frac{dG\left( \phi \right) }{d\phi }\right] \xi
^{A}+\left( \partial ^{\rho }\phi \right) \left( 2\bar{k}_{A}F\left( \phi
\right) -k_{A}\frac{dF\left( \phi \right) }{d\phi }\phi ^{2}\right) \partial
_{\rho }\xi ^{A}-  \nonumber \\
&&\left. -\frac{1}{2}\mathcal{W}_{A}^{\mu \alpha \nu \beta }\mathcal{W}_{\mu
\alpha \nu \beta }^{A}\frac{dv\left( \phi \right) }{d\phi }\left( \partial
^{\rho }\phi \right) \left( C_{AA}^{A}\eta _{\rho }^{A}+k_{C}\eta _{\rho
}^{C}\right) \right\} +  \nonumber \\
&&+2\left[ \left( \partial _{\mu }\phi \right) \left( \partial ^{\mu }\phi
\right) \right] ^{2}\xi ^{A}\left[ k_{A}\left( \frac{dJ\left( \phi \right) }{%
d\phi }\phi +4J\left( \phi \right) \right) +\bar{k}_{A}\frac{dJ\left( \phi
\right) }{d\phi }\right] +  \nonumber \\
&&+8\left( \partial _{\mu }\phi \right) \left( \partial ^{\mu }\phi \right)
\left( \partial _{\nu }\phi \right) \left( \partial ^{\nu }\xi ^{A}\right)
J\left( \phi \right) \left( \bar{k}_{A}+k_{A}\phi \right) .  \label{yyy48}
\end{eqnarray}
must be of the form
\begin{equation}
\hat{\Theta}=\gamma \hat{\theta}+\partial _{\mu }\hat{\chi}^{\mu },
\label{yzy48}
\end{equation}
for some local $\hat{\theta}$ and $\hat{\chi}^{\mu }$. Assume that (\ref
{yzy48}) holds. This implies the equation
\begin{equation}
\gamma \hat{\Theta}=\partial _{\mu }\left( \gamma \hat{\chi}^{\mu }\right) .
\label{yyz48}
\end{equation}
Acting with $\gamma $ on the formula (\ref{yyy48}), we are led to
\begin{equation}
\gamma \hat{\Theta}=\partial _{\mu }\left( 2\bar{k}_{A}\bar{k}_{A}\xi
^{A}\partial ^{\mu }\xi ^{A}\right) .  \label{zyy48}
\end{equation}
Comparing (\ref{yyz48}) with (\ref{zyy48}), we get that
\begin{equation}
2\bar{k}_{A}\bar{k}_{A}\xi ^{A}\partial ^{\mu }\xi ^{A}=\gamma \hat{\chi}%
^{\mu }+\partial _{\nu }\hat{\lambda}^{\nu \mu },  \label{uvx}
\end{equation}
where $\hat{\lambda}^{\nu \mu }=-\hat{\lambda}^{\mu \nu }$. The relation (%
\ref{uvx}) indicates that the equations (\ref{yyz48}) and (\ref{zyy48}) are
compatible. On behalf of the formula (\ref{wzq5}) where we make the changes
\begin{equation}
\xi \rightarrow \xi ^{A},\qquad \eta _{\nu }\rightarrow \eta _{\nu
}^{A},\qquad h\rightarrow h^{A},\qquad \mathcal{K}\rightarrow \mathcal{K}%
^{A},  \label{vux}
\end{equation}
we find that the equations (\ref{yyz48})--(\ref{zyy48}) are compatible if
\begin{equation}
\bar{k}_{A}=0.  \label{xyy48}
\end{equation}
Under these circumstances, the consistency of the first-order deformation
implies that the quantity
\begin{eqnarray}
\hat{\Theta}^{\prime } &=&2\left\{ \left[ k_{A}\left( \frac{dG\left( \phi
\right) }{d\phi }\phi -4G\left( \phi \right) \right) +\right. \right.
\nonumber \\
&&\left. +\left( \frac{dF\left( \phi \right) }{d\phi }\left( \partial ^{\rho
}\phi \right) \partial _{\rho }\phi +\frac{1}{2}\mathcal{W}_{B}^{\mu \alpha
\nu \beta }\mathcal{W}_{\mu \alpha \nu \beta }^{B}\frac{dv\left( \phi
\right) }{d\phi }\right) k_{A}\phi \right] \xi ^{A}+  \nonumber \\
&&+\left( \partial ^{\rho }\phi \right) \left( 2\bar{k}_{A}F\left( \phi
\right) -k_{A}\frac{dF\left( \phi \right) }{d\phi }\phi ^{2}\right) \partial
_{\rho }\xi ^{A}-  \nonumber \\
&&\left. -\frac{1}{2}\mathcal{W}_{A}^{\mu \alpha \nu \beta }\mathcal{W}_{\mu
\alpha \nu \beta }^{A}\frac{dv\left( \phi \right) }{d\phi }\left( \partial
^{\rho }\phi \right) \left( C_{AA}^{A}\eta _{\rho }^{A}+k_{C}\eta _{\rho
}^{C}\right) \right\} +  \nonumber \\
&&+2k_{A}\left[ \left( \partial _{\mu }\phi \right) \left( \partial ^{\mu
}\phi \right) \right] ^{2}\xi ^{A}\left( \frac{dJ\left( \phi \right) }{d\phi
}\phi +4J\left( \phi \right) \right) +  \nonumber \\
&&+8k_{A}\left( \partial _{\mu }\phi \right) \left( \partial ^{\mu }\phi
\right) \left( \partial _{\nu }\phi \right) \left( \partial ^{\nu }\xi
^{A}\right) \phi J\left( \phi \right)  \label{cba}
\end{eqnarray}
must read as
\begin{equation}
\hat{\Theta}^{\prime }=\gamma \hat{\theta}^{\prime }+\partial _{\mu }\hat{%
\chi}^{\prime \mu }.  \label{acb}
\end{equation}
Using again (\ref{wyq5}) adapted in agreement with (\ref{vux}), we infer
that $\hat{\Theta}^{\prime }$ given in (\ref{cba}) may be expressed like in (%
\ref{acb}) if the functions $J\left( \phi \right) $, $F\left( \phi \right) $%
, $v\left( \phi \right) $, and $G\left( \phi \right) $ read like in (\ref
{new1}), (\ref{qw1}), (\ref{aab}), and respectively (\ref{egal}). In this
case it follows that
\begin{equation}
\hat{\Theta}^{\prime }=0,  \label{abb}
\end{equation}
which further yields
\begin{eqnarray}
\beta _{0}^{\left( \mathrm{int}\right) } &=&-\frac{1}{2}k_{A}k_{A}\left[
\frac{1}{4}\left( \partial _{\rho }\phi \right) \left( \partial ^{\rho }\phi
\right) \left( h_{\mu \nu }^{A}h^{A\mu \nu }-\frac{1}{2}h^{A}h^{A}\right)
-\right.  \nonumber \\
&&-\left( \partial ^{\mu }\phi \right) \left( \partial ^{\nu }\phi \right)
\left( \sigma ^{\rho \lambda }h_{\mu \rho }^{A}h_{\nu \lambda }^{A}-\frac{1}{%
2}h^{A}h_{\mu \nu }^{A}\right) +  \nonumber \\
&&+\frac{1}{3}\phi ^{2}\left( \mathcal{R}_{\mu \nu }^{A}h^{A\mu \nu }-\frac{1%
}{4}\mathcal{R}^{A}h^{A}\right) -  \nonumber \\
&&\left. -\frac{1}{6}\sigma ^{\mu \nu }\sigma ^{\rho \lambda }\sigma
^{\alpha \beta }\phi ^{2}\left( \stackrel{(1)}{\Gamma }_{\mu \rho \alpha
}^{A}\stackrel{(1)}{\Gamma }_{\nu \lambda \beta }^{A}-\stackrel{(1)}{\Gamma }%
_{\mu \rho \lambda }^{A}\stackrel{(1)}{\Gamma }_{\nu \alpha \beta
}^{A}\right) \right] -  \nonumber \\
&&-\frac{1}{2}\kappa k_{A}\phi ^{4}h^{A}.  \label{yyx48}
\end{eqnarray}

If $q=0$, from (\ref{ww47}) we infer that $\bar{k}_{A}=0$, such that the
interaction terms will be parametrized only by the constants $k_{A}$.
Consequently, the relations (\ref{ww29}) and (\ref{ww45}) produce the
equations
\begin{equation}
k_{B}k_{C}=0\qquad \mathrm{if}\;B\neq C.  \label{ww48z}
\end{equation}
In this situation we have that $\beta _{2}^{\left( \mathrm{int}\right) }$
and $\beta _{1}^{\left( \mathrm{int}\right) }$ are still expressed by the
formulas (\ref{xw48})--(\ref{zw48}). By particularizing (\ref{yw48}) to the
case $\bar{k}_{A}=0$, we find that the consistency of the first-order
deformation requires that $\hat{\Theta}^{\prime }$ from (\ref{cba}) must
satisfy again the relation (\ref{acb}), so we obtain that $\beta
_{0}^{\left( \mathrm{int}\right) }$ still reads like in (\ref{yyx48}). The
relations (\ref{xw48})--(\ref{zw48}) and (\ref{yyx48}) clearly emphasize
that there are no indirect couplings between different Weyl gravitons via a
scalar field, irrespective of the value of $q$.\ Consider now the solution (%
\ref{ww48}). If $p\neq 0$, from (\ref{ww29}), (\ref{ww45})--(\ref{ww46}),
and (\ref{ww48}), we obtain again the relations (\ref{ww48y}), which
automatically lead to the second-order deformation given by the formulas (%
\ref{xw48})--(\ref{zw48}) and (\ref{yyx48}). In consequence, in the case $%
p\neq 0$ there are no indirect cross-couplings among different Weyl
gravitons, so only the case $p=0$ remains to be discussed below. If $p=0$,
then the equations (\ref{ww46}) and (\ref{ww48}) take the form
\begin{equation}
\bar{k}_{A}C_{BC}^{A}=0,  \label{ww51}
\end{equation}
and respectively
\begin{equation}
k_{A}=0.  \label{ww52}
\end{equation}
In this situation, with the help of (\ref{ww33}) and (\ref{ww37}), we deduce
that
\begin{equation}
\beta _{2}^{\left( \mathrm{int}\right) }=0,\qquad \beta _{1}^{\left( \mathrm{%
int}\right) }=0.  \label{ww53}
\end{equation}
Inserting (\ref{ww51})--(\ref{ww52}) in (\ref{ww33x}), we have that
\begin{eqnarray}
\hat{\Lambda}_{0}^{\left( \mathrm{int}\right) } &=&\gamma \left[
C_{AA}^{A}v\left( \phi \right) \mathcal{W}_{A}^{\mu \alpha \nu \beta }\left(
8\mathcal{K}_{\mu \nu }^{A}h_{\alpha \beta }^{A}+4\sigma ^{\rho \lambda }%
\stackrel{(1)}{\Gamma }_{\rho \mu \nu }^{A}\stackrel{(1)}{\Gamma }_{\lambda
\alpha \beta }^{A}+\right. \right.  \nonumber \\
&&\left. \left. +4\mathcal{W}_{\mu \rho \nu \beta }^{A}h_{\alpha }^{A\rho }-%
\frac{1}{2}\mathcal{W}_{\mu \alpha \nu \beta }^{A}h^{A}\right) \right] -
\nonumber \\
&&-2\bar{k}_{A}\left\{ \left[ \bar{k}_{B}\mathcal{K}^{B}-\left( \frac{dF}{%
d\phi }\left( \partial _{\rho }\phi \right) \partial ^{\rho }\phi +\frac{dG}{%
d\phi }\right) \right] \xi ^{A}-\right.  \nonumber \\
&&-2F\left( \phi \right) \left( \partial _{\rho }\phi \right) \partial
^{\rho }\xi ^{A}-\frac{1}{2}\mathcal{W}_{B}^{\mu \alpha \nu \beta }\mathcal{W%
}_{\mu \alpha \nu \beta }^{B}\frac{dv\left( \phi \right) }{d\phi }\xi ^{A}+
\nonumber \\
&&+8\left( \partial _{\mu }\phi \right) \left( \partial ^{\mu }\phi \right)
\left( \partial _{\nu }\phi \right) \left( \partial ^{\nu }\xi ^{A}\right)
J\left( \phi \right) +  \nonumber \\
&&\left. +2\left[ \left( \partial _{\mu }\phi \right) \left( \partial ^{\mu
}\phi \right) \right] ^{2}\xi ^{A}\frac{dJ\left( \phi \right) }{d\phi }%
\right\} -  \nonumber \\
&&-\mathcal{W}_{A}^{\mu \alpha \nu \beta }\mathcal{W}_{\mu \alpha \nu \beta
}^{A}\frac{dv\left( \phi \right) }{d\phi }\left( \partial ^{\rho }\phi
\right) C_{AA}^{A}\eta _{\rho }^{A}  \label{ww54}
\end{eqnarray}
Taking into account the second relation in (\ref{ww53}), we observe that the
equation (\ref{ww38}) requires that $\hat{\Lambda}_{0}^{\left( \mathrm{int}%
\right) }$ given by (\ref{ww54}) should be $\gamma $-exact modulo $d$%
\begin{equation}
\hat{\Lambda}_{0}^{\left( \mathrm{int}\right) }=-2\gamma \beta _{0}^{\left(
\mathrm{int}\right) }+\partial _{\mu }q_{0}^{\mu }.  \label{ww55}
\end{equation}
This takes place if
\begin{eqnarray}
\omega &=&-2\bar{k}_{A}\left\{ \left[ \bar{k}_{B}\mathcal{K}^{B}-\left(
\frac{dF}{d\phi }\left( \partial _{\rho }\phi \right) \partial ^{\rho }\phi +%
\frac{dG}{d\phi }\right) \right] \xi ^{A}-\right.  \nonumber \\
&&-2F\left( \phi \right) \left( \partial _{\rho }\phi \right) \partial
^{\rho }\xi ^{A}-\frac{1}{2}\mathcal{W}_{B}^{\mu \alpha \nu \beta }\mathcal{W%
}_{\mu \alpha \nu \beta }^{B}\frac{dv\left( \phi \right) }{d\phi }\xi ^{A}+
\nonumber \\
&&+8\left( \partial _{\mu }\phi \right) \left( \partial ^{\mu }\phi \right)
\left( \partial _{\nu }\phi \right) \left( \partial ^{\nu }\xi ^{A}\right)
J\left( \phi \right) +  \nonumber \\
&&\left. +2\left[ \left( \partial _{\mu }\phi \right) \left( \partial ^{\mu
}\phi \right) \right] ^{2}\xi ^{A}\frac{dJ\left( \phi \right) }{d\phi }%
\right\} -  \nonumber \\
&&-\mathcal{W}_{A}^{\mu \alpha \nu \beta }\mathcal{W}_{\mu \alpha \nu \beta
}^{A}\frac{dv\left( \phi \right) }{d\phi }\left( \partial ^{\rho }\phi
\right) C_{AA}^{A}\eta _{\rho }^{A},  \label{wwx55}
\end{eqnarray}
allows to be put in the form
\begin{equation}
\omega =\gamma \pi +\partial _{\mu }q_{0}^{\mu }.  \label{wwy55}
\end{equation}
Acting with $\gamma $ on the last equation, we get that
\begin{equation}
\gamma \omega =\partial _{\mu }\left( \gamma q_{0}^{\mu }\right) .
\label{ww56}
\end{equation}
Applying $\gamma $ on (\ref{wwx55}), we infer that
\begin{equation}
\gamma \omega =\partial _{\mu }\left( 2\bar{k}_{A}\bar{k}_{B}\xi
^{A}\partial ^{\mu }\xi ^{B}\right) .  \label{ww57}
\end{equation}
From (\ref{ww56}) and (\ref{ww57}) we arrive at
\begin{equation}
2\bar{k}_{A}\bar{k}_{B}\xi ^{A}\partial ^{\mu }\xi ^{B}=\gamma q_{0}^{\mu
}+\partial _{\nu }\varkappa ^{\nu \mu },  \label{wwx57}
\end{equation}
with $\varkappa ^{\nu \mu }=-\varkappa ^{\mu \nu }$. The formula (\ref{wwx57}%
) states nothing but the compatibility between (\ref{ww56}) and (\ref{ww57}%
). Simple computations lead to
\begin{eqnarray}
2\xi ^{A}\partial ^{\mu }\xi ^{B} &=&\gamma \left[ -\frac{1}{2}\left( \frac{1%
}{2}h^{A}\partial ^{\mu }\xi ^{B}+\eta _{\nu }^{A}\mathcal{K}^{B\mu \nu
}\right) \right] +  \nonumber \\
&&+\partial _{\nu }\left( -\frac{1}{4}\eta ^{A\left[ \nu \right. }\partial
^{\left. \mu \right] }\xi ^{B}\right) +\partial _{\nu }\left( -\frac{1}{4}%
\eta ^{A\left( \nu \right. }\partial ^{\left. \mu \right) }\xi ^{B}\right) .
\label{wwz57}
\end{eqnarray}
The presence of the term $\partial _{\nu }\left( -\frac{1}{4}\eta ^{A\left(
\nu \right. }\partial ^{\left. \mu \right) }\xi ^{B}\right) $ in the
right-hand side of (\ref{wwz57}) indicates that the equations (\ref{ww56})
and (\ref{ww57}) are compatible if
\begin{equation}
\bar{k}_{A}=0.  \label{ww58}
\end{equation}
Inserting (\ref{ww58}) in (\ref{wwx55}), we find that $\hat{\Lambda}%
_{0}^{\left( \mathrm{int}\right) }$ can be expressed like in (\ref{ww55}) if
the quantity
\begin{equation}
\omega ^{\prime }=-\mathcal{W}_{A}^{\mu \alpha \nu \beta }\mathcal{W}_{\mu
\alpha \nu \beta }^{A}\frac{dv\left( \phi \right) }{d\phi }\left( \partial
^{\rho }\phi \right) C_{AA}^{A}\eta _{\rho }^{A},  \label{wwz58}
\end{equation}
is of the form
\begin{equation}
\omega ^{\prime }=\gamma \pi ^{\prime }+\partial _{\mu }q_{0}^{\mu }.
\label{wwy58}
\end{equation}
It is clear that $\omega ^{\prime }$ given in (\ref{wwz58}) cannot be
written like in (\ref{wwy58}), so $\omega ^{\prime }$ must vanish. Its
vanishing then implies that the function $v\left( \phi \right) $ must be
constant. Reprising the same arguments like in the subsection \ref{secondord}%
, we can take $v\left( \phi \right) $ to vanish and thus (\ref{ww54}) and (%
\ref{ww58}) produce
\begin{equation}
\beta _{0}^{\left( \mathrm{int}\right) }=0.  \label{ww59}
\end{equation}
Consequently, in the case $p=0$ we have that
\begin{eqnarray}
\hat{S}_{1}^{\left( \mathrm{int}\right) } &=&\int d^{4}x\left( J\left( \phi
\right) \left( \partial _{\mu }\phi \right) \left( \partial ^{\mu }\phi
\right) \left( \partial _{\rho }\phi \right) \partial ^{\rho }\phi +\right.
\nonumber \\
&&\left. +F\left( \phi \right) \left( \partial _{\rho }\phi \right) \partial
^{\rho }\phi +G\left( \phi \right) \right) ,  \label{wwx60}
\end{eqnarray}
and
\begin{equation}
\hat{S}_{2}^{\left( \mathrm{int}\right) }=\cdots =\hat{S}_{k}^{\left( \mathrm{%
int}\right) }=\cdots =0,  \label{final}
\end{equation}
where $J\left( \phi \right) $, $F\left( \phi \right) $ and $G\left( \phi
\right) $ are now arbitrary functions of the undifferentiated scalar field.
It is simple to see that the term $\hat{S}_{1}^{\left( \mathrm{int}\right) }$
given in (\ref{wwx60}) does not describe interactions between a scalar field
and Weyl gravitons. Thus, the case $p=0$ corresponds to the more restrictive
situation where there are no interactions at all between the Weyl gravitons
and the scalar field, and the same holds for the indirect cross-couplings
among different Weyl gravitons.

In conclusion, the result announced in the beginning of this section has
been completely proved.

\section{Conclusion}

To conclude with, in this paper we have investigated the indirect couplings
between a collection of Weyl gravitons (described in the free limit by a sum
of linearized Weyl actions) in the presence of a scalar field by using the
powerful setting based on local BRST cohomology. Initially, we have obtained
the couplings between a single Weyl graviton and a scalar field, and then we
have proved, under the hypotheses of locality, smoothness of the
interactions in the coupling constant, Poincar\'{e} invariance, (background)
Lorentz invariance, and the preservation of the number of derivatives on
each field, together with the supplementary assumption that the internal
metric defined by the sum of Weyl lagrangians is positively defined, that
there are no consistent cross-interactions among different Weyl gravitons in
the presence of a scalar field.

\section*{Acknowledgment}

Two of the authors (C.B. and E.M.C.) are partially supported by the European
Commission FP6 program MRTN-CT-2004-005104 and by the type A grant 305/2004
with the Romanian National Council for Academic Scientific Research
(C.N.C.S.I.S.) and the Romanian Ministry of Education and Research (M.E.C.).
One of the authors (A.C.L.) was supported by the World Federation of
Scientists (WFS) National Scholarship Programme.

\appendix

\section{Solution to the equation $\gamma _{0}\bar{a}_{0}^{\prime \prime
\left( \mathrm{int}\right) }=\partial _{\mu }\bar{m}_{0}^{\left( \mathrm{int}%
\right) \mu }$\label{apphomog}}

In order to solve the equation
\begin{equation}
\gamma _{0}\bar{a}_{0}^{\prime \prime \left( \mathrm{int}\right) }=\partial
_{\mu }\bar{m}_{0}^{\left( \mathrm{int}\right) \mu },  \label{xww60}
\end{equation}
we start from the requirement that $\bar{a}_{0}^{\prime \prime \left(
\mathrm{int}\right) }$ may contain at most four derivatives. Then, $\bar{a}%
_{0}^{\prime \prime \left( \mathrm{int}\right) }$ can be decomposed like
\begin{equation}
\bar{a}_{0}^{\prime \prime \left( \mathrm{int}\right) }=\omega _{0}+\omega
_{1}+\omega _{2}+\omega _{3}+\omega _{4},  \label{ww60}
\end{equation}
where $\left( \omega _{i}\right) _{i=\overline{0,4}}$ contains $i$
derivatives. Since there is no Lorentz scalar that can be constructed out of
the first-order derivatives of the fields $h_{\mu \nu }$ and $\phi $, it
results that
\begin{equation}
\omega _{1}=0,\;\omega _{3}=0.  \label{ww61}
\end{equation}
Due to the different number of derivatives in the components $\omega _{0}$, $%
\omega _{2}$, and $\omega _{4}$, the equation (\ref{xww60}) leads to three
independent equations
\begin{eqnarray}
\gamma _{0}\omega _{0} &=&\partial _{\mu }j_{0}^{\mu },  \label{ww62} \\
\gamma _{0}\omega _{2} &=&\partial _{\mu }j_{2}^{\mu },  \label{ww63} \\
\gamma _{0}\omega _{4} &=&\partial _{\mu }j_{4}^{\mu }.  \label{wwxy}
\end{eqnarray}
Since $\omega _{0}$ is derivative-free, it can be represented as
\begin{equation}
\omega _{0}\left( \phi ,h_{\mu \nu }\right) =f_{1}\left( \phi \right)
f_{2}\left( h_{\mu \nu }\right) .  \label{ww64}
\end{equation}
Inserting (\ref{ww64}) in (\ref{ww62}), we obtain that
\begin{equation}
f_{1}\left( \phi \right) \frac{\partial f_{2}\left( h_{\mu \nu }\right) }{%
\partial h_{\mu \nu }}\partial _{(\mu }\eta _{\nu )}=\partial _{\mu
}j_{0}^{\mu }.  \label{ww65}
\end{equation}
The left-hand side of the above relation can be written like a full
divergence if
\begin{equation}
\partial _{\mu }\left( f_{1}\left( \phi \right) \frac{\partial f_{2}\left(
h_{\mu \nu }\right) }{\partial h_{\mu \nu }}\right) =0,  \label{ww66}
\end{equation}
which implies
\begin{equation}
f_{1}\left( \phi \right) \frac{\partial f_{2}\left( h_{\mu \nu }\right) }{%
\partial h_{\mu \nu }}=c^{\mu \nu },  \label{ww67}
\end{equation}
with $c^{\mu \nu }$ some arbitrary, symmetric constants. The only constants
with this property are
\begin{equation}
c^{\mu \nu }=c\sigma ^{\mu \nu },  \label{ww68}
\end{equation}
with $c$ an arbitrary, real constant. Accordingly, the equation (\ref{ww67})
leads to
\begin{equation}
f_{1}\left( \phi \right) =c^{\prime },\qquad f_{2}\left( h_{\mu \nu }\right)
=c^{\prime \prime }\sigma ^{\mu \nu }h_{\mu \nu },  \label{ww69}
\end{equation}
with $c^{\prime }$ and $c^{\prime \prime }$ two constants related by $%
c=c^{\prime }c^{\prime \prime }$. The solution (\ref{ww69}) provides no
interactions between the scalar field and the Weyl graviton, but merely
reduces to the cosmological term
\begin{equation}
\omega _{0}\left( \phi ,h_{\mu \nu }\right) =c\sigma ^{\mu \nu }h_{\mu \nu },
\label{ww70}
\end{equation}
so we can set $c=0$, and thus $\omega _{0}\left( \phi ,h_{\mu \nu }\right)
=0 $. Let us analyze now the equation (\ref{ww63}). If we use the notation
\begin{equation}
D^{\mu \nu }=\frac{\delta \omega _{2}}{\delta h_{\mu \nu }},  \label{ww70a}
\end{equation}
then we get that
\begin{equation}
\gamma _{0}\omega _{2}=-2\left( \partial _{\mu }D^{\mu \nu }\right) \eta
_{\nu }+\partial _{\mu }u^{\mu },  \label{ww71}
\end{equation}
with $u^{\mu }$ a local current. The relation (\ref{ww71}) expresses the
fact that $\omega _{2}$ is solution to (\ref{ww63}) if
\begin{equation}
\partial _{\mu }D^{\mu \nu }=0.  \label{ww72}
\end{equation}
The solution to the last equation reads as
\begin{equation}
D^{\mu \nu }=\partial _{\alpha }\partial _{\beta }U^{\mu \alpha \nu \beta },
\label{ww73}
\end{equation}
where $U^{\mu \alpha \nu \beta }$ displays the symmetry properties of the
Riemann tensor and involves only the undifferentiated fields $\phi $ and $%
h_{\mu \nu }$.

Let $N$ be a derivation in the algebra of the fields $h_{\mu \nu }$ and of
their derivatives that counts the powers of the fields and their
derivatives, defined by
\begin{equation}
N=\sum\limits_{k\geq 0}\left( \partial _{\mu _{1}\cdots \mu _{k}}h_{\mu \nu
}\right) \frac{\partial }{\partial \left( \partial _{\mu _{1}\cdots \mu
_{k}}h_{\mu \nu }\right) }.  \label{ww74}
\end{equation}
Then, it is easy to see that for every nonintegrated density $\chi $, we
have that
\begin{equation}
N\chi =h_{\mu \nu }\frac{\delta \chi }{\delta h_{\mu \nu }}+\partial _{\mu
}s^{\mu },  \label{ww75}
\end{equation}
where $\delta \chi /\delta h_{\mu \nu }$ denotes the variational derivative
of $\chi $. If $\chi ^{\left( l\right) }$ is a homogeneous polynomial of
order $l>0$ in the fields and their derivatives, then
\begin{equation}
N\chi ^{\left( l\right) }=l\chi ^{\left( l\right) }.  \label{ww76}
\end{equation}
Using (\ref{ww70a}), (\ref{ww73}), and (\ref{ww75}), we find that
\begin{equation}
N\omega _{2}=-\frac{1}{2}\mathcal{R}_{\mu \alpha \nu \beta }U^{\mu \alpha
\nu \beta }+\partial _{\mu }v^{\mu }.  \label{ww76a}
\end{equation}
We expand $\omega _{2}$ like
\begin{equation}
\omega _{2}=\sum\limits_{l>0}\omega _{2}^{\left( l\right) },  \label{ww77}
\end{equation}
where $N\omega _{2}^{\left( l\right) }=l\omega _{2}^{\left( l\right) }$,
such that
\begin{equation}
N\omega _{2}=\sum\limits_{l>0}l\omega _{2}^{\left( l\right) }.  \label{ww78}
\end{equation}
Comparing (\ref{ww76a}) with (\ref{ww78}), we reach the conclusion that the
decomposition (\ref{ww77}) induces a similar decomposition with respect to $%
U^{\mu \alpha \nu \beta }$, i.e.
\begin{equation}
U^{\mu \alpha \nu \beta }=\sum\limits_{l>0}U_{\left( l-1\right) }^{\mu
\alpha \nu \beta }.  \label{ww79}
\end{equation}
Substituting (\ref{ww79}) into (\ref{ww76a}) and comparing the resulting
expression with (\ref{ww78}), we obtain that
\begin{equation}
\omega _{2}^{\left( l\right) }=-\frac{1}{2l}\mathcal{R}_{\mu \alpha \nu
\beta }U_{\left( l-1\right) }^{\mu \alpha \nu \beta }+\partial _{\mu }\bar{v}%
_{(l)}^{\mu }.  \label{prform}
\end{equation}
Introducing (\ref{prform}) in (\ref{ww77}), we arrive at
\begin{equation}
\omega _{2}=-\frac{1}{2}\mathcal{R}_{\mu \alpha \nu \beta }\bar{U}^{\mu
\alpha \nu \beta }+\partial _{\mu }\bar{v}^{\mu },  \label{ww81}
\end{equation}
where
\begin{equation}
\bar{U}^{\mu \alpha \nu \beta }=\sum\limits_{l>0}\frac{1}{l}U_{\left(
l-1\right) }^{\mu \alpha \nu \beta }.  \label{ww82}
\end{equation}
On behalf of (\ref{ww81}), we find that
\begin{equation}
\gamma _{0}\omega _{2}=\eta _{\lambda }\partial _{\rho }\left( \mathcal{R}%
_{\mu \alpha \nu \beta }\frac{\partial \bar{U}^{\mu \alpha \nu \beta }}{%
\partial h_{\rho \lambda }}\right) +\partial _{\mu }j^{\mu }.  \label{ww83}
\end{equation}
The last relation shows that $\omega _{2}$ satisfies the equation (\ref{ww63}%
) if
\begin{equation}
\partial _{\rho }\left( \mathcal{R}_{\mu \alpha \nu \beta }\frac{\partial
\bar{U}^{\mu \alpha \nu \beta }}{\partial h_{\rho \lambda }}\right) =0.
\label{ec}
\end{equation}
Taking into account the fact that $\mathcal{R}_{\mu \alpha \nu \beta
}\partial \bar{U}^{\mu \alpha \nu \beta }/\partial h_{\rho \lambda }$ is
symmetric with respect to the indices $\rho $ and $\lambda $ and
second-order in the derivatives, we obtain that
\begin{equation}
\mathcal{R}_{\mu \alpha \nu \beta }\frac{\partial \bar{U}^{\mu \alpha \nu
\beta }}{\partial h_{\rho \lambda }}=\partial _{\gamma }\partial _{\delta
}V^{\rho \gamma \lambda \delta },  \label{ww84}
\end{equation}
where the functions $V^{\rho \gamma \lambda \delta }$\ exhibit the symmetry
properties of the Riemann tensor and depend only on the undifferentiated
fields $\phi $ and $h_{\mu \nu }$. By computing the left-hand side of (\ref
{ww84}), we arrive at
\begin{eqnarray}
\mathcal{R}_{\mu \alpha \nu \beta }\frac{\partial \bar{U}^{\mu \alpha \nu
\beta }}{\partial h_{\rho \lambda }} &=&\partial _{\gamma }\partial _{\delta
}\left( 2h_{\alpha \nu }\frac{\partial \bar{U}^{\gamma \alpha \nu \delta }}{%
\partial h_{\rho \lambda }}\right) -\frac{1}{2}\frac{\partial ^{2}\bar{U}%
^{\mu \alpha \nu \beta }}{\partial h_{\rho \lambda }\partial h_{\gamma
\delta }}\times  \nonumber \\
&&\times \left( \partial _{[\nu }h_{\beta ][\mu }\partial _{\alpha
]}h_{\gamma \delta }+\partial _{[\mu }h_{\alpha ][\nu }\partial _{\beta
]}h_{\gamma \delta }-\right.  \nonumber \\
&&\left. -h_{\nu [\mu }\partial _{\alpha ]}\partial _{\beta }h_{\gamma
\delta }+h_{\beta [\mu }\partial _{\alpha ]}\partial _{\nu }h_{\gamma \delta
}\right) -  \nonumber \\
&&-\frac{1}{2}\frac{\partial ^{2}\bar{U}^{\mu \alpha \nu \beta }}{\partial
h_{\rho \lambda }\partial \phi }\left( \partial _{[\nu }h_{\beta ][\mu
}\partial _{\alpha ]}\phi +\partial _{[\mu }h_{\alpha ][\nu }\partial
_{\beta ]}\phi -\right.  \nonumber \\
&&\left. -h_{\nu [\mu }\partial _{\alpha ]}\partial _{\beta }\phi +h_{\beta
[\mu }\partial _{\alpha ]}\partial _{\nu }\phi \right) -  \nonumber \\
&&-\frac{1}{2}\frac{\partial ^{3}\bar{U}^{\mu \alpha \nu \beta }}{\partial
h_{\rho \lambda }\partial h_{\gamma \delta }\partial h_{\varepsilon \omega }}%
h_{\gamma \delta ,[\nu }h_{\beta ][\mu }\partial _{\alpha ]}h_{\varepsilon
\omega }-  \nonumber \\
&&-\frac{1}{2}\frac{\partial ^{3}\bar{U}^{\mu \alpha \nu \beta }}{\partial
h_{\rho \lambda }\partial \phi \partial \phi }\phi _{,[\nu }h_{\beta ][\mu
}\partial _{\alpha ]}\phi -\frac{1}{2}\frac{\partial ^{3}\bar{U}^{\mu \alpha
\nu \beta }}{\partial h_{\rho \lambda }\partial h_{\gamma \delta }\partial
\phi }\times  \nonumber \\
&&\times \left( h_{\gamma \delta ,[\nu }h_{\beta ][\mu }\partial _{\alpha
]}\phi +\phi _{,[\nu }h_{\beta ][\mu }\partial _{\alpha ]}h_{\gamma \delta
}\right) ,  \label{ww85}
\end{eqnarray}
where we made the notations
\begin{equation}
\phi _{,\nu }=\partial _{\nu }\phi ,\qquad h_{\gamma \delta ,\nu }=\partial
_{\nu }h_{\gamma \delta }.  \label{ww86}
\end{equation}
The right-hand side of (\ref{ww85}) can be expressed like in the right-hand
side of (\ref{ww84}) if
\begin{equation}
\frac{\partial ^{2}\bar{U}^{\mu \alpha \nu \beta }}{\partial h_{\gamma
\delta }\partial h_{\rho \lambda }}=0,\qquad \frac{\partial ^{2}\bar{U}^{\mu
\alpha \nu \beta }}{\partial \phi \partial h_{\rho \lambda }}=0,
\label{ww87}
\end{equation}
whose general solution reads as
\begin{equation}
\bar{U}^{\mu \alpha \nu \beta }=-\sigma ^{\nu [\mu }\sigma ^{\alpha ]\beta
}f\left( \phi \right) +\tilde{c}\left( \sigma ^{\nu [\mu }h^{\alpha ]\beta
}-\sigma ^{\beta [\mu }h^{\alpha ]\nu }-\sigma ^{\nu [\mu }\sigma ^{\alpha
]\beta }h\right) ,  \label{u}
\end{equation}
with $f\left( \phi \right) $ an arbitrary, smooth function of the
undifferentiated scalar field and$\;\tilde{c}$ an arbitrary, real
constant. Substituting (\ref{u}) in (\ref{ww81}), we deduce that, up
to a total derivative, $\omega _{2}$ can be written like
\begin{equation}
\omega _{2}=\mathcal{R}f\left( \phi \right) -2\tilde{c}\left( \mathcal{R}%
_{\mu \nu }-\frac{1}{2}\sigma _{\mu \nu }\mathcal{R}\right) h^{\mu \nu }.
\label{sol}
\end{equation}
We remark that the terms proportional with $\tilde{c}$ from (\ref{sol}) are
not interacting terms, and therefore we will omit them by taking $\tilde{c}%
=0 $, such that
\begin{equation}
\omega _{2}=\mathcal{R}f\left( \phi \right) .  \label{ww88}
\end{equation}
In the end we analyze the equation (\ref{wwxy}). By means of the notation
\begin{equation}
E^{\mu \nu }=\frac{\delta \omega _{4}}{\delta h_{\mu \nu }},  \label{qp1}
\end{equation}
we find that
\begin{equation}
\gamma _{0}\omega _{4}=-2\left( \partial _{\mu }E^{\mu \nu }\right) \eta
_{\nu }+\partial _{\mu }\bar{u}^{\mu },  \label{qp2}
\end{equation}
with $\bar{u}^{\mu }$ a local current. With the help of (\ref{qp2}), we
observe that $\omega _{4}$ satisfies the equation (\ref{wwxy}) if
\begin{equation}
\partial _{\mu }E^{\mu \nu }=0.  \label{qp3}
\end{equation}
The solution to the last equation reads as
\begin{equation}
E^{\mu \nu }=\partial _{\alpha }\partial _{\beta }H^{\mu \alpha \nu \beta },
\label{qp4}
\end{equation}
where $H^{\mu \alpha \nu \beta }$ depends on $\phi $ and $h_{\mu \nu }$,
exhibits the symmetries of the Riemann tensor, and contains just two
derivatives. Acting similarly to $\omega _{2}$, we arrive at
\begin{equation}
\omega _{4}=-\frac{1}{2}\mathcal{R}_{\mu \alpha \nu \beta }\tilde{H}^{\mu
\alpha \nu \beta }+\partial _{\mu }\tilde{u}^{\mu },  \label{qp5}
\end{equation}
where
\begin{equation}
\tilde{H}^{\mu \alpha \nu \beta }=\sum_{l>0}\frac{1}{l}H_{\left( l-1\right)
}^{\mu \alpha \nu \beta }.  \label{qp6}
\end{equation}
It is clear that $\tilde{H}^{\mu \alpha \nu \beta }$ presents the symmetry
of the Riemann tensor and includes only two derivatives. In order to
preserve the differential order of the scalar field equation, it results
that $\tilde{H}^{\mu \alpha \nu \beta }$ must be of the form
\begin{equation}
\tilde{H}^{\mu \alpha \nu \beta }=\bar{f}\left( \phi \right) \bar{H}^{\mu
\alpha \nu \beta },  \label{qp7}
\end{equation}
where $\bar{H}^{\mu \alpha \nu \beta }$ depends only on $h_{\mu \nu }$, has
the symmetry properties of the Riemann tensor, and has precisely two
derivatives. Substituting (\ref{qp7}) into (\ref{qp5}), we deduce that
\begin{equation}
\omega _{4}=-\frac{1}{2}\bar{f}\left( \phi \right) \mathcal{R}_{\mu \alpha
\nu \beta }\bar{H}^{\mu \alpha \nu \beta }+\partial _{\mu }\tilde{u}^{\mu }.
\label{qp7a}
\end{equation}
Acting with $\gamma _0$ on (\ref{qp7a}), we obtain
\begin{eqnarray}
\gamma _{0}\omega _{4} &=&\eta ^{\rho }\partial ^{\tau }\left[ \bar{f}\left(
\phi \right) \mathcal{R}_{\mu \alpha \nu \beta }\frac{\partial \bar{H}^{\mu
\alpha \nu \beta }}{\partial h^{\tau \rho }}-\right.  \nonumber \\
&&-\partial ^{\lambda }\left( \bar{f}\left( \phi \right) \mathcal{R}_{\mu
\alpha \nu \beta }\frac{\partial \bar{H}^{\mu \alpha \nu \beta }}{\partial
\left( \partial ^{\lambda }h^{\tau \rho }\right) }\right) +  \nonumber \\
&&\left. +\partial ^{\lambda }\partial ^{\sigma }\left( \bar{f}\left( \phi
\right) \mathcal{R}_{\mu \alpha \nu \beta }\frac{\partial \bar{H}^{\mu
\alpha \nu \beta }}{\partial \left( \partial ^{\lambda }\partial ^{\sigma
}h^{\tau \rho }\right) }\right) \right] +\partial _{\mu }\bar{j}^{\mu }.
\label{qp8}
\end{eqnarray}
The last formula shows that $\omega _{4}$ given in (\ref{qp7a}) satisfies
the equation (\ref{wwxy}) if
\begin{eqnarray}
&&\partial ^{\tau }\left[ \bar{f}\left( \phi \right) \mathcal{R}_{\mu \alpha
\nu \beta }\frac{\partial \bar{H}^{\mu \alpha \nu \beta }}{\partial h^{\tau
\rho }}-\partial ^{\lambda }\left( \bar{f}\left( \phi \right) \mathcal{R}%
_{\mu \alpha \nu \beta }\frac{\partial \bar{H}^{\mu \alpha \nu \beta }}{%
\partial \left( \partial ^{\lambda }h^{\tau \rho }\right) }\right) +\right.
\nonumber \\
&&\left. +\partial ^{\lambda }\partial ^{\sigma }\left( \bar{f}\left( \phi
\right) \mathcal{R}_{\mu \alpha \nu \beta }\frac{\partial \bar{H}^{\mu
\alpha \nu \beta }}{\partial \left( \partial ^{\lambda }\partial ^{\sigma
}h^{\tau \rho }\right) }\right) \right] =0.  \label{qp9}
\end{eqnarray}
The equation (\ref{qp9}) further leads to
\begin{eqnarray}
&&\bar{f}\left( \phi \right) \mathcal{R}_{\mu \alpha \nu \beta }\frac{%
\partial \bar{H}^{\mu \alpha \nu \beta }}{\partial h^{\tau \rho }}-\partial
^{\lambda }\left( \bar{f}\left( \phi \right) \mathcal{R}_{\mu \alpha \nu
\beta }\frac{\partial \bar{H}^{\mu \alpha \nu \beta }}{\partial \left(
\partial ^{\lambda }h^{\tau \rho }\right) }\right) +  \nonumber \\
&&+\partial ^{\lambda }\partial ^{\sigma }\left( \bar{f}\left( \phi \right)
\mathcal{R}_{\mu \alpha \nu \beta }\frac{\partial \bar{H}^{\mu \alpha \nu
\beta }}{\partial \left( \partial ^{\lambda }\partial ^{\sigma }h^{\tau \rho
}\right) }\right) =\partial ^{\lambda }\partial ^{\sigma }\bar{V}_{\lambda
\tau \sigma \rho },  \label{qp10}
\end{eqnarray}
where $\bar{V}_{\lambda \tau \sigma \rho }$ depends on $\phi $ and $h_{\mu
\nu }$, presents the symmetry of $\mathcal{R}_{\lambda \tau \sigma \rho }$,
and has two derivatives. On the other hand, the most general form of $\bar{H}%
^{\mu \alpha \nu \beta }$ can be represented like
\begin{eqnarray}
\bar{H}^{\mu \alpha \nu \beta } &=&a^{\gamma \delta \delta ^{\prime }\xi
\chi \chi ^{\prime }\mu \alpha \nu \beta }\left( h_{\mu \nu }\right) \left(
\partial _{\gamma }h_{\delta \delta ^{\prime }}\right) \partial _{\xi
}h_{\chi \chi ^{\prime }}+  \nonumber \\
&&+b^{\gamma \delta \xi \xi ^{\prime }\mu \alpha \nu \beta }\left( h_{\mu
\nu }\right) \partial _{\gamma }\partial _{\delta }h_{\xi \xi ^{\prime }},
\label{qp11}
\end{eqnarray}
where the functions $a^{\gamma \delta \delta ^{\prime }\xi \chi \chi
^{\prime }\mu \alpha \nu \beta }\left( h_{\mu \nu }\right) $ and $b^{\gamma
\delta \xi \xi ^{\prime }\mu \alpha \nu \beta }\left( h_{\mu \nu }\right) $
have appropriate symmetry properties. Inserting (\ref{qp11}) in (\ref{qp10}%
), we infer that the left-hand side of the latter reads as $\partial
^{\lambda }\partial ^{\sigma }\bar{V}_{\lambda \tau \sigma \rho }$ if
\begin{eqnarray}
a^{\gamma \delta \delta ^{\prime }\xi \chi \chi ^{\prime }\mu \alpha \nu
\beta }\left( h_{\mu \nu }\right) &=&0,  \label{qp12} \\
b^{\gamma \delta \xi \xi ^{\prime }\mu \alpha \nu \beta }\left( h_{\mu \nu
}\right) &=&C^{\gamma \delta \xi \xi ^{\prime }\mu \alpha \nu \beta },
\label{qp13}
\end{eqnarray}
where $C^{\gamma \delta \xi \xi ^{\prime }\mu \alpha \nu \beta }$ are some
constants, and, moreover, the quantity $\frac{\partial \bar{H}^{\mu \alpha
\nu \beta }}{\partial \left( \partial ^{\lambda }\partial ^{\sigma }h^{\tau
\rho }\right) }$ inherits the symmetries of $\mathcal{R}_{\lambda \tau
\sigma \rho }$. Based on the last considerations, it is simple to see that
\begin{eqnarray}
\bar{H}^{\mu \alpha \nu \beta } &=&\bar{c}_{1}\mathcal{R}^{\mu \alpha \nu
\beta }+\bar{c}_{2}\left( \sigma ^{\mu \nu }\mathcal{R}^{\alpha \beta
}+\sigma ^{\alpha \beta }\mathcal{R}^{\mu \nu }-\sigma ^{\alpha \nu }%
\mathcal{R}^{\mu \beta }-\sigma ^{\mu \beta }\mathcal{R}^{\alpha \nu
}\right) +  \nonumber \\
&&+\bar{c}_{3}\left( \sigma ^{\mu \beta }\sigma ^{\alpha \nu }-\sigma ^{\mu
\nu }\sigma ^{\alpha \beta }\right) \mathcal{R},  \label{qp14}
\end{eqnarray}
where $\bar{c}_{1}$, $\bar{c}_{2}$, and $\bar{c}_{3}$ are some real
constants. Replacing (\ref{qp14}) in (\ref{qp7a}), we get, up to a total
derivative, that $\omega _{4}$ takes the form
\begin{equation}
\omega _{4}=\left( c_{1}\mathcal{R}_{\mu \alpha \nu \beta }\mathcal{R}^{\mu
\alpha \nu \beta }+c_{2}\mathcal{R}_{\mu \nu }\mathcal{R}^{\mu \nu }+c_{3}%
\mathcal{R}^{2}\right) \bar{f}\left( \phi \right) ,  \label{qp15}
\end{equation}
with $c_{1}=-\frac{1}{2}\bar{c}_{1}$, $c_{2}=-2\bar{c}_{2}$, and $c_{3}=\bar{%
c}_{3}$. By employing (\ref{ww88}) and (\ref{qp15}), we then find that the
solution to the equation (\ref{xww60}) becomes
\begin{equation}
\bar{a}_{0}^{\prime \prime \left( \mathrm{int}\right) }=\left( c_{1}\mathcal{R}%
_{\mu \alpha \nu \beta }\mathcal{R}^{\mu \alpha \nu \beta }+c_{2}\mathcal{R}%
_{\mu \nu }\mathcal{R}^{\mu \nu }+c_{3}\mathcal{R}^{2}\right) \bar{f}\left(
\phi \right) +\mathcal{R}f\left( \phi \right) ,  \label{qp16}
\end{equation}
as stated in the subsection~\ref{firstord}.

\end{document}